\documentclass[a4paper,10pt]{elsarticle}
\usepackage{jcomp}
\usepackage{framed,multirow}
\usepackage[utf8]{inputenc}
\usepackage{bm}
\usepackage{amsmath}
\usepackage{mathrsfs}
\usepackage{graphicx}
\usepackage{epsfig, setspace}
\usepackage{url}
\usepackage{graphicx, transparent, color}
\usepackage{algorithm} 
\usepackage{algorithmicx}
\usepackage{etex}
\usepackage[bookmarks=true]{hyperref}
\usepackage{amsmath}
\usepackage{xcolor, soul}
\usepackage{rotating} % Rotating the figures but keeps the page in portrait mode
\usepackage{pdflscape} % Rotates single page into Landscape mode
\usepackage{silence}
\usepackage{ulem,cancel}
\newtheorem{remark}{\bf Remark}[section]
\newtheorem{example}{Example}[section]
\usepackage{graphicx}
\usepackage{float}
\usepackage{subfigure}% subcaptions for subfigures
\usepackage{subfigmat}% matrices of similar subfigures,

\usepackage{tikz}
\usepackage{capt-of}
\usepackage{setspace,lipsum}
\usepackage{lineno}
\newcommand{\half}{\frac{1}{2}}
\usepackage{setspace,lipsum}
\usepackage{listings}
\usepackage{xcolor}
\definecolor{aquamarine}{rgb}{0.5, 1.0, 0.83}
\definecolor{OliveGreen}{rgb}{0,0.6,0}

\sethlcolor{aquamarine}

\definecolor{codegreen}{rgb}{0,0.6,0}
\definecolor{codegray}{rgb}{0.5,0.5,0.5}
\definecolor{codepurple}{rgb}{0.58,0,0.82}
\definecolor{backcolour}{rgb}{0.95,0.95,0.92}

\lstdefinestyle{mystyle}{
    backgroundcolor=\color{backcolour},   
    commentstyle=\color{codegreen},
    keywordstyle=\color{magenta},
    numberstyle=\tiny\color{codegray},
    stringstyle=\color{codepurple},
    basicstyle=\ttfamily\footnotesize,
    breakatwhitespace=false,         
    breaklines=true,                 
    captionpos=b,                    
    keepspaces=true,                 
    numbers=left,                    
    numbersep=5pt,                  
    showspaces=false,                
    showstringspaces=false,
    showtabs=false,                  
    tabsize=2
}
\begin{document}

\begin{frontmatter}
\title{Consistent  Interface Capturing Adaptive Reconstruction Approach for Viscous Compressible  Multicomponent Flows}

\author[AA_address]{Amareshwara Sainadh Chamarthi \cortext[cor1]{Corresponding author. \\ 
E-mail address: sainath@caltech.edu (Amareshwara Sainadh  Ch.).}}
\address[AA_address]{Division of Engineering and Applied Science, California Institute of Technology, Pasadena, CA, USA}

\begin{abstract}
The paper proposes a physically consistent numerical discretization approach for simulating viscous compressible multicomponent flows. It has two main contributions. First, a contact discontinuity (and material interface) detector is developed. In those regions of contact discontinuities, the THINC (Tangent of Hyperbola for INterface Capturing) approach is used for reconstructing appropriate variables (phasic densities). For other flow regions, the variables are reconstructed using the Monotonicity-preserving (MP) scheme (or  Weighted essentially non-oscillatory scheme (WENO)). For reconstruction in the characteristic space, the THINC approach is used only for the contact (or entropy) wave and volume fractions and for the reconstruction of primitive variables, the THINC approach is used for phasic densities and volume fractions only, offering an effective solution for reducing dissipation errors near contact discontinuities. The second contribution is the development of an algorithm that uses a central reconstruction scheme for the tangential velocities, as they are continuous across material interfaces in viscous flows. In this regard, the Ducros sensor (a shock detector that cannot detect material interfaces) is employed to compute the tangential velocities using a central scheme across material interfaces. Using the central scheme does not produce any oscillations at the material interface. The proposed approach is thoroughly validated with several benchmark test cases for compressible multicomponent flows, highlighting its advantages. The numerical results of the benchmark tests show that the proposed method captured the material interface sharply compared to existing methods.
\end{abstract}

\begin{keyword}
THINC, Multicomponent flows, Interface-capturing, Tangential velocities, Viscous flows.
\end{keyword}

\end{frontmatter}
\section{Introduction}\label{sec:intro}
Compressible flows can manifest two primary types of discontinuities: shocks and contact discontinuities, the latter often representing the boundaries between different materials or phases. Density, pressure, and normal velocity are discontinuous across shocks, and only the density and volume fractions are discontinuous across a material interface \cite{hirschvol2}. On the other hand, while tangential velocities are continuous in viscous flow simulations \cite{batchelor1967introduction}, they are discontinuous across contact discontinuities in inviscid simulations \cite{hirschvol2}. Addressing these discontinuities with one approach has become common in numerical simulations. Various methodologies have been devised to obtain oscillation-free results for these discontinuities, including Weighted Essentially Non-Oscillatory (WENO) schemes \cite{Jiang1995,coralic2014finite, Hu2010}, the Monotonicity Preserving (MP) approach \cite{suresh1997accurate,chamarthi2021high,chamarthi2023gradient}, and localized artificial dissipation methods \cite{kawai2010assessment,kawai2011high}. This study's focus is treating these discontinuities with different approaches rather than a single approach while considering the physics of these discontinuities.

Even for smooth initial data, in contrast with problems with discontinuous initial conditions, the flow may develop sudden changes in density, which may lead to oscillations. During simulations, these discontinuities will result in spurious oscillations, which can be mitigated by introducing additional dissipation. The shock/contact-capturing techniques typically used for the simulation of compressible flows identify and characterize all types of waves, including those inherently discontinuous \cite{saurel2018diffuse}. Of the many approaches proposed in the literature to overcome oscillations near discontinuities are the WENO schemes proposed by Liu et al. \cite{liu1994weighted} and significantly improved by Jiang and Shu \cite{Jiang1995}. WENO schemes are widely used to capture discontinuities \cite{Hu2011,adams1996high,Liu2015,Balsara2016} and material interfaces \cite{Wong2017,Nonomura2017,Nonomura2012}. Apart from WENO schemes, other shock-capturing approaches have also been extensively studied in the literature. The Monotonicity-Preserving (MP) scheme proposed by Suresh and Hyunh can also effectively suppress numerical oscillations across discontinuities \cite{suresh1997accurate}. The MP approach is also used in conjunction with the WENO scheme to enhance its robustness \cite{Balsara2000a}. It is well-known from the literature that contact discontinuities are captured with excessive numerical dissipation than the shocks using these schemes.\\

Several approaches are proposed in the literature to reduce the numerical dissipation around contact discontinuities. Shukla et al. \cite{shukla2010interface} have proposed an interface compression approach to minimise numerical diffusion. Another proposed approach in the literature is to improve the limiters typically employed to capture the shocks and remove oscillations. Chiapolino et al. \cite{chiapolino2017sharpening} have proposed an Overbee limiter that captured the interfaces sharply compared to the Superbee limiter. A different approach to reducing the numerical diffusion is Harten's artificial compression method (ACM) for single-component flows \cite{harten1977artificial}. Yang further improved the ACM approach \cite{yang1990artificial}, and the improved version was used in conjunction with the WENO scheme by Balsara and Shu \cite{Balsara2000a}. The ACM approach improved the steepness of the linearly degenerate characteristic fields by adding an extra piecewise profile. He et al. \cite{he2017characteristic} used the ACM approach in multicomponent flows to capture the material interfaces sharply. Harten \cite{harten1989eno} also developed a subcell resolution approach by modifying the linearly degenerate characteristic field to improve the resolution of contact discontinuities. Huynh developed a slope steepening approach for single-component flows that reduced the numerical dissipation around the contact discontinuities \cite{huynh1995accurate}. The steepening approach is used only for the contact/entropy wave. Huynh's approach to detecting contact discontinuities is based on the wave strengths of characteristic waves with ad-hoc parameters. Harten's and Huynh's approach to modifying the linearly degenerate fields is limited to one-dimensional single-species cases.

On the other hand, Shyue and Xiao \cite{shyue2014eulerian} proposed an interface-sharpening approach for multiphase flows using the THINC scheme. The key idea of their interface sharpening approach is to replace the standard shock-capturing schemes with the THINC scheme (a non-polynomial reconstruction approach). Shyue and Xiao \cite{shyue2014eulerian} applied THINC in any computational cell containing the material interface, which is defined as any cell satisfying the condition $\epsilon$ $<$ $\alpha_i$ $\leq$ 1 - $\epsilon$,  where $\epsilon$ = ${10^{-5}}$ and $\alpha$ is the volume fraction of the fluid, and a monotonicity constraint ($\alpha_{i+1}$ - $\alpha_i$)($\alpha_i$ - $\alpha_{i-1}$) $>$ 0. They developed a homogeneous-equilibrium-consistent reconstruction scheme for interface sharpening. Inspired by their work, Garrick et al. \cite{garrick2017interface} improved the interface capturing approach by using the THINC only for the volume fraction and phasic densities, and the other variables are reconstructed by using the standard MUSCL or WENO schemes. Zhang et al. \cite{zhang2024hybrid} further improved the approach of Garrick et al. by using the THINC scheme subjected to the same criterion $\epsilon$ $<$ $\alpha_i$ $\leq$ 1 - $\epsilon$ and ($\alpha_{i+1}$ - $\alpha_i$)($\alpha_i$ - $\alpha_{i-1}$) $>$ 0 and away from the interfaces the WENOIS approach is used. The main drawback of this approach in these studies is that the detection criterion checks for the material interfaces using volume fractions. This type of checking can become tedious if there are several species/gases in the domain, and it will also fail to detect the contact discontinuity within a material (jump in density within the material, for example, the compressible triple point test case considered in this study). Recently, Sun et al. \cite{sun2016boundary} proposed a Boundary Variation Diminishing algorithm (BVD) approach. The key idea behind the approach is to choose the reconstruction algorithm with minimum jump or dissipation at the interfaces. The variables of interest are computed with two different reconstruction schemes, and the scheme with minimum total boundary variation is eventually employed for reconstruction. The BVD approach typically uses the THINC scheme as one of the reconstruction procedures and, therefore, can capture sharply the contact discontinuities \cite{deng2019fifth}. The BVD approach was further extended in simulating multicomponent flows and can capture the material interface with minimum dissipation compared with the WENO scheme \cite{deng2018high}. Chamarthi and Frankel also reported similar observations using the THINC approach (see Appendix B of \cite{chamarthi2021high}). Furthermore, Takagi et al. \cite{takagi2022novel} has developed a TENO-based discontinuity sensor such that all the discontinuities are computed with the THINC scheme as opposed to that of Garrick et al. \cite{garrick2017interface}, who computed only phasic densities and volume fractions using THINC. \textbf{A significant flaw in Takagi et al.'s approach is that the proposed discontinuity sensor detects both shocks and contact discontinuities, and the THINC scheme is applied in those regions. However, across the contact discontinuities, the pressure and velocity are continuous, and applying THINC (a discontinuity capturing and interface sharpening approach) could lead to the failure of the simulation. Concerned TENO-THINC approach is also applied for multiphase flows \cite{li2024high}.} The key takeaways from the above discussion are:\\

\begin{enumerate}
	\item The critical observation here is that every scheme has its strengths and disadvantages, and employing them in the flow regions that are more suitable to them can reduce the overall numerical dissipation \cite{shyue2014eulerian,garrick2017interface,zhang2024hybrid,deng2018high}.
\item Researchers have exploited the characteristic wave structure of the
Euler equations and the reconstruction of contact waves are often used to improve the 
resolution of the contact discontinuities %
\cite{harten1989eno,huynh1995accurate}.
	\item Material interfaces are identified based on the volume fractions, but such an approach might not be able to detect a contact discontinuity within a material. For test cases with several species, it can be computationally intensive to check for all the material interfaces \cite{shyue2014eulerian,garrick2017interface,zhang2024hybrid}.
	\item While some researchers applied THINC for only phasic densities and volume fractions, others have applied THINC to all the discontinuities \cite{takagi2022novel}. It needs to be better understood whether the THINC scheme can be used for all the discontinuities, as it is well known that the standard shock-capturing schemes can resolve the shocks sharply within a few cells. The contact discontinuities are the ones that are typically smeared, as discussed by Harten \cite{harten1989eno}, and require improvement. Furthermore, If a discontinuity sensor detects shocks and contact discontinuities, and the THINC scheme is applied to both of them, then it could lead to the failure of the simulation or lead to inconsistent results. The pressure and velocity are continuous across the contact discontinuities and discontinuous across shocks. Pressure and velocity may not be reconstructed using the THINC scheme across a contact discontinuity.
	\item Majority of the simulations conducted in the literature are \textit{inviscid}. An algorithm that can account for the viscous effects needs to be developed.
\end{enumerate}

Therefore, it is desirable to develop an adaptive interface capturing scheme to employ schemes suitable to the flow physics. Based on the above discussion, the objectives of this study are:

\begin{itemize}
\item Develop an approach to detect contact discontinuities (that do not necessarily depend on volume fractions) and reconstruct those regions using the THINC scheme to minimise the numerical dissipation.
 \item Apply THINC to the appropriate variables according to the physics of the equations being solved. The approach should not lead to erroneous results when there are no discontinuities.
 \item Develop an algorithm to reconstruct tangential velocities using a central scheme across a material interface in viscous flow simulations.
\end{itemize}

This work is also related to and can be considered an extension of the wave-appropriate discontinuity detector approach presented in Ref. \cite{chamarthi2023wave} to multicomponent flows. In \cite{chamarthi2023wave}, In that study, Chamarthi et al. used the Ducros sensor \cite{ducros1999large}, a shock sensor for acoustic and shear waves and a density-based sensor for the entropy wave to simulate compressible turbulent flows with shocks. It is well known that the Ducros sensor cannot detect contact discontinuities, and the wave-appropriate sensor overcame the Ducros sensor's deficiency. Using only a density-based sensor can detect shocks and contact discontinuities, but it would be too dissipative for turbulent flows. Using two different discontinuity sensors depending on the wave structure significantly improved the numerical results. In \cite{chamarthi2023wave}, the \textit{reconstruction approach} does not change based on the waves, and only one type of reconstruction approach is used. However, the present study has two ``different reconstruction'' approaches one for the shockwaves and the other for the material interfaces and contact discontinuities. The approach in \cite{chamarthi2023wave} was further improved in \cite{hoffmann2024centralized} and could predict transition to turbulence in hypersonic flows in practical geometries. Ref. \cite{chamarthi2023wave} was titled \textit{``wave appropriate discontinuity sensor''} and similarly, the present methodology can be called  \textit{``wave appropriate reconstruction approach''}, as either MP/WENO or THINC are used, depending on the characteristics of the wave structure. In the algorithm presented below, the reconstruction scheme choice depends on the Euler equation's characteristic waves. Likewise, if the primitive variables are reconstructed, then the appropriate physics will again be considered. The objective of Ref. \cite{chamarthi2023wave} was different from those of the present study; Ref. \cite{chamarthi2023wave} addressed the simulation of single-species hypersonic turbulent flows, and the present study targets the simulation of high-speed multicomponent flows with material interfaces. \textbf{The present approach does not consider mass transfer, reactive flows, gas-liquid flows and phase change, which will have additional physical constraints and require different algorithms.}%\\

The rest of the paper is organised as follows. The governing equations of the compressible multicomponent flows are present in Section \ref{sec:eqns-gov}. Details of the numerical discretisation of the equations, including the details of the physically consistent novel adaptive interface capturing scheme, are described in Section \ref{sec:num}. Several one- and two-dimensional test cases for compressible multicomponent flows are presented in Section \ref{sec:results}, and Section \ref{sec:conclusions} summarises the findings.

\section{Governing equations}\label{sec:eqns-gov}

In this study, the viscous compressible multi-component flows as described by the quasi-conservative five equation model of Allaire et al. \cite{allaire2002five} including viscous effects \cite{coralic2014finite} are considered. A system consisting of two fluids has two continuity equations, one momentum and one energy equation. In addition, an advection equation for the volume fraction of one of the two fluids is considered. The governing equations are:
\begin{equation}\label{5eqn-base}
\frac{\partial \mathbf{Q}}{\partial t}+\frac{\partial \mathbf{F^c}}{\partial x}+\frac{\partial \mathbf{G^c}}{\partial y}+\frac{\partial \mathbf{F^v}}{\partial x}+\frac{\partial \mathbf{G^v}}{\partial y}=\mathbf{S},
\end{equation}
where the state vector, flux vectors and source term, $\mathbf{S}$, are given by:
\begin{equation}
\mathbf{Q}=\left[\begin{array}{c}
\alpha_{1} \rho_{1} \\
\alpha_{2} \rho_{2} \\
\rho u \\
\rho v \\
E \\
\alpha_{1}
\end{array}\right], \quad \mathbf{F^c}=\left[\begin{array}{c}
\alpha_{1} \rho_{1} u \\
\alpha_{2} \rho_{2} u \\
\rho u^{2}+p \\
\rho v u \\
(E+p) u \\
\alpha_{1} u
\end{array}\right], \quad \mathbf{G^c}=\left[\begin{array}{c}
\alpha_{1} \rho_{1} v \\
\alpha_{2} \rho_{2} v \\
\rho u v \\
\rho v^{2}+p \\
(E+p) v \\
\alpha_{1} v
\end{array}\right],  \quad \mathbf{S}=\left[\begin{array}{c}
0 \\
0 \\
0 \\
0 \\
0 \\
\alpha_{1} \nabla \cdot \mathbf{u}
\end{array}\right],
\end{equation}
\textcolor{black}{
\begin{equation}
\mathbf{F^v}=\left[\begin{array}{c}
0 \\
0 \\
-\tau_{x x} \\
-\tau_{y x} \\
-\tau_{x x} u-\tau_{x y} v\\
0 \\
\end{array}\right],
\mathbf{G^v}=\left[\begin{array}{c}
0 \\
0 \\
-\tau_{x y} \\
-\tau_{y y} \\
-\tau_{y x} u-\tau_{y y} v\\
0 \\
\end{array}\right],
\end{equation}}
where $\rho_1$ and $\rho_2$ correspond to the densities of fluids $1$ and $2$, $\alpha_{1}$ and $\alpha_{2}$ are the volume fractions of the fluids $1$ and $2$, $\rho$, $u$,$v$, $p$ and $E$ are the density, $x-$ and $y-$ velocity components, pressure, total energy per unit volume of the mixture, respectively. When employing a diffuse interface approach for mathematical modelling, the five-equation model becomes incomplete near the material interface, as the fluids exist in a mixed state. A set of mixture rules for various fluid properties must be established to close the model. These rules include the mixture rules for the volume fractions of the two fluids, denoted as $\alpha_1$ and $\alpha_2$, the density, and the mixture rules for the ratio of specific heats ($\gamma$) of the mixture. The mixture rules are as follows:
\begin{equation}
\alpha_{2}=1-\alpha_{1},
\end{equation}
\begin{equation}
\rho=\rho_{1} \alpha_{1}+\rho_{2} \alpha_{2},
\end{equation}
\begin{equation}
\frac{1}{\gamma-1}=\frac{\alpha_{1}}{\gamma_{1}-1}+\frac{\alpha_{2}}{{\gamma}_{2}-1},
\end{equation}
where $\gamma_1$ and $\gamma_2$ are the specific heat ratios of fluids $1$ and $2$, respectively. Under the isobaric assumption, the equation of state used to close the system is as follows:
\begin{equation}\label{eqn:pressure}
p = (\gamma -1) (E - \rho \frac{(u^2+v^2+w^2)}{2}).
\end{equation}
The viscous terms are given by:
\begin{equation}\label{eqn:5-stress}
\tau_{x x}=\frac{2}{3} \mu\left(2 \frac{\partial u}{\partial x}-\frac{\partial v}{\partial y}\right), \quad \tau_{x y}=\tau_{y x}=\mu\left(\frac{\partial u}{\partial y}+\frac{\partial v}{\partial x}\right), \quad {\color{black}\tau_{y y}=\frac{2}{3} \mu\left(2 \frac{\partial v}{\partial y}-\frac{\partial u}{\partial x}\right).}
\end{equation}

\section{Numerical discretization}\label{sec:num}
Equations (\ref{5eqn-base}) are discretized using a finite volume method on a uniform Cartesian grid of cell sizes $\Delta x$ and $\Delta y$ in the $x-$ and $y-$ directions, respectively. The conservative variables $\mathbf{ Q}$ are stored at the center of the cell $I_{i, j}$ and the indices $i$ and $j$ denote the $i-$th cell in $x-$ direction and $j-$th cell in $y-$ direction. The time evolution of the cell-centered conservative variables $\mathbf{Q}_{i, j}$ is given by the following semi-discrete equation:\\
%The final model is discretized using the finite volume method on a uniform  Cartesian
%grid. The resulting semi-discrete form of the equations is given by:

\begin{equation}\label{eqn-differencing_residual}
\begin{aligned}
\frac{d \mathbf{Q}_{i, j}}{d t} & =-\left[\frac{\left(\mathbf {\hat{F}^c}_{i+ \frac{1}{2}, j}-\mathbf {\hat{F}^c}_{i- \frac{1}{2}, j}\right) - \left(\mathbf {\hat{F}^v}_{i+ \frac{1}{2}, j}-\mathbf {\hat{F}^v}_{i-\frac{1}{2}, j}\right)}{\Delta x}\right]-\left[\frac{\left(\mathbf {\hat{G}^c}_{i,j+\frac{1}{2}}-\mathbf {\hat{G}^c}_{i,j-\frac{1}{2}}\right) - \left(\mathbf {\hat{G}^v}_{i, j+ \frac{1}{2}}-\mathbf {\hat{G}^v}_{i,j-\frac{1}{2}}\right)}{\Delta y}\right]+\mathbf{S}_{i, j} \\
& = \mathbf{Res}\left(\mathbf{ Q}_{i, j}\right),
\end{aligned}
\end{equation}
where $\mathbf {\hat{F}^c}$, $\mathbf {\hat{G}^c}$ and $\mathbf {\hat{F}^{\textcolor{black}{v}}}$, $\mathbf {\hat{G}^{\textcolor{black}{v}}}$ are the numerical approximations of the convective and viscous fluxes in the $x-$, and $y-$ directions, respectively, at the cell interfaces, $i\pm\frac{1}{2}$ and $j\pm\frac{1}{2}$. $\mathbf{Res}\left(\mathbf{ Q}_{i, j}\right)$ is the residual function. The conserved variables are then integrated in time using the following third order strong stability-preserving (SSP) Runge-Kutta scheme\cite{Jiang1995}:
\begin{eqnarray}\label{rk}
\mathbf{ Q}_{i, j}^{(1)}&=&\mathbf{ Q}_{i, j}^{n}+\Delta t \ \mathbf{Res}\left(\mathbf{ Q}_{i, j}^{n}\right) \\
\mathbf{ Q}_{i, j}^{(2)}&=&\frac{3}{4} \mathbf{ Q}_{i, j}^{n}+\frac{1}{4} \mathbf{ Q}_{i, j}^{(1)}+\frac{1}{4} \Delta t \ \mathbf{Res}\left(\mathbf{ Q}_{i, j}^{(1)}\right) \\
\mathbf{ Q}_{i, j}^{n+1}&=&\frac{1}{3} \mathbf{ Q}_{i, j}^{n}+\frac{2}{3} \mathbf{ Q}_{i, j}^{(2)}+\frac{2}{3} \Delta t \ \mathbf{Res}\left(\mathbf{ Q}_{i, j}^{(2)}\right).
\end{eqnarray}
 The superscripts ${(1)-(2)}$ denote the intermediate steps, and the superscripts ${n}$ and ${n+1}$ denote the current and the next time-steps. The time-step, $\Delta t$, is computed as:

\begin{equation}
\Delta t= \text{CFL} \cdot \min \left(\frac{1}{\alpha_v} \min _{i, j}\left(\frac{\Delta \textcolor{black}{x^{2}}}{\mu_{i, j}}, \frac{\Delta \textcolor{black}{y^{2}}}{\mu_{i, j}}\right),  \min _{i, j}\left(\frac{\Delta\textcolor{black}{x}}{\left|u_{i, j}\right|+c_{i, j}}, \frac{\Delta \textcolor{black}{y}}{\left|v_{i, j}\right|+c_{i, j}}\right)\right),
\end{equation}
where $\alpha_v = 3$, $\mu$ is the dynamic viscosity, and $c$ is the speed of sound. The following sections present the computation of numerical approximations of viscous and convective fluxes.

\subsection{Spatial discretization of viscous fluxes}\label{subsec:visc}

In this section, the discretization of the viscous fluxes is presented. For simplicity, a one-dimensional scenario is considered. The grid is discretized on a uniform grid with $N$ cells on a spatial domain spanning $x \in \left[x_l, x_r \right]$. The cell center locations are at $x_i = x_l + (i - 1/2) \Delta x$, $\forall j \in \{1, \: 2, \: \dots, \: N\}$, where $\Delta x = (x_r - x_l)/N$. In one-dimension, the viscous flux at the interface is as follows:
\begin{equation}\label{eqn:visc-interface}
\mathbf{\hat F^v}_{i+\frac{1}{2}}=\left[\begin{array}{c}
0 \\
-\tau_{i+\frac{1}{2}} \\
-\tau_{i+\frac{1}{2}} u_{i+\frac{1}{2}}
\end{array}\right],\tau_{i+1 / 2}=\frac{4}{3} \mu_{i+1 / 2}\left(\frac{\partial u}{\partial x}\right)_{i+1 / 2}
\end{equation}
As it can be seen from Equation (\ref{eqn:visc-interface}) the viscous fluxes at cell interfaces $x_{i+\frac{1}{2}}$, $\forall i \in \{ 0, \: 1, \: 2, \: \dots, \: N \}$, has to be evaluated. For this purpose, we consider the $\alpha$-damping approach of Nishikawa \cite{Nishikawa2011a}. The following equation computes the velocity gradient at the cell interface,
\begin{equation}\label{eqn:alpha-damping}
\begin{aligned}
\left(\frac{\partial u}{\partial x}\right)_{i+\frac{1}{2}}=\frac{1}{2}\left[\left(\frac{\partial u}{\partial x}\right)_{i}+\left(\frac{\partial u}{\partial x}\right)_{i+1}\right]+\frac{\alpha}{2 \Delta x}\left({u}_{i+ \frac{1}{2}}^{R}-{u}_{i+ \frac{1}{2}}^{L}\right),
\end{aligned}
\end{equation}
where,
\textcolor{black}{\begin{equation}\label{eqn:vel-visc}
\begin{aligned}
{u}_{i+ \frac{1}{2}}^{L} &={\hat {u}}_{i}+\frac{\Delta x}{2} \left(\frac{\partial u}{\partial x}\right)_{i},{u}_{i+ \frac{1}{2}}^{R} &={\hat {u}}_{i+1}-\frac{\Delta x}{2} \left(\frac{\partial u}{\partial x}\right)_{i+1}  \quad 
\end{aligned}.
\end{equation}}
By substituting the Equations (\ref{eqn:vel-visc}) in the Equation (\ref{eqn:alpha-damping}) we get the following equation:
\textcolor{black}{\begin{equation}
\begin{aligned}
\left(\frac{\partial u}{\partial x}\right)_{i+1 / 2}&=\frac{1}{2}\left( u_{i}^{\prime}+ u_{i+1}^{\prime}\right)+\frac{\alpha}{2 \Delta x}\left(\hat u_{i+1}-\frac{\Delta x}{2}  u_{i+1}^{\prime}-\hat u_{i}-\frac{\Delta x}{2}  u_{i}^{\prime}\right),
\end{aligned}
\end{equation}}
where $u_{i}^{\prime}$ represents $\left(\frac{\partial u}{\partial x}\right)_{i}$. The gradients at the cell-\textcolor{black}{centers}, $\left(\frac{\partial {\phi}}{\partial x}\right)_{i}$, are computed by the following second-order formula:
\begin{eqnarray}\label{sixth-ordergrad}
\left(\frac{\partial u}{\partial x}\right)_{i}
 =
 \frac{  1 }{2} \left[ 
           \frac{  \hat u_{i+1} - \hat u_{i-1}   }{\Delta x}
 \right].
 \end{eqnarray}
 By substituting $\alpha$ = 3 in the cell interface gradients, the second derivative can be explicitly written as follows:
\begin{equation}
\begin{aligned}
\left(\frac{\partial^2 u}{\partial x^2}\right)_{i} =  \frac{\left(\frac{\partial u}{\partial x}\right)_{i+1 / 2} - \left(\frac{\partial u}{\partial x}\right)_{i-1 / 2}} {\Delta x}=\frac{-\hat u_{i-2}+12 \hat u_{i-1}-22 \hat u_i+12 \hat u_{i+1}-\hat u_{i+2}}{8 \Delta x^2}.
\end{aligned}
\end{equation}
The second derivatives thus computed are cell-averaged second derivatives as explained in \cite{buchmuller2014improved}.

\subsection{Convective Flux Discretization}\label{sec:inviscid flux discretization}

In this section, the discretization of the viscous convective is presented. The determination of convective fluxes involves two essential stages: first, a reconstruction phase where the solution vector at the cell centre is reconstructed to the cell interfaces, and second, an evolution of approximate Riemann solver phase in which the average fluxes at each interface are assessed using a procedure that considers the directions of the ``waves.'' The convective flux at the interface can then be expressed as:
\begin{equation}
	\mathbf {\hat{F}^c}_{i+\frac{1}{2}} = F_{i+\frac{1}{2}}^{Riemann} \left(\mathbf{U}_{i+\frac{1}{2}}^L,\mathbf{U}_{i+\frac{1}{2}}^R\right),
\end{equation}
where $\mathbf{U}$ = $( \alpha_{1}\rho_{1}, \alpha_{2}\rho_{2}, u, v, p,\alpha_1)^T$ is the primitive variable vector (in the two-dimensional scenario), and the superscripts $L$, and $R$ denote the left and right-sided reconstructed solution vectors respectively.  The Hartex-Lax-van Leer-Contact (HLLC) \cite{toro2009riemann} approximate Riemann solver is used in this study.

The objective of the paper is to compute the values of $\mathbf{U}_{i+\frac{1}{2}}^L,\mathbf{U}_{i+\frac{1}{2}}^R$ such that they are physically consistent and the contact discontinuities are captured sharply. Before presenting the details of the numerical discretization, the physics is briefly discussed. The computation of $\mathbf{U}_{i+\frac{1}{2}}^L,\mathbf{U}_{i+\frac{1}{2}}^R$ is typically carried out in two ways. The primitive variables are directly reconstructed at the interfaces, or the primitive variables are transformed to characteristic space, and characteristic variables are reconstructed (they are transformed back to primitive variable space). For conversion from physical to characteristic space, the primitive variables are multiplied with the left eigenvectors obtaining the characteristic variables, denoted by $\mathbf{W}$, where ${\bm{W}}_{m} = \bm{L}_{\bm{n}} \bm{ \mathbf{U}}$. Once the shock-capturing procedure is completed, the characteristic variables are multiplied with the right eigenvectors, thereby recovering the primitive variables. The left and right eigenvectors of the two-dimensional multi-component equations, denoted by $\bm{L_n}$ and $\bm{R_n}$, used for characteristic variable projection are as follows:

\begin{align}\label{eqn:leftright-multi}
\onehalfspacing
	\bm{R_n} = \begin{bmatrix} 
   		\frac{\alpha_{1} \rho_{1}}{c^{2} \rho} & 1 & 0 & 0 & 0 & \frac{\alpha_{1} \rho_{1}}{c^{2} \rho} \\ 
\\
\frac{\alpha_{2} \rho_{2}}{c^{2} \rho} & 0 & 1 & 0 & 0 & \frac{\alpha_{2} \rho_{2}}{c^{2} \rho}\\
\\
- \frac{n_x}{c \rho} & 0 & 0 & n_y & 0 & \frac{n_x}{c \rho}\\
\\
- \frac{n_y}{c \rho} & 0 & 0 & n_x & 0 & \frac{n_y}{c \rho}\\
\\
1 & 0 & 0 & 0 & 0 & 1\\
\\
0 & 0 & 0 & 0 & 1 & 0
   			 \end{bmatrix}, \enskip
	\bm{L_n} = \begin{bmatrix}
   		0 & 0 & - \frac{n_x c \rho}{2} & - \frac{n_y c \rho}{2} & \frac{1}{2} & 0\\
\\
1 & 0 & 0 & 0 & - \frac{\alpha_{1} \rho_{1}}{c^{2} \rho} & 0\\
\\
0 & 1 & 0 & 0 & - \frac{\alpha_{2} \rho_{2}}{c^{2} \rho} & 0\\
\\
0 & 0 & n_y & n_x & 0 & 0\\
\\
0 & 0 & 0 & 0 & 0 & 1\\
\\
0 & 0 & \frac{n_x c \rho}{2} &  \frac{n_y c \rho}{2} & \frac{1}{2} & 0
   			 \end{bmatrix}, &&\\\nonumber
\end{align}
where $\bm{n}$ = $[n_x \ n_y]^t$ and $[l_x \ l_y]^t$ is a tangent vector (perpendicular to $\bm{n}$) such as $[l_x \ l_y]^t$ = $[-n_y \ n_x]^t$. By taking $\bm{n}$ = $[1, 0]^t$ and $[0, 1]^t$ we obtain the corresponding eigenvectors in $x$ and $y$ directions. Let $\mathbf{W_b}$, where $b = 1, 2, 3, 4, 5, 6$, represent the vector of characteristic variables for the multi-species system; and \textcolor{black}{these variables} have the following features:

\begin{itemize}
	\item  The first and sixth characteristic variables, $\mathbf{W_1}$ and $\mathbf{W_6}$, corresponds to acoustic waves (shocks or rarefactions) \cite{hirschvol2,chargy1990comparisons}. 
	\item The second and third characteristic variables , $\mathbf{W_2}$ and $\mathbf{W_3}$, corresponds to what is known as the entropy or contact wave. Across the contact discontinuity, there will be a jump in density, but the pressure and the normal velocity are continuous.
	\item The fourth characteristic variable, $\mathbf{W_4}$, corresponds to what is known as the shear wave. When there is a contact discontinuity, there will be a jump in  tangential velocity \cite{hirschvol2} in inviscid scenario but the tangential velocities are continuous in viscous scenario \cite{batchelor1967introduction}. In $x-$ direction the left eigenvector is as follows (and it can be observed that $\mathbf{W_4}$ is $v$ velocity):
\begin{equation}
\bm{W_x}=\mathbf{L_xU}, \text{where}\
\mathbf{L_x}=\left[\begin{array}{ccccccc}
0 & 0  & \frac{-c\rho}{2} & 0 & \frac{1}{2}  & 0 \\
\\
1 & 0  & 0 & 0 & -\frac{\alpha_1 \rho_1}{\rho c^2}   & 0 \\
\\
0 & 1  & 0 & 0 & -\frac{\alpha_1 \rho_1}{\rho c^2}   & 0 \\
\\
0 & 0 & 0 & 1  & 0 & 0 \\
\\
0 & 0 & 0 & 0  & 0 & 1 \\
\\
0 & 0  & \frac{c\rho}{2} & 0 & \frac{1}{2}  & 0 \\
\end{array}\right], \text{and} \ \mathbf{U}=\left[\begin{array}{c}
\alpha_1 \rho_1 \\
\\
\alpha_2 \rho_2 \\
\\
u \\
\\
v \\
\\
p \\
\\
\alpha_1
\end{array}\right].
\end{equation}
	\item  Finally, the volume fraction $\mathbf{\alpha_{1}}$ or characteristic variable 
$\mathbf{W_{5}}$ is constant across shocks or rarefactions, and only varies
across the contact discontinuity \cite{chargy1990comparisons}.
As the volume fractions are
multiplied by \textcolor{black}{unity}, the shock-capturing is \textcolor{black}{performed}
directly on the physical values, i.e. there is no characteristic
transformation. 
\end{itemize}

In primitive variable space, shocks lead to discontinuities in phasic densities, pressure, and normal velocity, while material interfaces cause discontinuities in phasic densities and volume fractions \cite{hirschvol2}. In viscous flow simulations, tangential velocities remain continuous \cite{batchelor1967introduction}, but they become discontinuous across contact discontinuities in inviscid simulations \cite{hirschvol2}. The following subsections present three algorithms that account for these physical phenomena in order to obtain $\mathbf{U}_{i+\frac{1}{2}}^L,\mathbf{U}_{i+\frac{1}{2}}^R$.

\begin{itemize}
	\item In section \ref{invis-char}, algorithm with \textit{characteristic} variable reconstruction in \textit{inviscid} scenario is presented.
	\item In section \ref{vis-char}, algortihm with \textit{characteristic} variable reconstruction in \textit{viscous} scenario is presented.	
	\item In section \ref{vis-prim}, algortihm with \textit{primitive} variable reconstruction in \textit{viscous} scenario is presented.
\end{itemize}

The main difference between inviscid and viscous algorithms is how tangential velocities are computed, taking appropriate physics into account. In all the algorithms, the same contact discontinuity sensor will be used. The details are presented below. 

\subsubsection{Algorithm for characteristic variable reconstruction (inviscid scenario):} \label{invis-char}

 The complete numerical algorithm for characteristic variable reconstruction for inviscid flow simulations is summarized below, which includes the transforming of primitive variables into characteristics variables necessary for capturing discontinuities. The computations are shown for $\mathbf{U}_{i+\frac{1}{2}}^L$.

\begin{enumerate}

\item Compute the arithmetic or Roe averages at the interface $(x_{i+\frac{1}{2}})$ by using neighbouring cells, $(x_i)$ and $(x_{i+1})$. Compute the left $\bm{L_{n}}$ and right $\bm{R_{n}}$ eigenvectors. Transform the variables, $\bm{\mathbf{U}}$, into characteristic space by multiplying with the left eigenvectors\\

\begin{equation}
	{\bm{W}}_{m,b} = \bm{L}_{\bm{n}_{i+\frac{1}{2}}} \bm{ \mathbf{U}}_{m,b}
\end{equation}
The transformed variables are denoted as ${\bm{W}}_{m,b}$, where $m$ is $\{i-2, i-1, i, i+1, i+2, i+3 \}$ and  $b$  is $\{1, 2, 3, 4, 5, 6 \}$.\\

\item Carry out the appropriate reconstruction procedure for each variable, as explained below, and obtain the interface values denoted by ${\bm{W}}_{i+\frac{1}{2},b}^{L}$. \\

\begin{equation}
    \mathbf{W}^{L}_{i+\frac{1}{2},b} = 
    \left\{
    \begin{array}{ll}
        \text{if } b = 2,3\text{:} & \begin{cases}
            \mathbf{W}^{L,Non-Linear}_{i+\frac{1}{2},b} & \text{if } \left( \mathbf{W}^{L,Linear}_{i+\frac{1}{2}} - \mathbf{W}_i \right) \left( \mathbf{W}^{L,Linear}_{i+\frac{1}{2}} - \mathbf{W}^{L,MP}_{i+\frac{1}{2}} \right) \geq 10^{-40},
            \\[20pt]
            \mathbf{W}^{L,Linear}_{i+\frac{1}{2},b} & \text{otherwise}.
            \\[20pt]
{\bm{W}}_{i+\frac{1}{2},b}^{L, T} & \text{if } \min \left(\psi_{i-1}, \psi_{i}, \psi_{i+1}\right)<\psi_{c}.
        \end{cases}\\ 
        \\[10pt]
        \text{if } b = 5\text{:} & \begin{cases}
            \mathbf{W}^{L,T}_{i+\frac{1}{2},b} .
    \end{cases}
    \end{array}
    \right.
    \label{eqn:contact}
\end{equation}

\begin{equation}
    \mathbf{W}^{L}_{i+\frac{1}{2},b} = 
    \left\{
    \begin{array}{ll}
         \text{if } b = 1,6\text{:} & \begin{cases}
           \mathbf{W}^{L,Non-Linear}_{i+\frac{1}{2},b} & \text{if } \left( \mathbf{W}^{L,Linear}_{i+\frac{1}{2}} - \mathbf{W}_i \right) \left( \mathbf{W}^{L,Linear}_{i+\frac{1}{2}} - \mathbf{W}^{L,MP}_{i+\frac{1}{2}} \right) \geq 10^{-40},
            \\[20pt]
            \mathbf{W}^{L,Linear}_{i+\frac{1}{2},b} & \text{otherwise}.
        \end{cases}
    \end{array}
    \right.
    \label{eqn:acoustic}
\end{equation}

\begin{equation}
    \mathbf{W}^{L}_{i+\frac{1}{2},b} = 
    \left\{
    \begin{array}{ll}
        \text{if } b = 4\text{:} & \begin{cases}
           \mathbf{W}^{L,Non-Linear}_{i+\frac{1}{2},b} & \text{if } \left( \mathbf{W}^{L,Linear}_{i+\frac{1}{2}} - \mathbf{W}_i \right) \left( \mathbf{W}^{L,Linear}_{i+\frac{1}{2}} - \mathbf{W}^{L,MP}_{i+\frac{1}{2}} \right) \geq 10^{-40},
            \\[20pt]
            \mathbf{W}^{L,Linear}_{i+\frac{1}{2},b} & \text{otherwise}.
        \end{cases}
    \end{array}
    \right.
    \label{eqn:contact_inv}
\end{equation}

\item After obtaining interface values, the reconstructed states are then recovered by projecting the characteristic variables back to physical fields: \\

\textcolor{black}{
\begin{equation}
\begin{aligned}\label{IG-right-transform}
	{{\bm{\mathbf{U}}}}_{i+\frac{1}{2}}^{L} &= \bm{R}_{\bm{n}_{i+\frac{1}{2}}} {\bm{W}}_{i+\frac{1}{2}}^{L}, \\
  {{\bm{\mathbf{U}}}}_{i+\frac{1}{2}}^{R} &= \bm{R}_{\bm{n}_{i+\frac{1}{2}}} {\bm{W}}_{i+\frac{1}{2}}^{R}.
\end{aligned}
\end{equation}}
\end{enumerate}

In the above-presented algorithm, there are two different reconstruction procedures. The first reconstruction approach is the MP5 scheme ($\mathbf{W}^{L,Non-Linear}$ and $\mathbf{W}^{L, Linear}$) of Suresh and Hyunh \cite{suresh1997accurate}. The computation of the MP5 scheme is as follows:

\begin{equation}\label{eqn:alpha}
\begin{aligned}
 \mathbf{W}_{i+\frac{1}{2}}^{L, Linear}&=\frac{1}{30} \mathbf{W}_{i-2}-\frac{13}{60} \mathbf{W}_{i-1}+\frac{47}{60} \mathbf{W}_{i+0}+\frac{9}{20} \mathbf{W}_{i+1}-\frac{1}{20} \mathbf{W}_{i+2},\\
  \mathbf{W}_{j+1 / 2}^{\text {Non-Linear }} &=\mathbf{W}_{j+1 / 2}^{\text {L, Linear }}+\operatorname{minmod}\left(\mathbf{W}_{j+1 / 2}^{\min }-\mathbf{W}_{j+1 / 2}^{\text {L, Linear }}, \mathbf{W}_{j+1 / 2}^{\max }-\mathbf{W}_{j+1 / 2}^{\text {L, Linear}}\right), \\
\mathbf{W}_{j+1 / 2}^{M P} &=\mathbf{W}_{j}+\operatorname{minmod}\left[\mathbf{\hat{W}}_{j+1}-\mathbf{\hat{W}}_{j}, 4\left(\mathbf{\hat{W}}_{j}-\mathbf{\hat{W}}_{j-1}\right)\right], \\
\mathbf{W}_{j+1 / 2}^{\min } &=\max \left[\min \left(\mathbf{\hat{W}}_{j}, \mathbf{\hat{W}}_{j+1}, \mathbf{W}_{j+1 / 2}^{M D}\right), \min \left(\mathbf{\hat{W}}_{j}, \mathbf{W}_{j+1 / 2}^{U L}, \mathbf{W}_{j+1 / 2}^{L C}\right)\right], \\
\mathbf{W}_{j+1 / 2}^{\max } &=\min \left[\max \left(\mathbf{\hat{W}}_{j}, \mathbf{\hat{W}}_{j+1}, \mathbf{W}_{j+1 / 2}^{M D}\right), \max \left(\mathbf{\hat{W}}_{j}, \mathbf{W}_{j+1 / 2}^{U L}, \mathbf{W}_{j+1 / 2}^{L C}\right)\right], \\
\mathbf{W}_{j+1 / 2}^{M D} &=\frac{1}{2}\left(\mathbf{\hat{W}}_{j}+\mathbf{\hat{W}}_{j+1}\right)-\frac{1}{2} d_{j+1 / 2}^{M}, \\
\mathbf{W}_{j+1 / 2}^{U L} &=\mathbf{\hat{W}}_{j}+4\left(\mathbf{\hat{W}}_{j}-\mathbf{\hat{W}}_{j-1}\right), \\
\mathbf{W}_{j+1 / 2}^{L C} &=\frac{1}{2}\left(3 \mathbf{\hat{W}}_{j}-\mathbf{\hat{W}}_{j-1}\right)+\frac{4}{3} d_{j-1 / 2}^{M}, \\
d_{j+1 / 2}^{M} &=\operatorname{minmod}\left(4 d_{j}-d_{j+1}, 4 d_{j+1}-d, d_{j},d_{j+1}\right), \\
d_{j} &=\mathbf{\hat{W}}_{j-1}-2 \mathbf{\hat{W}}_{j}+\mathbf{\hat{W}}_{j+1},
\end{aligned}
\end{equation}
\textcolor{black} {where,
\begin{equation}
minmod(a,b) = \half \left ( sign(a)+sign(b) \right ) min(|a|,|b|).
\end{equation}}
\begin{remark}\label{eqn:other}
\normalfont One can also use other shock-capturing methods like WENO \cite{Borges2008}, Montonocity preserving explicit or implicit gradient based methods \cite{chamarthi2023gradient,chamarthi2023implicit,chamarthi2023efficient}, MUSCL scheme \cite{vanleer1979} apart from the MP5 scheme presented above. The contact discontinuity sensor will work without any modifications, including all the parameters. Results using the WENO \cite{Borges2008} scheme instead of the MP scheme for certain test cases are shown in the Appendix of this paper.
\end{remark}
\noindent The second candidate reconstruction function is the THINC reconstruction ($\mathbf{W}^{L, T}$), a differentiable and monotone Sigmoid function \cite{xiao2011revisit}. Unlike the MP5 discussed above, the THINC scheme is a non-polynomial function. The explicit formula for the left and right interface for the THINC function are as follows \cite{wakimura2022symmetry}:
\begin{equation}\label{THINC}
\textcolor{black}{\mathbf{W}_{i+1 / 2}^{L, T}=\left\{\begin{array}{l}
\textcolor{black}{\mathbf{{u}_{a}}}+\textcolor{black}{\mathbf{{u}_{d}}}\frac{K_1 + (K_2/K_1)}{1+K_2} \text { if }\left({\mathbf{W}}_{i+1}-{\mathbf{W}}_{i}\right)\left({\mathbf{W}}_{i}-{\mathbf{W}}_{i-1}\right)>0, \\
\mathbf{U}_{i} \text { otherwise }.
\end{array}\right.}
\end{equation}

\begin{equation}
\textcolor{black}{\mathbf{W}_{i-1 / 2}^{R, T}=\left\{\begin{array}{ll}
\textcolor{black}{\mathbf{{u}_{a}}}-\textcolor{black}{\mathbf{{u}_{d}}}\frac{K_1 - (K_2/K_1)}{1-K_2} & \text { if }\left({\mathbf{W}}_{i+1}-{\mathbf{W}}_{i}\right)\left({\mathbf{W}}_{i}-{\mathbf{W}}_{i-1}\right)>0, \\
\mathbf{U}_{i} \text { otherwise },
\end{array}\right.}
\end{equation}
where
\begin{eqnarray*}
&&K_1=\tanh\left(\frac{\beta}{2}\right),\ K_2=\tanh\left(\frac{ \textcolor{black}{\bm{\alpha}_i}\beta}{2}\right),\ \textcolor{black}{\bm{\alpha}_i}=\frac{{\mathbf{W}}_{i}-\textcolor{black}{\mathbf{{u}_{a}}}}{\textcolor{black}{\mathbf{{u}_{d}}}},\textcolor{black}{\mathbf{{u}_{a}}}=\frac{{\mathbf{W}}_{i+1}+{\mathbf{W}}_{i-1}}{2},\ \textcolor{black}{\mathbf{{u}_{d}}}=\frac{{\mathbf{W}}_{i+1}-{\mathbf{W}}_{i-1}}{2}.
\end{eqnarray*}

The performance of the THINC function depends on the value of the steepness parameter $\beta$ as discussed in \cite{deng2018high,deng2019fifth}. The parameter $\beta$ controls the jump thickness, i.e., a small value of $\beta$ leads to a smooth profile, while a large one leads to a sharp jump-like distribution. When $\beta$ is set to 1.8, the reconstruction function becomes closer to a step-like profile, and the discontinuous solution can be resolved within about four mesh cells \cite{xiao2011revisit}. In this \textcolor{black}{study}, the value of $\beta$ is set to 1.8 for one-dimensional cases and 1.9 for multi-dimensional test cases. The minimum value is 1.6, and the maximum is 2.0 with the proposed sensor. \\

\noindent Finally, the proposed contact discontinuity sensor for the five-equation model is as follows: 

\begin{equation}\label{psi-mp}
\begin{aligned}
&\psi_{i}=\frac{2ab + \varepsilon}{\left(a^2+b^2+\varepsilon\right)}, \ \text{where}
&\varepsilon=\frac{0.9 \psi_{c}}{1-0.9 \psi_{c}} \xi, \quad \xi=10^{-2}, \quad \psi_{c}=0.35,
\end{aligned}
\end{equation}

\begin{equation}\label{detector-new}
\begin{aligned}
&a  = \frac{13}{12} \left|s_{i-2} - 2 s_{i-1} + s_{i}\right| + \frac{1}{4} \left|s_{i-2} - 4 s_{i-1} + 3 s_{i}\right|,\\
&b  = \frac{13}{12} \left|s_{i} - 2 s_{i+1} + s_{i+2}\right| + \frac{1}{4} \left|3 s_{i} - 4 s_{i+1} + s_{i+2}\right|, \ \text{where} \ s=\frac{p}{\rho^{\gamma}}, \ \text{and}\ \rho=\rho_{1} \alpha_{1}+\rho_{2} \alpha_{2}.
\end{aligned}
\end{equation}

The variables $a$ and $b$ in the Equation (\ref{detector-new}) are (inspired by) the smoothness indicators of the WENO scheme (Equation (\ref{eq:smoothness})). $a$ and $b$ are infact $\beta_0$ and $\beta_2$ without the squares and different choice of variable. To the author's knowledge, these smoothness indicators are not used to detect contact discontinuities in this manner, so the THINC scheme can be applied robustly to improve the resolution of contact discontinuities. Furthermore, the word ``smoothness indicator'' is a misnomer (they are discontinuity detectors) as explained in \cite{balsara2017higher}. The sensor is based on \textit{discontintuity sensor} of \cite{chamarthi2023efficient} with modifications in variable (density is used in \cite{chamarthi2023efficient} and here it is $s$), the parameter $\xi$ is $10^{-2}$ instead of $10^{-6}$. Using a value of $10^{-6}$ for $\xi$ will detect regions that are not contact discontinuities, and the simulations will crash. The corresponding details of the sensor are presented in Example \ref{ex:shu}. It will be shown later that the variable $s$ better captures the contact discontinuities than density.

The variable $s$ in Equation (\ref{detector-new}), where $\ s=\frac{p}{\rho^{\gamma}}$, is loosely based on entropy. It has similar variables as that of the physical entropy, $S$=$ln\left(\frac{p}{\rho^{\gamma}}\right)$, but it is not the same. It was chosen based on the physics of the Euler equations (along with some numerical experiments) that the second (and also third in the five-equation model for two-species case) characteristic variable, $\mathbf{W_2}$, is known as the \textit{entropy wave}, which corresponds to the entropy change. As it is not exactly entropy, and there are several definitions for entropy in literature, it is not defined as entropy in this manuscript. It is referred to as the variable $s$ to avoid confusion. It was found that the variable $s$ has a large jump in value across a material interface, and the sensor could detect it.
\begin{remark}\label{eqn:some}
\normalfont  The proposed contact discontinuity sensor consistently detected contact discontinuities in all the cases, and in some cases, it detected shockwaves. Still, it cannot be called a \textit{discontinuity sensor}. Because a discontinuity detector detected both shockwaves and contact discontinuities, one cannot apply THINC (a steepening approach) \textit{for all the variables}. Some variables are still continuous across discontinuities. The difficulty is not in reconstructing shockwaves (acoustic waves of Euler equations wave structure) but in applying THINC for continuous variables. Density, pressure, and normal velocity are discontinuous across shocks, and only the density and volume fractions are discontinuous across a material interface \cite{hirschvol2}.
On the other hand, while tangential velocities are continuous in viscous flow simulations \cite{batchelor1967introduction}, they are discontinuous across contact discontinuities in inviscid simulations \cite{hirschvol2}. It will be shown later that pressure and tangential velocities can be computed using a central scheme across material interfaces, and there will be no oscillations. Using the THINC scheme for all the variables leads to failure of the scheme (simulation getting crashed or oscillatory results) as it is physically \textbf{inconsistent}. \textbf{This is an important aspect of the paper.} Harten also mentioned in \cite{harten1989eno} that the ENO schemes highly resolve shocks, and the subcell resolution is applied only to the linearly degenerate characteristic field to improve the resolution of contact discontinuities. The objective of the present study is also to improve the resolution of contact discontinuity and material interfaces as shocks are well resolved by the MP scheme.
\end{remark}

\subsubsection{Algorithm for characteristic variable reconstruction (viscous scenario):}\label{vis-char}

In an inviscid scenario presented above, the tangential velocities are discontinuous across a contact discontinuity \cite{hirschvol2}. However, for a viscous flow scenario (the intended target for the proposed algorithm and the physically realistic scenario), the tangential velocities are continuous across a contact discontinuity \cite{batchelor1967introduction}. If a variable is continuous, one may use a central scheme. In this regard, the following approach is used for the fourth characteristic variable ($\mathbf{W_4}$) to ensure that a central scheme ($\mathbf{W}^{C,Linear}$) computes the tangential velocities:
\begin{equation}
    \mathbf{W}^{L}_{i+\frac{1}{2},b} = 
    \left\{
    \begin{array}{ll}
        \text{if } b = 4\text{:} & \begin{cases}
           \mathbf{W}^{L,Non-Linear}_{i+\frac{1}{2},b} & \text{if } {\Omega_{d}} > 0.01
            \\[20pt]
            \mathbf{W}^{C,Linear}_{i+\frac{1}{2},b} & \text{otherwise},
        \end{cases}
    \end{array}
    \right.
    \label{eqn:contact_visc}
\end{equation}
where
\begin{equation}
\mathbf{W}^{C,Linear}_{i+\frac{1}{2}}=\frac{1}{60}\left( \mathbf{W}_{i-2}-8  \mathbf{W}_{i-1}+37  \mathbf{W}_i+37  \mathbf{W}_{i+1}-8  \mathbf{W}_{i+2}+ \mathbf{W}_{i+3}\right).	
\end{equation}
${\Omega_{d}}$ is the Ducros sensor \cite{chamarthi2023wave,ducros1999large,chamarthi2024generalized}
\begin{equation}
    \Omega_d = \max \left( \Omega_{i+m} \right), \quad \text{for } m = -1,0,1,  
\end{equation}
and computed as follows: 
\begin{equation}
    \Omega_i = \frac{\left| -p_{i-2} + 16 p_{i-1} - 30 p_{i} + 16 p_{i+1} - p_{i+2} \right|}{\left| +p_{i-2} + 16 p_{i-1} + 30 p_{i} + 16 p_{i+1} + p_{i+2} \right|} \frac{ \left( \nabla \cdot \mathbf{u} \right)^2}{ \left( \nabla \cdot \mathbf{u} \right)^2 + \left| \nabla \times \mathbf{u} \right|^2},
    \label{eqn:ducros}
\end{equation}
\noindent where $\mathbf{u}$ is the velocity vector, and the derivatives of velocities are computed by the fourth-order implicit gradient approach of \cite{nishikawa2018green}, which is as follows:

\begin{equation}
    \frac{1}{6} \left. \frac{\partial u}{\partial x} \right|_{i-1} + \frac{2}{3} \left. \frac{\partial u}{\partial x} \right|_{i} + \frac{1}{6} \left. \frac{\partial u}{\partial x} \right|_{i+1} =  \frac{1}{2 \Delta x} \left( \hat u_{i+1} - \hat u_{i-1} \right).
    \label{eqn:firstDerivative}
\end{equation}
Ducros sensor is a shock detector and cannot detect contact discontinuities as shown in \cite{chamarthi2023wave}. In \cite{chamarthi2023wave}, authors have used a separate discontinuity sensor for shocks and contact discontinuities and showed that using the Ducros sensor in the presence of contact discontinuities will lead to oscillations. Therefore, a central scheme will always compute the tangential velocities ($\mathbf{W_4}$) because it cannot detect the contact discontinuities. In references \cite{chamarthi2023wave,hoffmann2024centralized}, the Ducros sensor is also considered for acoustic waves, but this manuscript does not consider such an approach. The paper aims to improve the contact discontinuities and material interfaces only. Finally, the above algorithm is denoted as HY-THINC-D and is used only for viscous test cases.
\begin{remark}\label{eqn:please}
\normalfont  One could argue that tangential velocities are continuous across a shockwave and should (or might ) always be computed using a central scheme. The current approach uses a dimension-by-dimension approach, and the shockwaves are not always aligned with the grid. If a shockwave is at an angle with the grid, there will be oscillations if a limiter is not applied. An approach that aligns the grid with a shockwave might (or can) compute tangential velocities always with a central scheme. The author tried using a central scheme for tangential velocities without any sensor; while it worked in some cases, it crashed in many cases (due to oscillations near shocks).
\end{remark}

\subsubsection{Algorithm for primitive variable reconstruction (viscous scenario):}\label{vis-prim}

Shock-capturing is typically performed on characteristic variables for coupled hyperbolic equations like the Euler equations to achieve the cleanest results, as explained by van Leer in \cite{van2006upwind}. When interface values are directly reconstructed using primitive variables, it can lead to small oscillations, particularly for high-resolution schemes. Reconstructing primitive variables $\mathbf{U}$ = $( \alpha_{1}\rho_{1}, \alpha_{2}\rho_{2}, u, v, p,\alpha_1)^T$ implies that the THINC scheme is applied to phasic densities and volume fractions. However, the proposed algorithm can be used with primitive variable reconstruction, especially in viscous flow scenarios. The algorithm for primitive variable reconstruction for viscous flow simulations is as follows:

In $x$-direction:
\begin{equation}\label{eqn:centralScheme_x}
    \mathbf{U}^{L}_{i+\frac{1}{2},b} = 
    \left\{
    \begin{array}{ll}
        \text{if } b = 3\text{:} & \begin{cases}
           {\bm{U}}^{L,Non-Linear}_{i+\frac{1}{2},b} & \text{if } \left( {\bm{U}}^{L,Linear}_{i+\frac{1}{2}} - {\bm{U}}_i \right) \left( {\bm{U}}^{L,Linear}_{i+\frac{1}{2}} - {\bm{U}}^{L,MP}_{i+\frac{1}{2}} \right) \geq 10^{-40},
            \\[10pt]
            {\bm{U}}^{L,Linear}_{i+\frac{1}{2},b} & \text{otherwise}.
        \end{cases}\\
        \\[10pt]
        \text{if } b = 4\text{:} &\begin{cases}
           {\bm{U}}^{L,Non-Linear}_{i+\frac{1}{2},b} & \text{if } {\Omega_{d}} > 0.01
            \\[10pt]
            {\bm{U}}^{C,Linear}_{i+\frac{1}{2},b} & \text{otherwise}.
        \end{cases}
        \end{array}
    \right.
\end{equation}
In $y$-direction:
\begin{equation}\label{eqn:centralScheme_y}
    \mathbf{U}^{L}_{i+\frac{1}{2},b} = 
    \left\{
    \begin{array}{ll}
        \text{if } b = 4\text{:} & \begin{cases}
           {\bm{U}}^{L,Non-Linear}_{i+\frac{1}{2},b} & \text{if } \left( {\bm{U}}^{L,Linear}_{i+\frac{1}{2}} - {\bm{U}}_i \right) \left( {\bm{U}}^{L,Linear}_{i+\frac{1}{2}} - {\bm{U}}^{L,MP}_{i+\frac{1}{2}} \right) \geq 10^{-40},
            \\[10pt]
            {\bm{U}}^{L,Linear}_{i+\frac{1}{2},b} & \text{otherwise}.
        \end{cases}\\
        \\[10pt]
        \text{if } b = 3\text{:} &\begin{cases}
           {\bm{U}}^{L,Non-Linear}_{i+\frac{1}{2},b} & \text{if } {\Omega_{d}} > 0.01
            \\[10pt]
            {\bm{U}}^{C,Linear}_{i+\frac{1}{2},b} & \text{otherwise}.
        \end{cases}        \end{array}
    \right.
\end{equation}

In all directions:

\begin{equation}
    {\bm{U}}^{L}_{i+\frac{1}{2},b} = 
    \left\{
    \begin{array}{ll}
         \text{if } b = 5\text{:} & 
         \begin{cases}
           {\bm{U}}^{L,Non-Linear}_{i+\frac{1}{2},b} & \text{if } \left( {\bm{U}}^{L,Linear}_{i+\frac{1}{2}} - {\bm{U}}_i \right) \left( {\bm{U}}^{L,Linear}_{i+\frac{1}{2}} - {\bm{U}}^{L,MP}_{i+\frac{1}{2}} \right) \geq 10^{-40},
            \\[10pt]
           {\bm{U}}^{L,Linear}_{i+\frac{1}{2},b} & \text{otherwise}.
        \end{cases}\\
        \\[10pt]
        \text{if } b = 1,2\text{:} & 
        \begin{cases}
            {\bm{U}}^{L,Non-Linear}_{i+\frac{1}{2},b} & \text{if } \left( {\bm{U}}^{L,Linear}_{i+\frac{1}{2}} - {\bm{U}}_i \right) \left( {\bm{U}}^{L,Linear}_{i+\frac{1}{2}} - {\bm{U}}^{L,MP}_{i+\frac{1}{2}} \right) \geq 10^{-40},
            \\[10pt]
            {\bm{U}}^{L,Linear}_{i+\frac{1}{2},b} & \text{otherwise}.
            \\[10pt]
			{\bm{U}}_{i+\frac{1}{2},b}^{L, T} & \text{if } \min \left(\psi_{i-1}, \psi_{i}, \psi_{i+1}\right)<\psi_{c}.
        \end{cases}\\ 
        \\[10pt]
        \text{if } b = 6\text{:} & 
        \begin{cases}
            {\bm{U}}^{L,T}_{i+\frac{1}{2},b} .
        \end{cases}
    \end{array}
    \right.
    \label{eqn:prim}
\end{equation}

The above algorithm is still denoted as HY-THINC-D, but it will be indicated in the results that the primitive variables are reconstructed. The similarity between the characteristic and primitive variable reconstruction is that in $x-$ direction, $v$ is reconstructed in the characteristic space and centralized if the Ducros sensor criterion is satisfied. Likewise, in the $y-$ direction, $ u$ is reconstructed in the characteristic space and centralized if the Ducros sensor criterion is satisfied. Similarly, the variable $v$ is centralized in primitive variable space in $x-$ direction (Equations (\ref{eqn:centralScheme_x})), and  $u$ is centralized in primitive variable space in $y-$ direction (Equations (\ref{eqn:centralScheme_y})), respectively. As explained earlier, the proposed contact discontinuity sensor may detect shockwaves in some test cases, and in those cases (if detected by the sensor), density is also reconstructed using THINC if primitive variables are recontructed. Such a reconstruction is also physically consistent as density is discontinuous across the shock. It must be repeated here that velocities and pressure are continuous across a contact discontinuity, and THINC should not be applied in these regions.

\section{Results and discussion}\label{sec:results}

In this section, the proposed spatial discretization schemes are tested for a set of benchmark cases to assess the performance of both single and multi-dimensional test cases.

\begin{example}\label{ex:multi-species}{Multi-species shock tube}
\end{example}

The first one-dimensional test case is the two-fluid modified shock tube of Abgrall and Karni \cite{abgrall2001computations}. The initial conditions of the test case are as follows:
\begin{equation}
\left(\alpha_{1} \rho_{1}, \alpha_{2} \rho_{2}, u, p, \alpha_{1}, \gamma \right)=\left\{\begin{array}{ll}
\left(1 , 0, 0, 1, 1, 1.4 \right) & \text { for } x<0 \\
\left(0,0.125,0,0.1, 0, 1.6 \right) & \text { for }  x \geq 0 .
\end{array}\right.
\end{equation}
Simulations are carried out on a uniformly spaced grid with $200$ cells on the spatial domain $-0.5 \leq x \leq 0.5$ with a constant CFL of $0.4$ and the final time is $t=0.2$. 

\begin{figure}[H]
\centering
\subfigure[Density]{\includegraphics[width=0.45\textwidth]{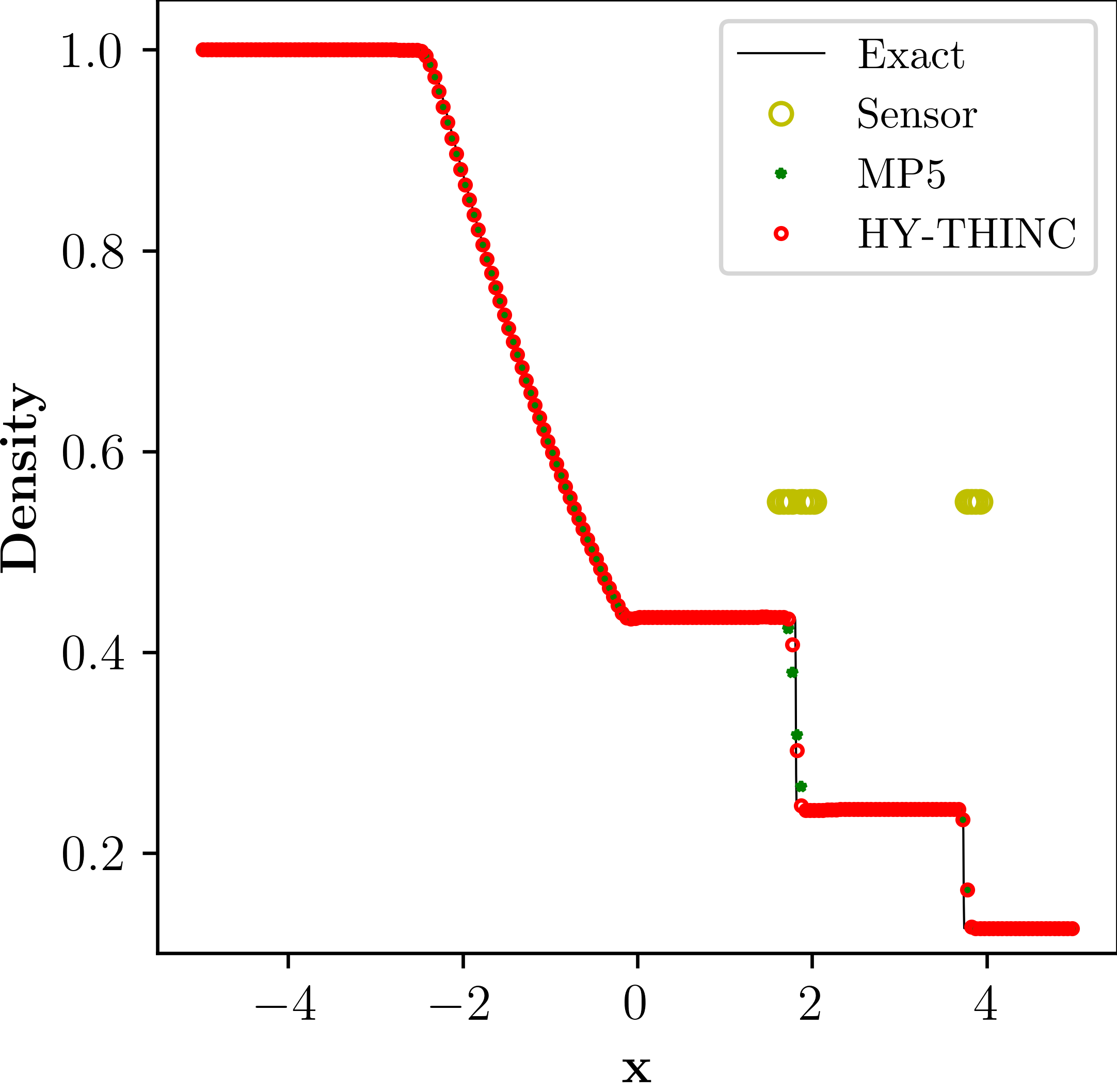}
\label{fig:multi_sod-den}}
\subfigure[Pressure]{\includegraphics[width=0.45\textwidth]{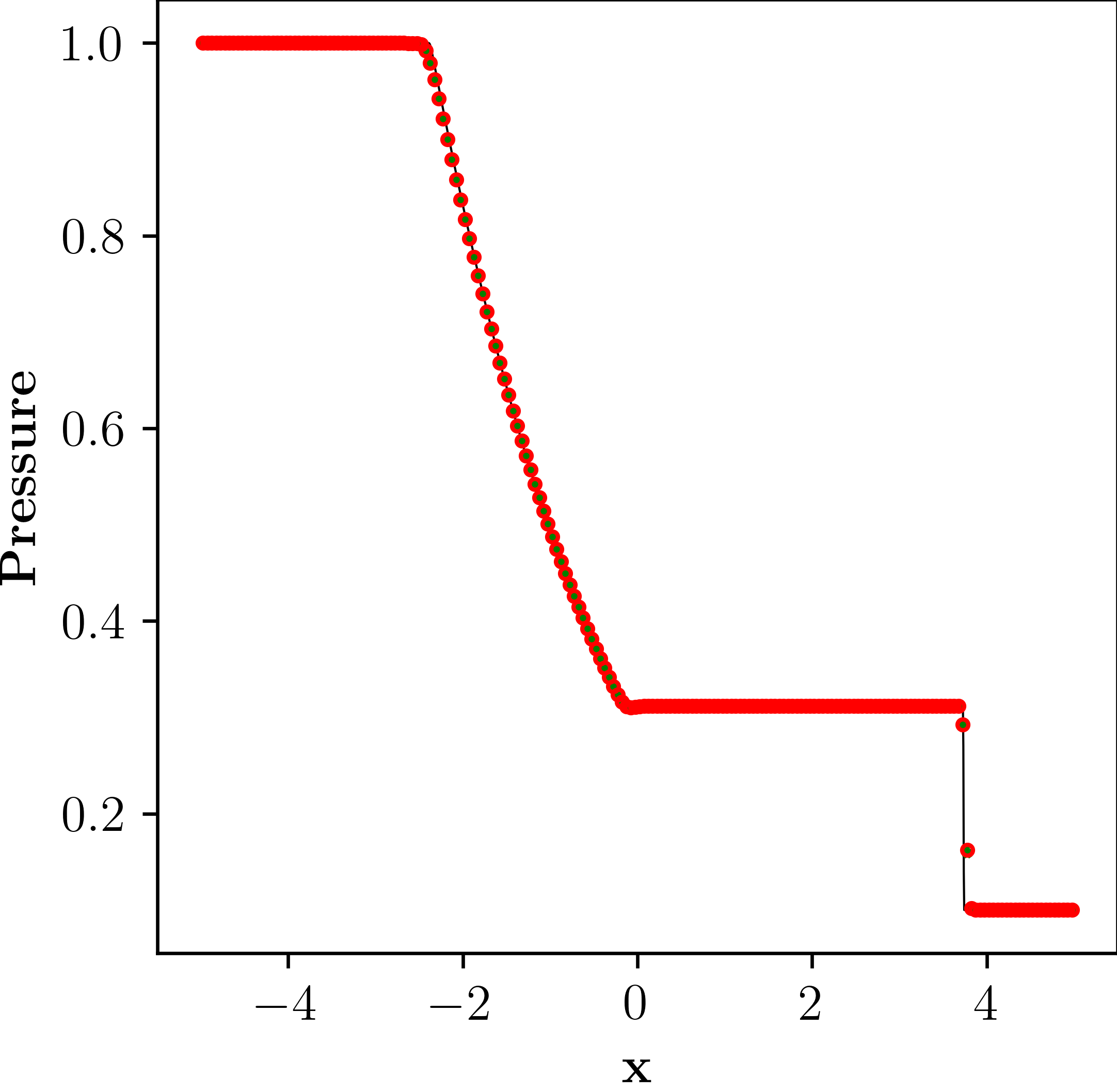}
\label{fig:multi_sod-pres}}
\subfigure[Volume fraction]{\includegraphics[width=0.31\textwidth]{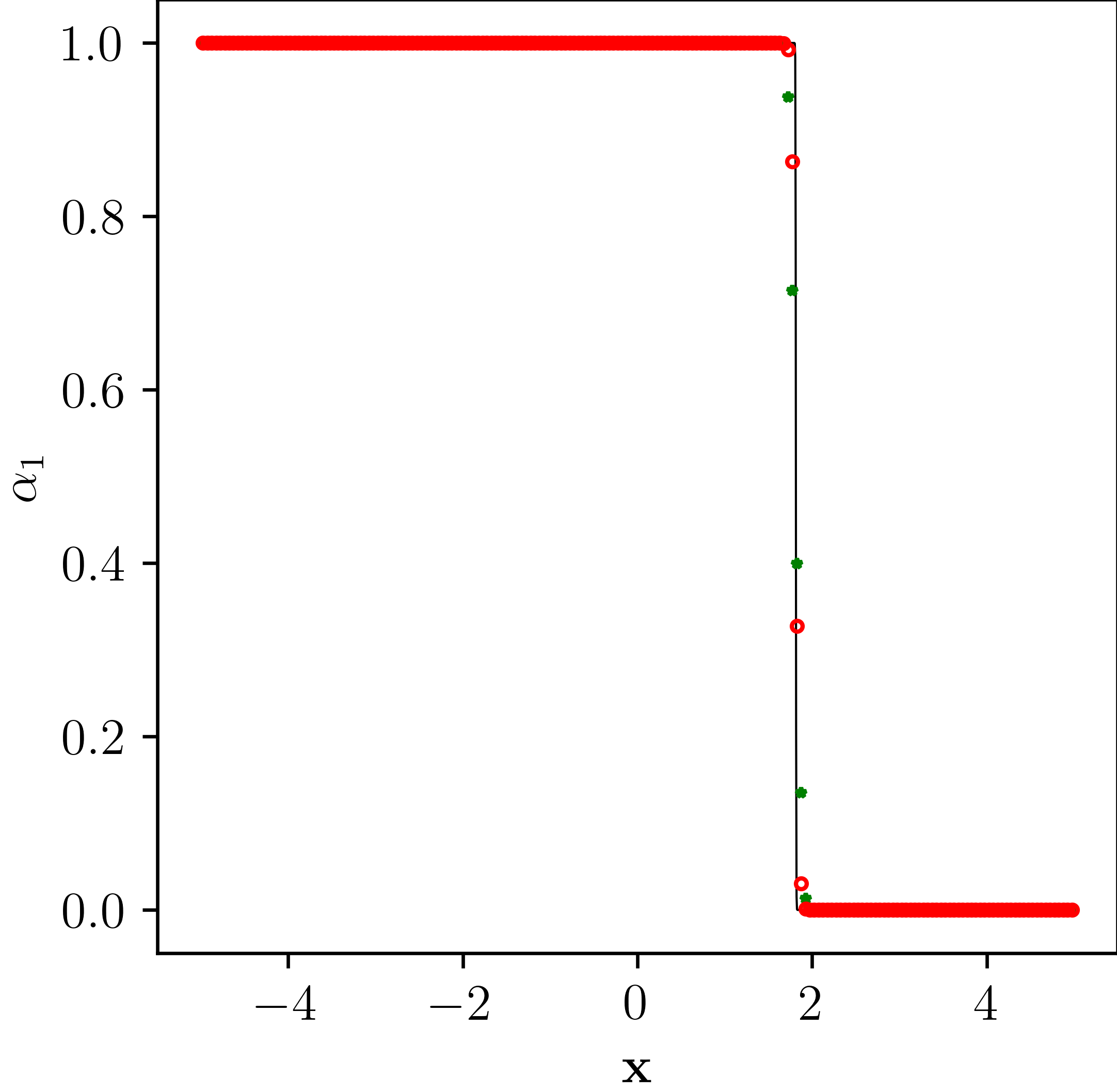}
\label{fig:multi_sod-volf}}
\subfigure[Sensor variables.]{\includegraphics[width=0.325\textwidth]{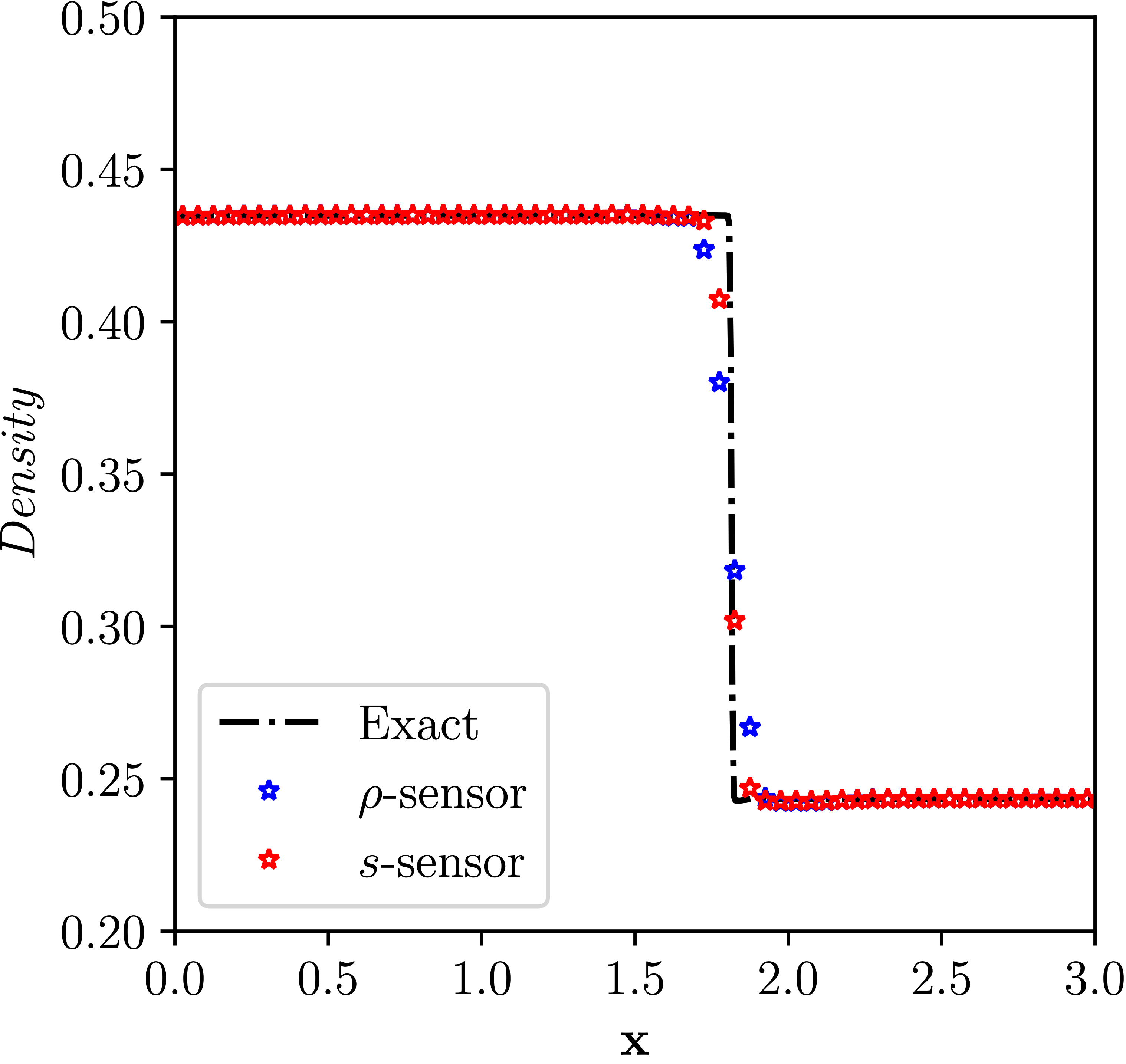}
\label{fig:sensor}}
\subfigure[Variable $s$]{\includegraphics[width=0.325\textwidth]{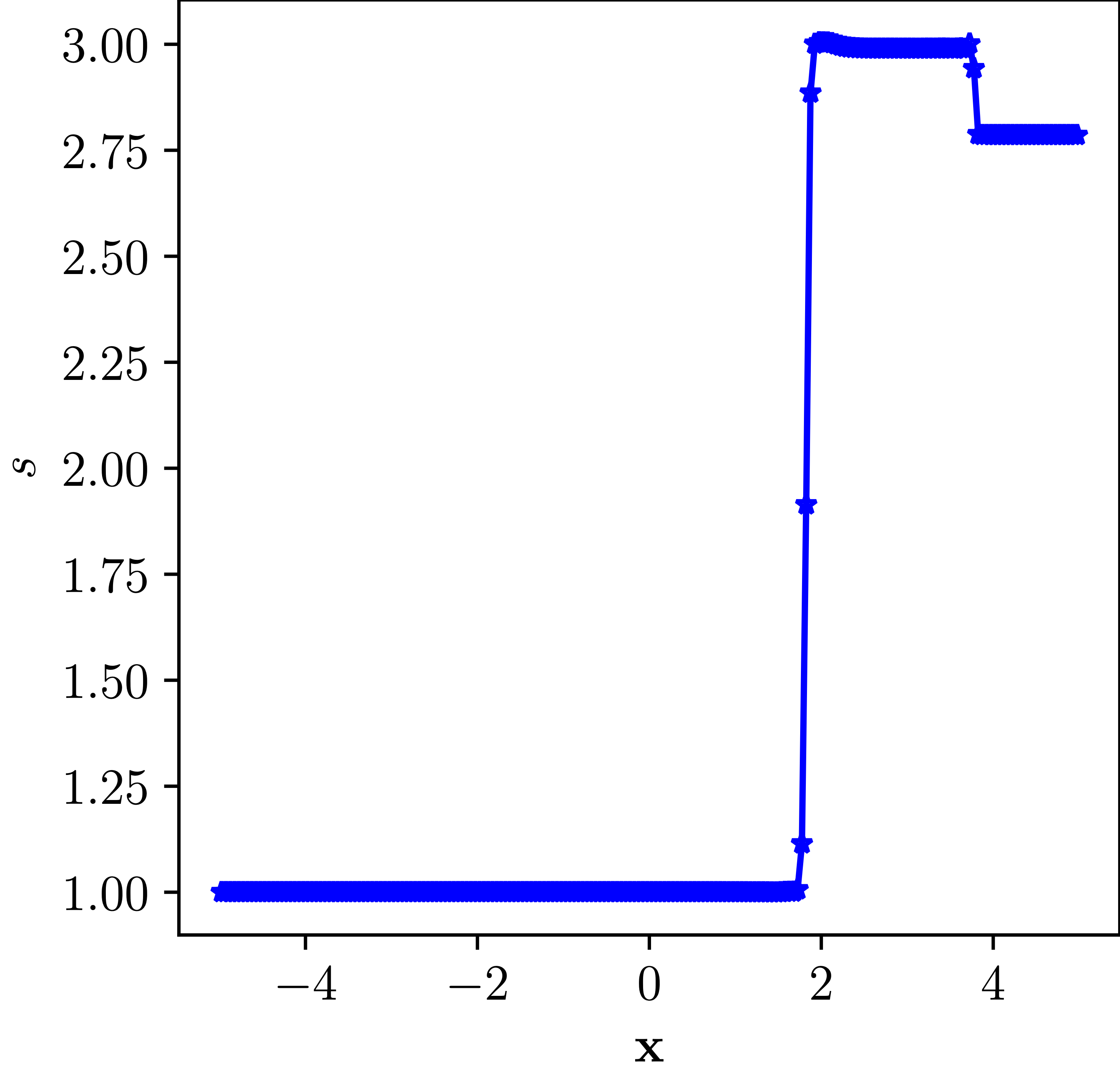}
\label{fig:multi_sod-ent}}
\caption{Numerical solution for multi-species shock tube problem in Example \ref{ex:multi-species} on a grid size of $N=200$. Solid line: Reference solution; green stars: MP5; red circles: HY-THINC.}
\label{fig_multisod}
\end{figure}

Fig. \ref{fig_multisod} shows various schemes' density, pressure, volume fraction profiles, sensor variables and the profile of variable $s$. The proposed scheme, HY-THINC, captures the material interface without oscillations and uses fewer points than the MP5 scheme, Fig. \ref{fig:multi_sod-den}. Fig. \ref{fig:multi_sod-pres} shows no visible difference in the pressure profiles. The volume fraction profiles shown in Fig. \ref{fig:multi_sod-volf} indicate that the HY-THINC scheme captures the volume fractions within a few points compared to MP5. Fig. \ref{fig:sensor} shows the effect of using different variables in the proposed contact discontinuity sensor. Fig. \ref{fig:sensor} indicates that using variable $s$ detects the contact discontinuity, whereas if the density is used as a variable, the contact discontinuity is not detected as indicated by the number of points used for resolving them. Finally, the variation of variable $s$ is shown in Fig. \ref{fig:multi_sod-ent}, indicating a significant jump in variable $s$ across the material interface.

\begin{figure}[H]
\centering
\subfigure[Density profile]{\includegraphics[width=0.45\textwidth]{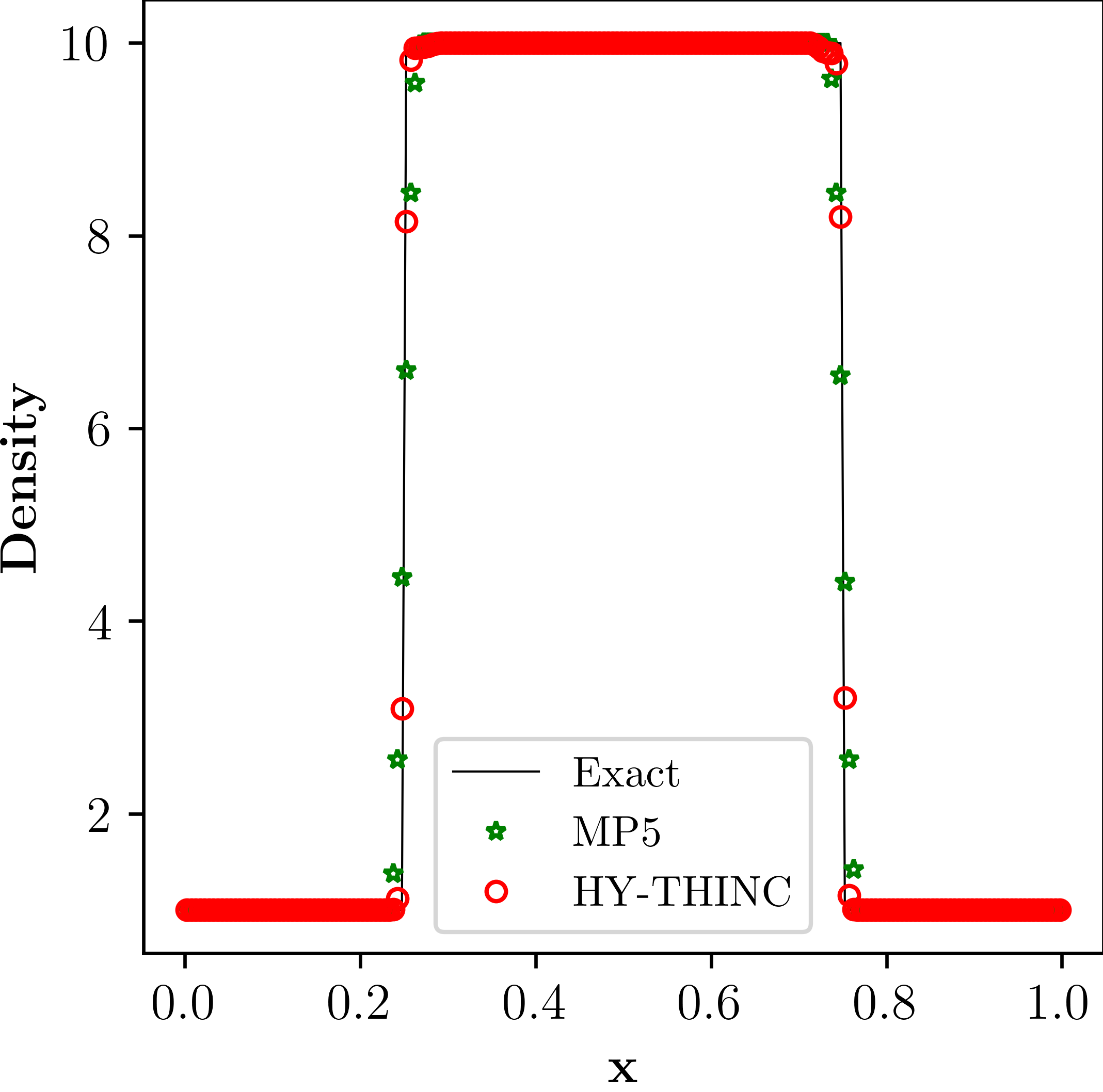}
\label{fig:iso-met_l}}
\subfigure[Pressure profile]{\includegraphics[width=0.45\textwidth]{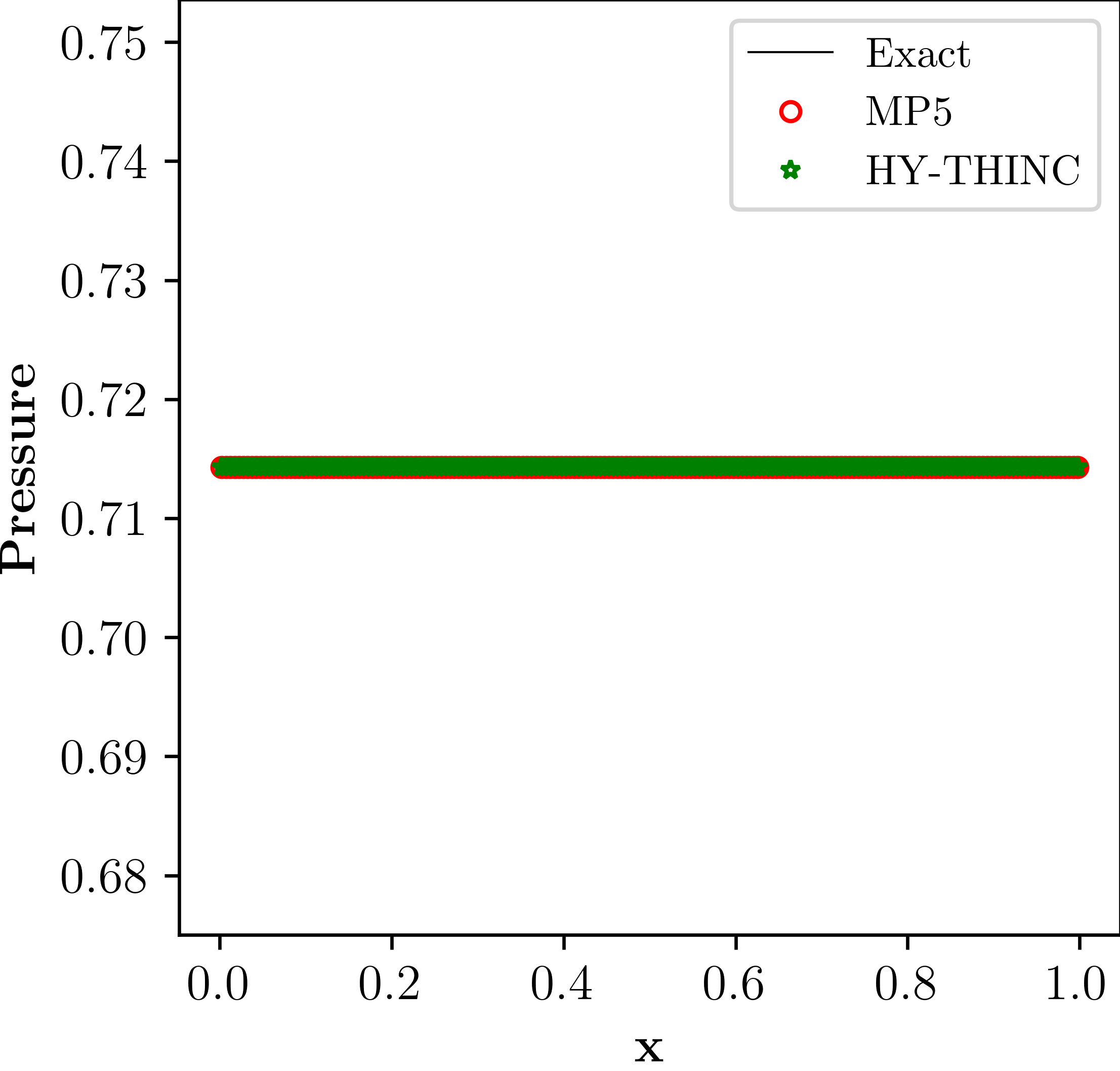}
\label{fig:iso-met2_l2}}
    \caption{Numerical solution for isolated contact test case using $N = 200$ grid points at $t = 1.0$,  Example \ref{ex:isol}, where Solid line: Reference solution; green stars: MP5; red circles: HY-THINC.}
    \label{fig_iso-mla}
\end{figure}

\begin{example}\label{ex:isol}{Material interface advection}
\end{example}
This two-species one-dimensional problem is the advection of an isolated material interface \cite{johnsen2006implementation,Wong2016}. The initial conditions for this test case are given by

\begin{equation}
\left(\alpha_{1} \rho_{1}, \alpha_{2} \rho_{2}, u, p, \alpha_{1}, \gamma \right)= \begin{cases}(10,0,0.5,1 / 1.4,1.0,1.6), & 0.25 \leq x<0.75 \\ (00,1,0.5,1 / 1.4,0.0,1.4), & x<0.25 \text { or } x \geq 0.75\end{cases}
\end{equation}
The simulation was conducted on a computational domain spanning from $x = 0$ to $x = 1$, utilizing a total of $N = 200$, as in \cite{Wong2016}, uniformly distributed grid points until reaching a final time of $t = 2.0$. Both boundaries were subject to periodic conditions. Fig. \ref{fig_iso-mla} illustrates the comparison between the exact and numerical solutions of various schemes, which all precisely captured the material interface without any unwanted oscillations. The HY-THINC scheme accomplished this with fewer points than the MP5 scheme, which indicates that the sensor reliably detected the material interface. Pressure remained constant without oscillation and remained close to machine precision. 

\begin{example}\label{ex:shu}{Shock/density wave interaction}
\end{example}
In this test case, the Shu-Osher problem  \cite{Shu1988} extended for a binary mixture of Helium and Nitrogen by \cite{lv2014discontinuous}, that  is initially separated by a shock, is considered. The initial conditions are as follow:

\begin{equation}
\left(\rho, u, p, \alpha_{\mathrm{He}}, \alpha_{\mathrm{N}_2}\right)= \left\{\begin{array}{ll}
\left(3.8571,2.6294,10.3333,0,1 \right) & \text { for } x< -4 \\
\left(1+0.2 \sin (5 x), 0,1,1,0 \right) & \text { for }  x \geq-4 .
\end{array}\right.
\end{equation}

\begin{figure}[H]
\centering
\subfigure[Density and $\alpha_1$, N=200]{\includegraphics[width=0.46\textwidth]{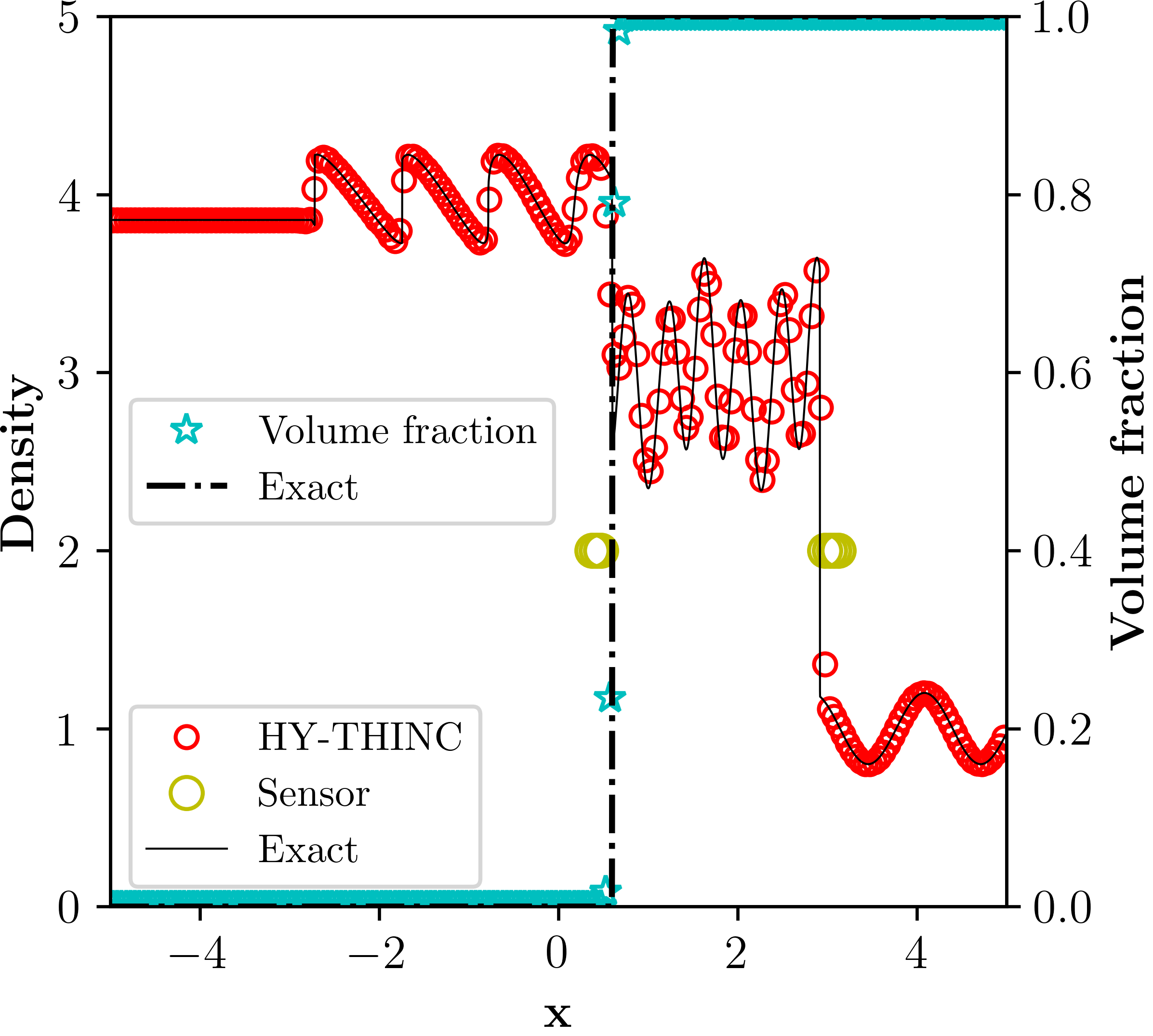}
\label{fig:mshu-met}}
\subfigure[Density and $\alpha_1$, N=800]{\includegraphics[width=0.46\textwidth]{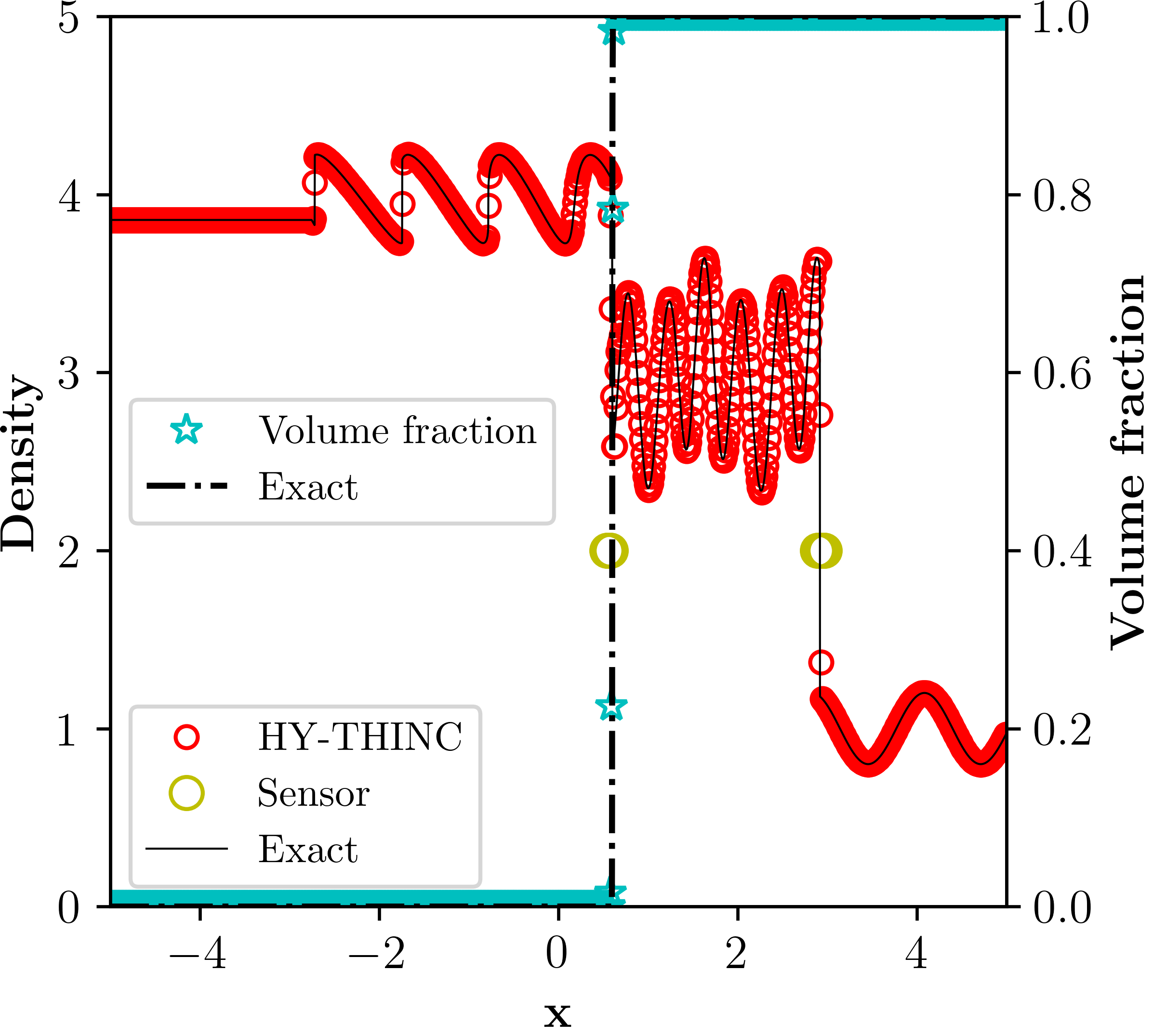}
\label{fig:mshu-met2}}
\caption{Numerical solution for multicomponent shock-density wave interaction using $N = 200$ and $800$ grid points at $t = 2.0$, for  Example \ref{ex:shu}, where Solid line: Reference solution; red circles: HY-THINC density; cyan stars: HY-THINC volume fractions; yellow circles: sensor location.}
\label{fig_shu-m}
\end{figure}

 The case was solved on a computational domain $x = [-5,5]$ with $N = 200$ and $800$ \textcolor{black}{, as in \cite{lv2014discontinuous},} uniformly distributed grid points until a final time, $t = 1.8$. The Shu-Osher problem is commonly used to evaluate a scheme's one-dimensional shock and density perturbation capturing capabilities. As shown in Fig. \ref{fig_shu-m}, the proposed HY-THINC scheme matches the ``exact'' solution well, computed on a grid size of 3200 by the WENO scheme, both for density and volume fractions. The proposed contact discontinuity sensor correctly identifies the material interface and does not alter the high-frequency regions, as they are not discontinuities at all.

The contact discontinuity sensor presented earlier, Equation (\ref{detector-new}), is also analyzed here in this example. Aspects of interest are the location of the sensor detection regions, the motivation for the current sensor and the differences from the sensor used in \cite{chamarthi2023efficient} are presented. \textcolor{black}{First, the modifications in the
proposed discontinuity detector compared to that of Li et al.
\cite{li2022class} and Chamarthi \cite{chamarthi2023efficient} are examined. The Li et al.
detector used fluxes as the variable (Equation (\ref{detector-density})) to
detect the discontinuities, and fluxes cannot be used for multicomponent
flows as primitive variables are to be reconstructed to avoid oscillations
\cite{abgrall2001computations,johnsen2006implementation,abgrall1996prevent}.
Li's objective for proposing the detector was to choose between linear and
WENO schemes, an objective which is different from that of the present study.
Chamarthi \cite{chamarthi2023efficient} used Li's detector but with density
as the variable in the detector to choose between variable or flux
reconstruction in the regions away from discontinuities for high-order
accuracy. Li's detector with density as the variable (as in
\cite{chamarthi2023efficient}) is given below:}
\textcolor{black}{\begin{equation}
\psi_{j}=\frac{2ab + \varepsilon}{\left(a^2+b^2+\varepsilon\right)}, \varepsilon=\frac{0.9 \psi_{c}}{1-0.9 \psi_{c}} \xi^{2}, \quad \xi=10^{-3}, \quad \psi_{c}=0.5,
\end{equation}}

\textcolor{black}{\begin{equation}\label{detector-density}
\begin{aligned}
&a=\left|\rho_{j}-\rho_{j-1}\right|+\left|\rho_{j}-2 \rho_{j-1}+\rho_{j-2}\right|, \ b=\left|\rho_{j}-\rho_{j+1}\right|+\left|\rho_{j}-2 \rho_{j+1}+\rho_{j+2}\right|.
\end{aligned}
\end{equation}}

\textcolor{black}{Li detector for reconstructing the entropy wave with THINC would be as follows:}

\textcolor{black}{
\begin{equation}\label{li}
\begin{aligned}
&{\bm{W}}_{i+\frac{1}{2},b}^{L, MP5}= 
\begin{cases}{\bm{W}}_{i+\frac{1}{2},b}^{L, T} & \text{if } b = 2 \text{ and } \underbrace {\min \left(\psi_{j-2},\psi_{j-1}, \psi_{j}, \psi_{j+1}, \psi_{j+2},\psi_{j+3}\right)<\psi_{c}}_{\text{Li sensor}} \\
\\
{\bm{W}}_{i+\frac{1}{2},b}^{L, MP5}  & \text { otherwise.}
\end{cases}
\end{aligned}
\end{equation}}

\textcolor{black}{A one-dimensional scenario is here considered. Therefore, only the entropy wave is reconstructed with THINC if  a discontinuity is detected. The current detector is also presented here for comparison.}

\textcolor{black}{\begin{equation}
\begin{aligned}
&\psi_{j}=\frac{2ab + \varepsilon}{\left(a^2+b^2+\varepsilon\right)}, \ \text{where}
&\varepsilon=\frac{0.9 \psi_{c}}{1-0.9 \psi_{c}} \xi, \quad \xi=10^{-2}, \quad \psi_{c}=0.35,
\end{aligned}
\end{equation}}

\textcolor{black}{\begin{equation}\label{detector}
\begin{aligned}
&a  = \frac{13}{12} \left|s_{i-2} - 2 s_{i-1} + s_{i}\right| + \frac{1}{4} \left|s_{i-2} - 4 s_{i-1} + 3 s_{i}\right|,\\
&b  = \frac{13}{12} \left|s_{i} - 2 s_{i+1} + s_{i+2}\right| + \frac{1}{4} \left|3 s_{i} - 4 s_{i+1} + s_{i+2}\right|, \ \text{where} \ s=\frac{p}{\rho^{\gamma}}.
\end{aligned}
\end{equation}}

\textcolor{black}{
\begin{equation}\label{sainath}
\begin{aligned}
&{\bm{W}}_{i+\frac{1}{2},b}^{L, MP5}= 
\begin{cases}{\bm{W}}_{i+\frac{1}{2},b}^{L, T} & \text{if } b = 2 \text{ and }  \underbrace {\min \left(\psi_{i-1},\psi_{i}, \psi_{i+1}\right)<\psi_{c}}_{\text{Current sensor}} \\
\\
{\bm{W}}_{i+\frac{1}{2},b}^{L, MP5}  & \text { otherwise. }
\end{cases}
\end{aligned}
\end{equation}}

\begin{figure}[H]
\centering
\subfigure[\textcolor{black}{Density profile}]{\includegraphics[width=0.45\textwidth]{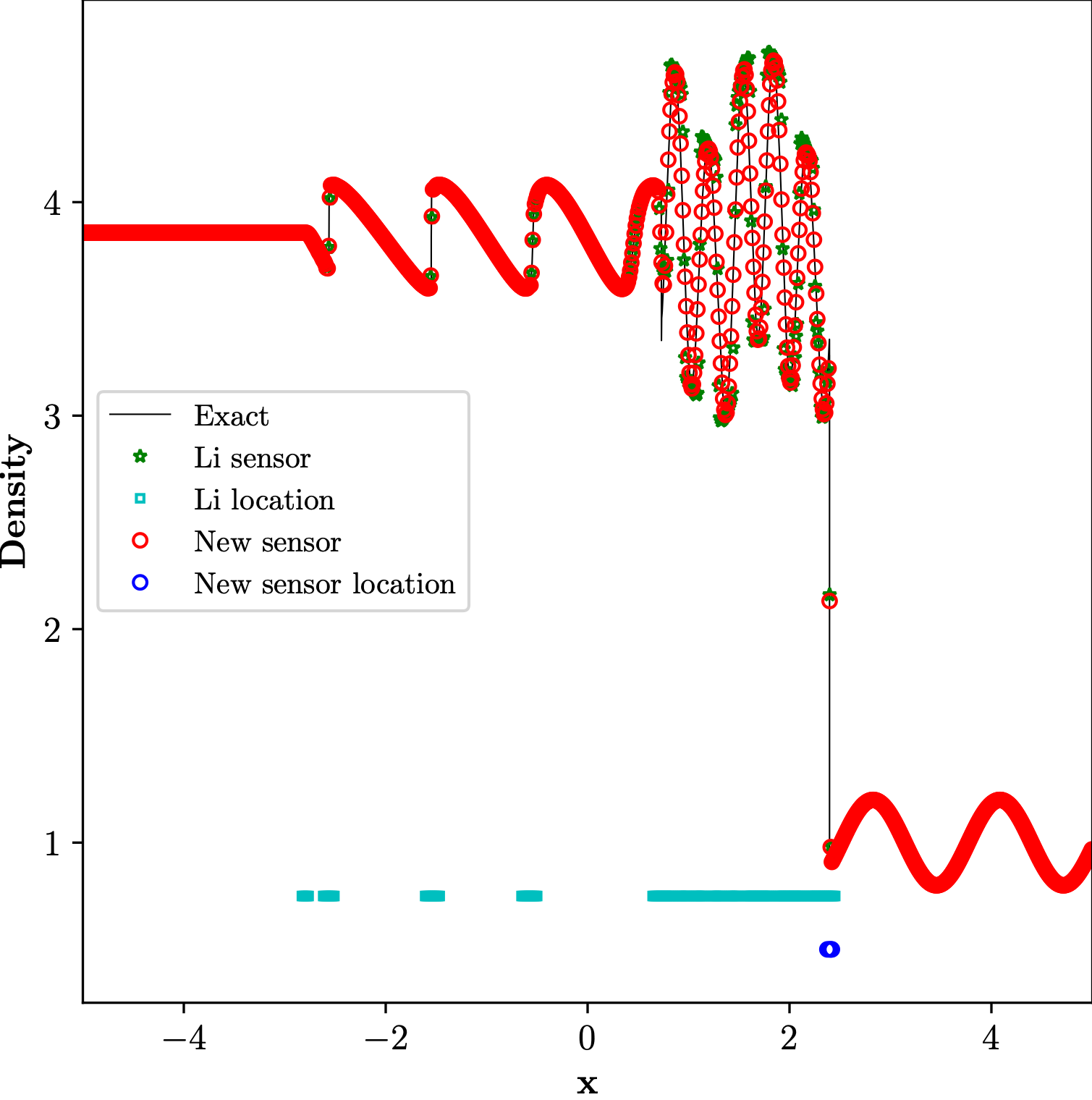}
\label{fig:seno-met}}
\subfigure[\textcolor{black}{Zoomed density profile}]{\includegraphics[width=0.45\textwidth]{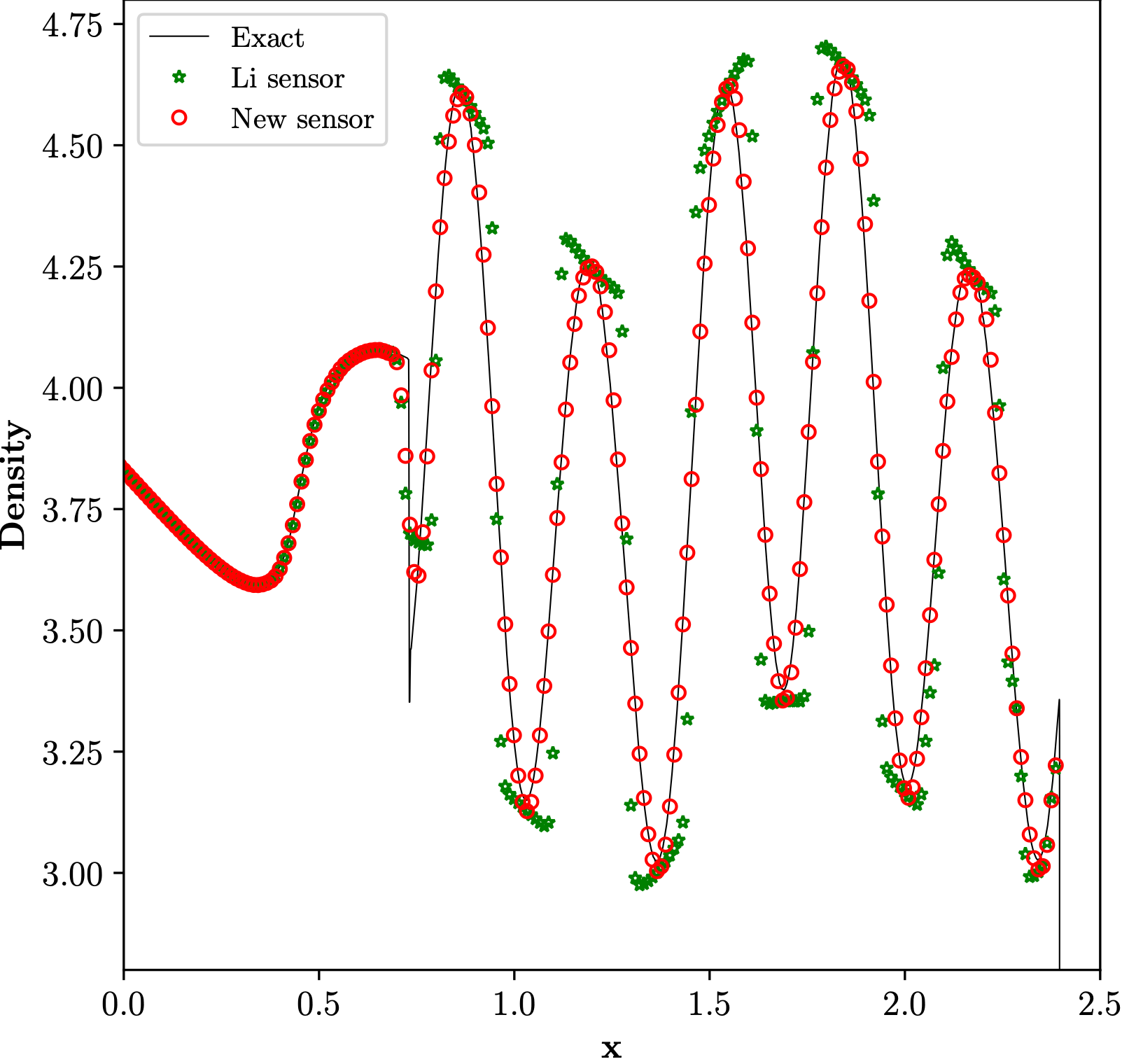}
\label{fig:seno-met-zoom}}
\caption{\textcolor{black}{Density profiles obtained for Shu-Osher test case using Li and new sensors. Solid line: Reference solution; green stars: density with Li sensor; red circles: density with new sensor; cyan squares: location of Li sensor's detection region and blue circles: location of new sensor's detection region.}}
\label{fig_sens}
\end{figure}

\textcolor{black}{The parameters $a$ and $b$ in Li's detector given by Equation (\ref{detector-density}) are different from the current detector given by Equation (\ref{detector}). The variable used in Li's sensor is density (as modified by Chamarthi in \cite{chamarthi2023efficient}), $\rho$, and the current sensor is $s$, and $s=\frac{p}{\rho^\gamma}$. It is not known if Li's sensor performs appropriately for multicomponent flows. As explained in Appendix \ref{append-a}, the variables $a$ and $b$ in the Equation (\ref{detector}) are inspired by the smoothness indicators of the WENO scheme (Equation \ref{eq:smoothness}). $a$ and $b$ are infact $\beta_0$ and $\beta_2$ without the squares and a different choice of variable.  The final difference between the two detectors is the number of cells that detect the discontinuity. \textcolor{black}{Increasing the value of $\psi_c$ detects the regions that might not be discontinuities, and reducing it below a threshold will not detect the discontinuities.} To understand the advantages and differences of the proposed sensor, the single-species shock/entropy-wave problem Shu and Osher \cite{Shu1988} is considered. The initial conditions of the test case are as follows:}
\textcolor{black}{\begin{equation}
        \left( \rho,u,p \right) =
        \begin{cases}
            (3.857,2.629,10.333), & \text{if } -5 \leq x < -4, \\
            (1+0.2\sin(5(x-5)),0,1), & \text{if } -4 \leq x \leq 5.
        \end{cases}
\end{equation}}
\textcolor{black}{ This test case is about shockwave interaction with high-frequency oscillating sinusoidal waves as it evaluates a scheme's capability to capture both shock and avoid high-frequency regions simultaneously. Li et al. \cite{li2022class} also devised their sensor to avoid activation of WENO in high-frequency regions (see Fig. 10 in \cite{li2022class} and the corresponding discussion).} \textcolor{black}{The computational domain of this test case is $x = [-5,5]$, and the final time is $t = 1.8$. Simulations are carried out on a grid size of $N = 900$, as in \cite{chamarthi2023gradient}, using both the sensors with the MP5 scheme. The \textit{exact solution} is obtained on a fine grid resolution of 40,000 points using the WENO scheme. As shown in Fig. \ref{fig:seno-met-zoom}, the proposed sensor matches the reference solution well, whereas Li's sensor with density as the variable modified the solution profile, squaring effect, compared to the reference result. Observing Fig. \ref{fig:seno-met}, Li's sensor, shown with cyan squares, flagged several regions as discontinuities, whereas the current sensor, shown with blue circles, detected only the shockwave at $x \approx 2.4$.}

\begin{figure}[H]
\centering
\subfigure[\textcolor{black}{TENO detector, figure taken from \cite{fu2019hybrid},  with permission from Global Science Press. Density profile with 200, 400 and 800 points.}]{\includegraphics[width=0.95\textwidth]{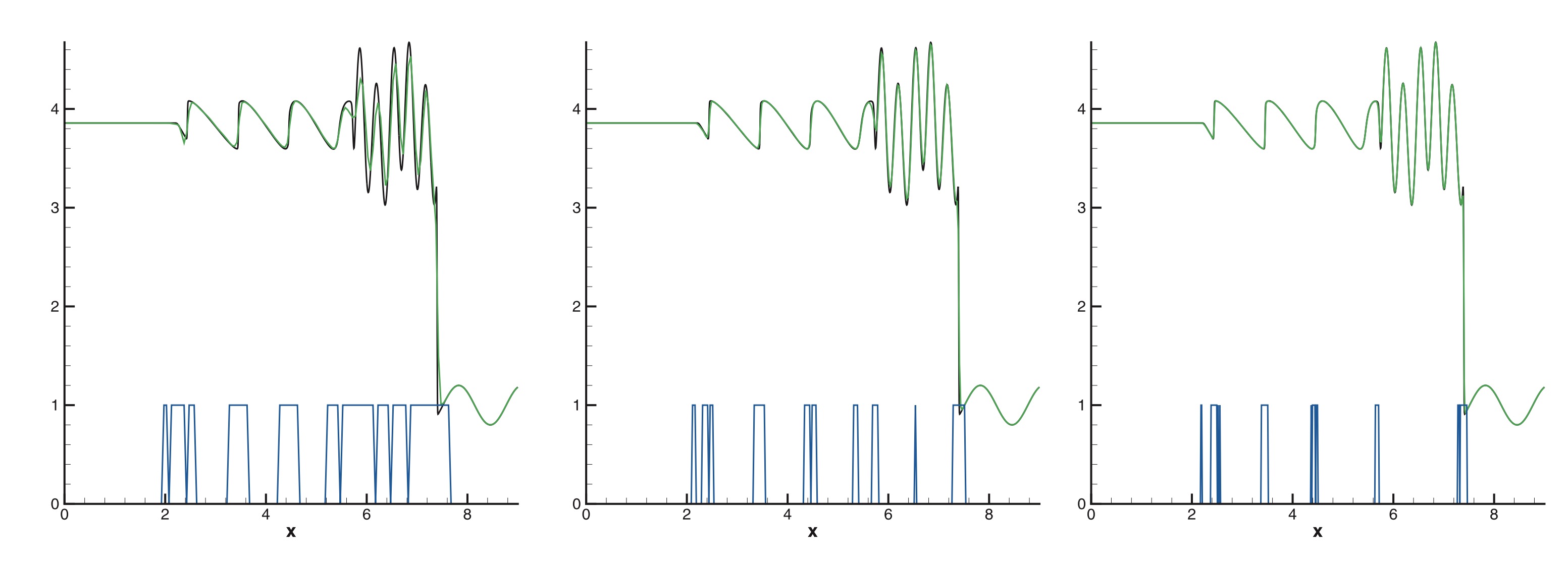}}
\caption{\textcolor{black}{ The TENO discontinuity sensor from \cite{fu2019hybrid}.}}
\label{final-nail}
\end{figure}

\begin{figure}[H]
\centering
\subfigure[\textcolor{black}{Density profile, 200 points}]{\includegraphics[width=0.32\textwidth]{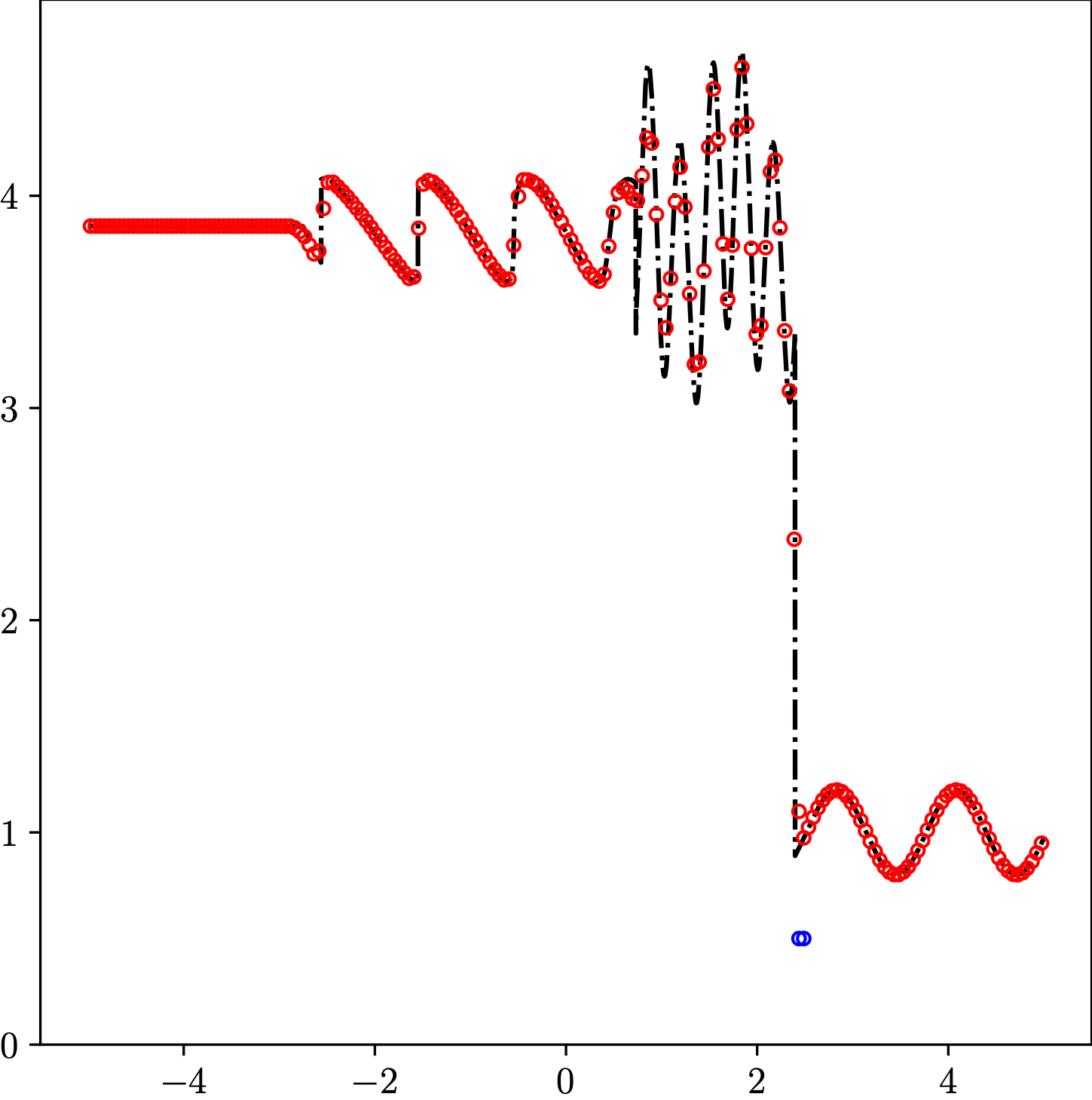}
\label{fig:c200}}
\subfigure[\textcolor{black}{Density profile, 400 points}]{\includegraphics[width=0.32\textwidth]{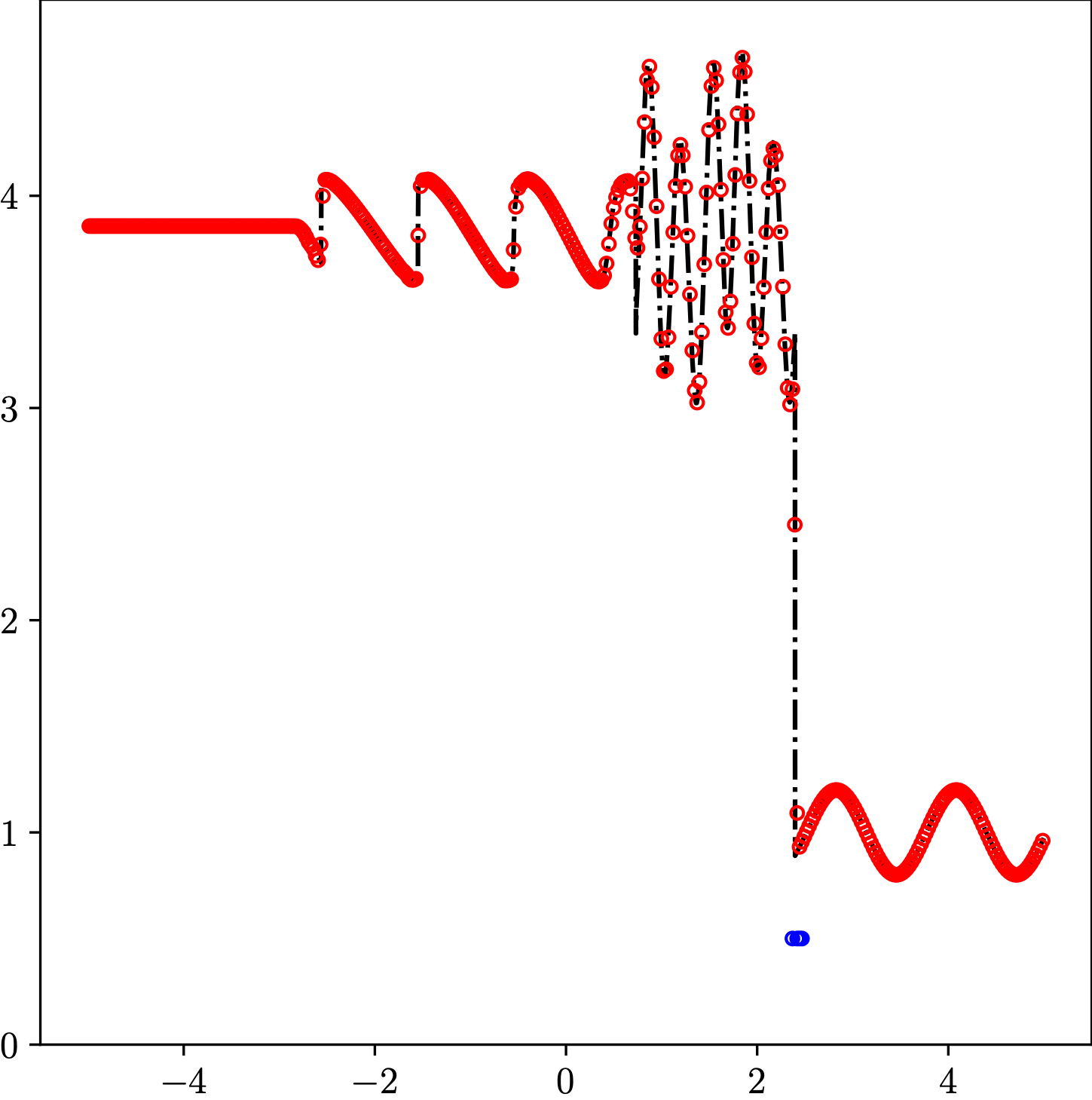}
\label{fig:c400}}
\subfigure[\textcolor{black}{Density profile, 800 points}]{\includegraphics[width=0.32\textwidth]{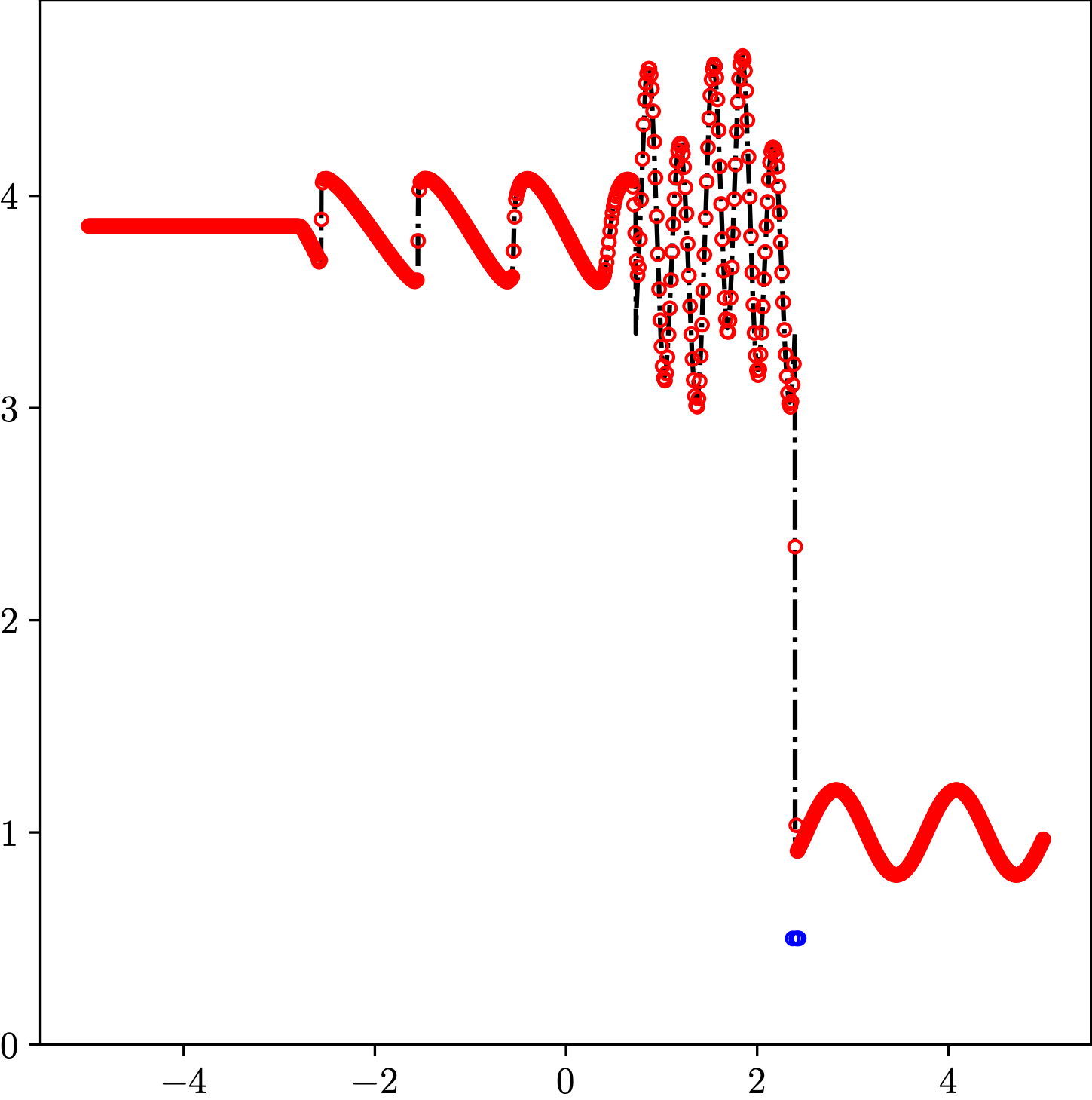}
\label{fig:c800}}
\caption{\textcolor{black}{Density profile for Shu-Osher test case with the proposed sensor using 200, 400 and 800 grid points and HY-THINC scheme: Figs. \ref{fig:c200}, \ref{fig:c400} and \ref{fig:c800}. Red circles: HY-THINC; blue circles: Sensor location; and dashed line: Reference solution. }}
\label{final-nail_hope}
\end{figure}

\textcolor{black}{The Li sensors' detection of high-frequency regions as discontinuities required further attention. It has been observed in the literature that some studies (and the corresponding discontinuity detectors) considered those regions not to be discontinuities. In \cite{fu2019hybrid}, one of the authors of the TENO-THINC scheme \cite{takagi2022novel} proposed a discontinuity detector based on the TENO scheme, and it has been mentioned in the corresponding paper that the number of cells detected as troubled cells decreases with increasing resolution. Fig. \ref{final-nail} shows results from \cite{fu2019hybrid} with TENO sensor. In the rightmost figure, with 800 grid points, the TENO discontinuity sensor did not detect the high-frequency region as discontinuities. The current approach did not alter the high-frequency region regardles of the grid resolution as shown in Fig. \ref{final-nail_hope}.} 

\begin{figure}[H]
\centering
\subfigure[\textcolor{black}{TENO-THINC detector \cite{takagi2022novel}}]{\includegraphics[width=0.3\textwidth]{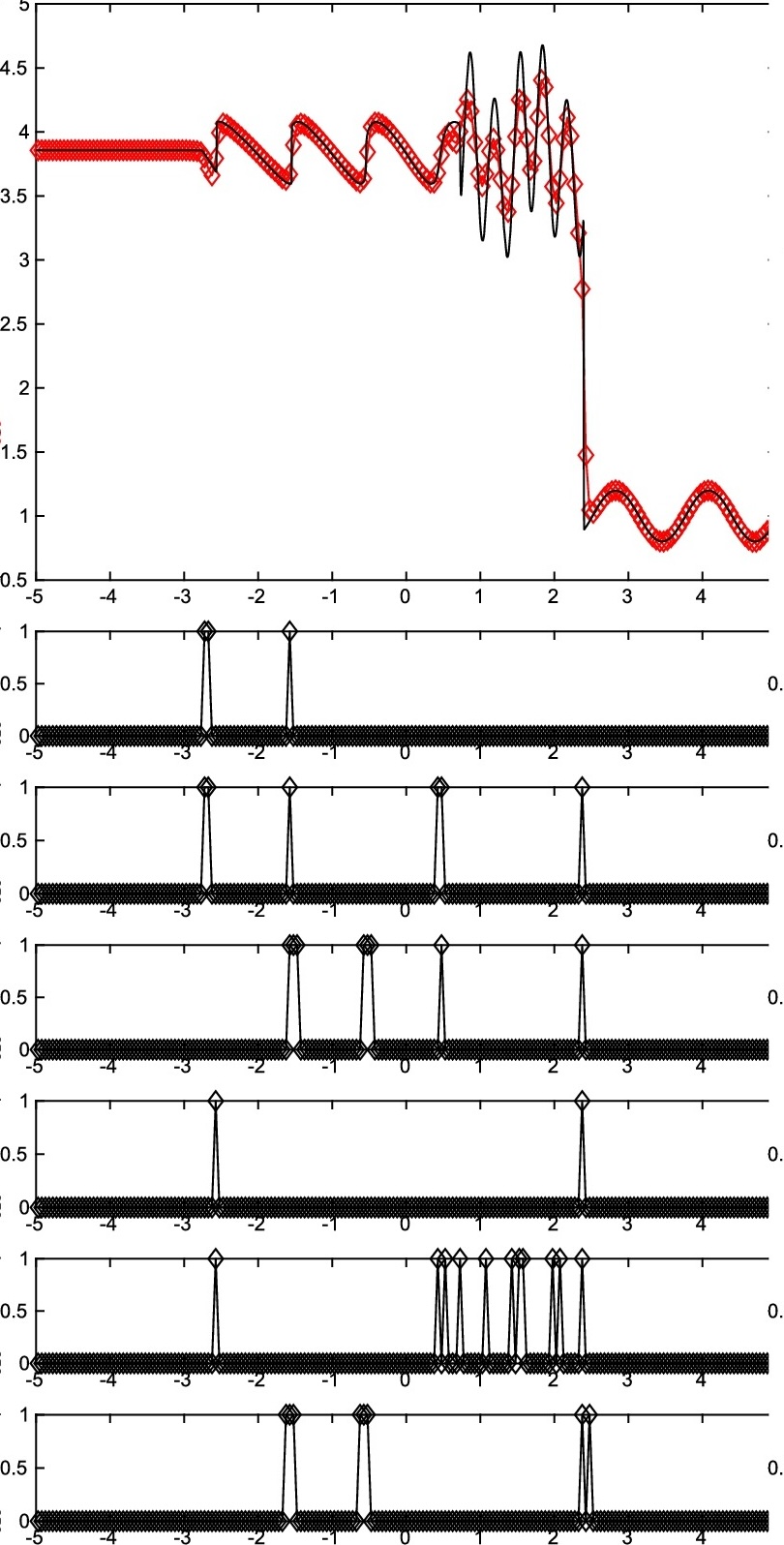}
\label{fig:shu-tt}}
\subfigure[\textcolor{black}{Krividonova detector \cite{krivodonova2004shock}}]{\includegraphics[width=0.37\textwidth]{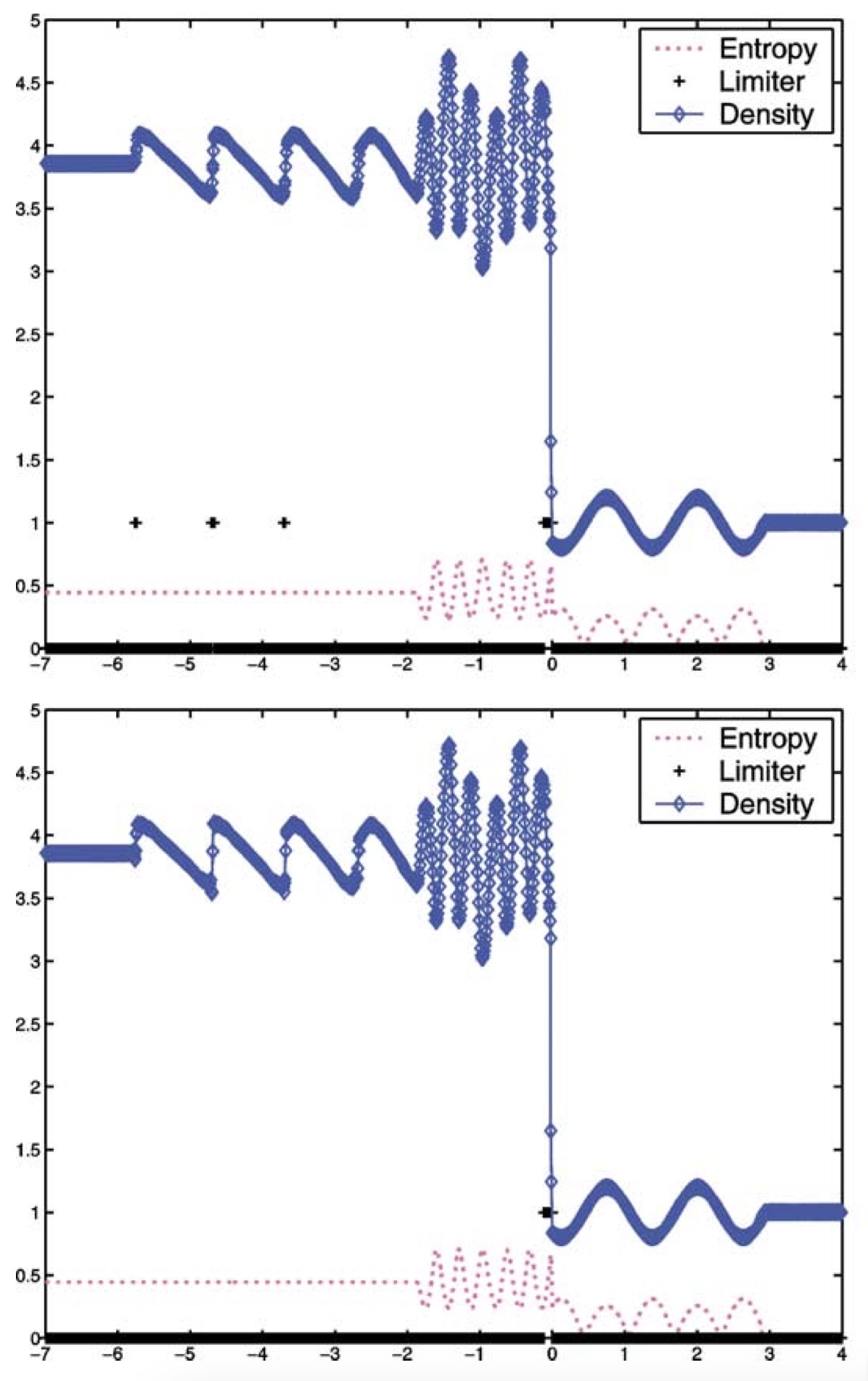}
\label{fig:shu-kriv}}
\subfigure[\textcolor{black}{Zhao detector \cite{zhao2020shock}}.]{\includegraphics[width=0.78\textwidth]{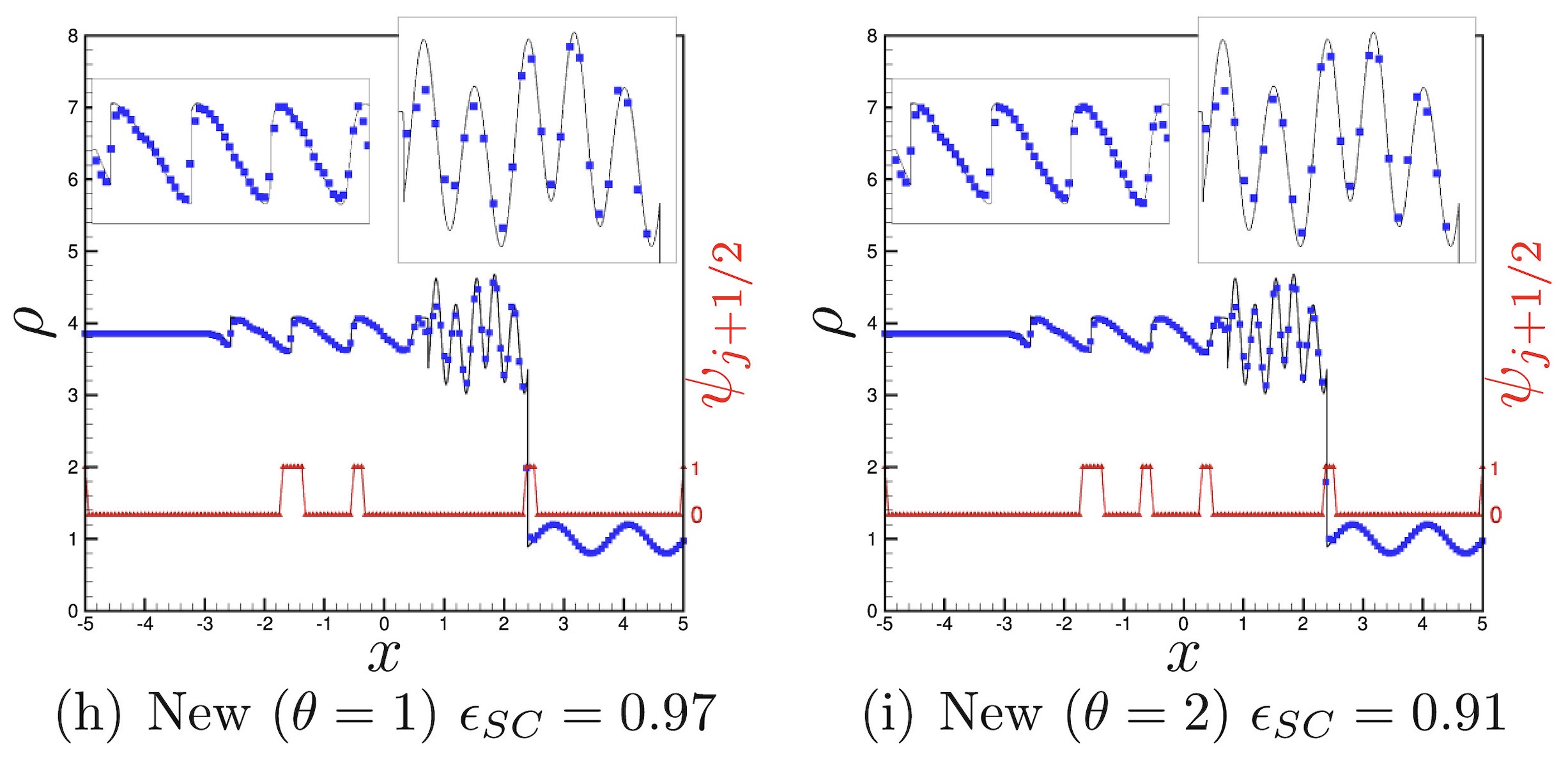}
\label{fig:shu-zhao}}
\caption{\textcolor{black}{Discontinuity detection locations in various papers from the literature. Fig. \ref{fig:shu-tt} is reproduced from \cite{takagi2022novel} with permission from Elsevier BV 2024, License number 5734370293424. Fig. \ref{fig:shu-kriv} is reproduced from \cite{krivodonova2004shock} with permission from Elsevier BV 2024, License number 5734370669124. Fig. \ref{fig:shu-zhao} is reproduced from  \cite{zhao2020shock} with permission from Elsevier BV 2024, License number 5734370461931.}}
\label{fig_shu-s}
\end{figure}

\begin{itemize}
\item Chamarthi and Frankel \cite{chamarthi2021high} also made similar observations in the work that the limiting process should be avoided in the high-frequency region by conducting simulations with a linear scheme in their paper (readers can refer to Fig. 14 in \cite{chamarthi2021high} and the corresponding discussion).
\item Furthermore, the TENO-THINC approach of \cite{takagi2022novel} detected the high-frequency region between $x \approx$ 0.7 and 2.4 as a discontinuity and applied THINC, as shown in Fig. \ref{fig:shu-tt}. Whereas the discontinuity detector of Krividonova et al. \cite{krivodonova2004shock} did not detect the high-frequency region with either density or entropy as a variable used in their shock detector. 
\item Finally, Zhao et al. \cite{zhao2020shock} studied several discontinuity sensors, and their new proposed sensor also did not alter the high-frequency region, as shown in Fig. \ref{fig:shu-zhao}. Figs. \ref{fig:c200}, \ref{fig:c400}, and \ref{fig:c800} show the density profile for the current sensor, and regardless of the resolution, the sensor did not detect the high-frequency region and applied THINC. The sensor detected only shockwave (but the THINC is applied only to the entropy wave and not to shockwave in this study), as indicated by the blue circles, similar to that of Krividonova et al. \cite{krivodonova2004shock}. 
\item Even for the multi-species case the sensor did not modify the regions of high-frequency, as shown in Fig. \ref{fig_shu-m}. \textbf{Using density as the variable to detect the interface also failed with the current sensor, as shown in Figure \ref{fig:sensor}.} All these observations played a role in devising the new sensor. Furthermore, the high-frequency region in the Shu-Osher test case is not a \textit{contact discontinuity at all.} 
\end{itemize}

\begin{figure}[H]
\centering
\subfigure[TENO-THINC, RKDG \cite{huang2025new}]{\includegraphics[width=0.4\textwidth]{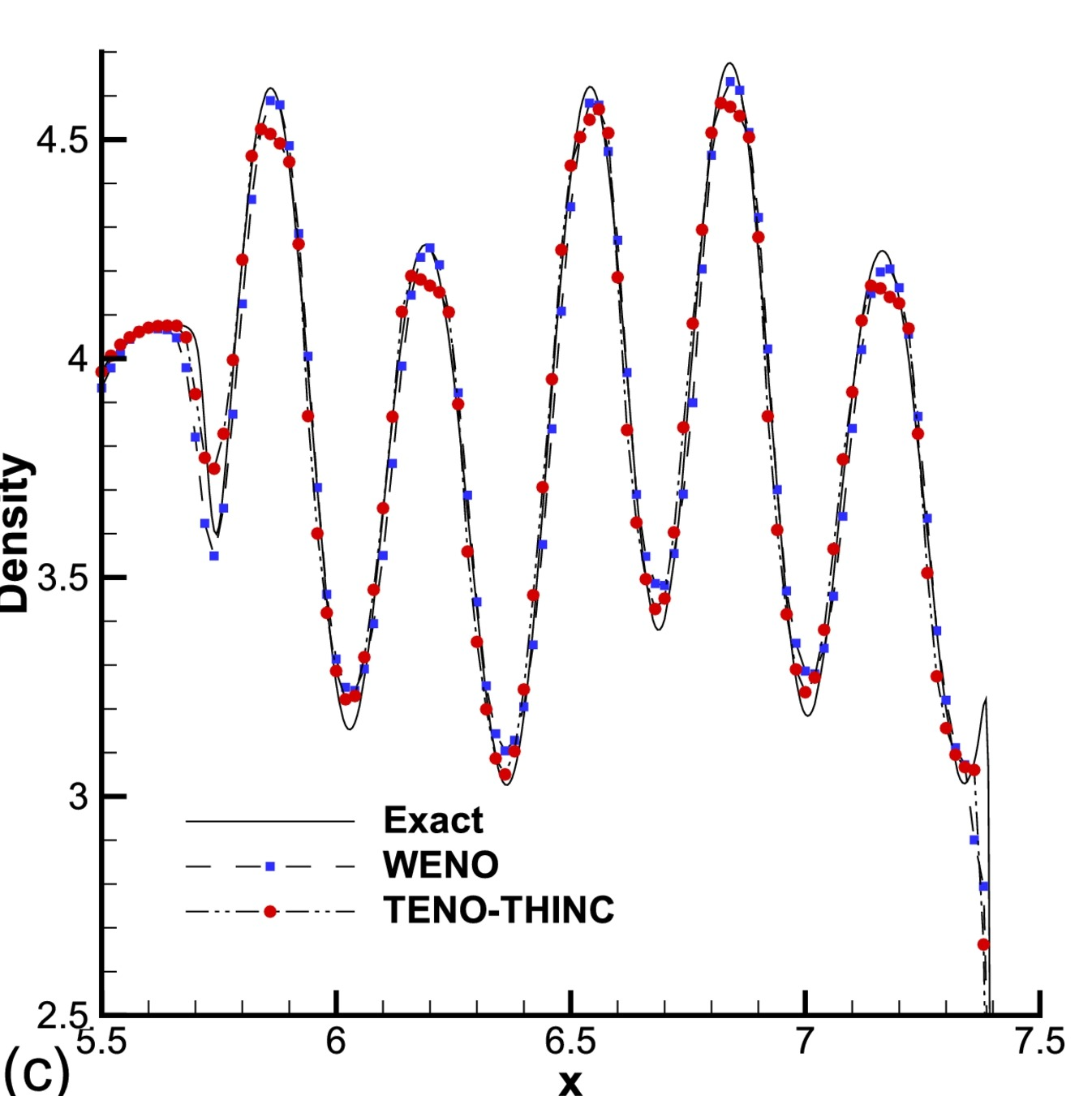}}
\caption{Results of Huang et al. \cite{huang2025new} using THINC   with permission from Elsevier BV 2024, License number 5936770943086.}
\label{fig_huang}
\end{figure}

The developers of the TENO scheme also integrated it with the THINC scheme using Discontinuous Galerkin methods. Although employing a distinct algorithm, the THINC scheme tended to activate in high-frequency regions, leading to less accurate results than those obtained with the WENO scheme, as illustrated in Fig. \ref{fig_huang}. The authors acknowledged that their proposed TENO-THINC limiter did not surpass the performance of the WENO limiter \cite{huang2025new}. In contrast, the current algorithm avoids using THINC in high-frequency regions. \textbf{These findings underscore the robustness and reliability of the present method while highlighting the potential issues of previous approaches.}

\subsection{Multi-dimensional test cases}

This section carries out numerical simulations for multi-dimensional test cases. Each example highlights the advantages of the proposed algorithms.

\begin{itemize}
	\item Examples \ref{ex:acc}, \ref{ex:TGV}, and \ref{ex:dsl} show that the proposed contact discontinuity sensor does not falsely detect and modify the results if there are no contact discontinuities. These test cases are single species to evaluate the proposed sensor.
\item In Example \ref{ex:kh}, it is demonstrated that tangential velocities across the contact discontinuity can be accurately reconstructed using a central scheme, effectively preventing oscillations. This example also illustrates that pressure can be accurately computed using a central scheme when reconstructing primitive variables.	
\item Example \ref{ex:triple} and \ref{ex:RM-viscous} show that even in the case of shock-material interface interaction, the tangential velocities can be computed using a central scheme.
	\item Example \ref{ex:multiple} is to show that the proposed sensor can detect contact discontinuities even if there are more than two species in the flow, and the sensor can be used for hypersonic flows.
\end{itemize}

\begin{example}\label{ex:acc} {Isentropic vortex (Inviscid case)}
\end{example}

\begin{figure}[H]
\centering
\subfigure[\textcolor{black}{Density profile}]{\includegraphics[width=0.4\textwidth]{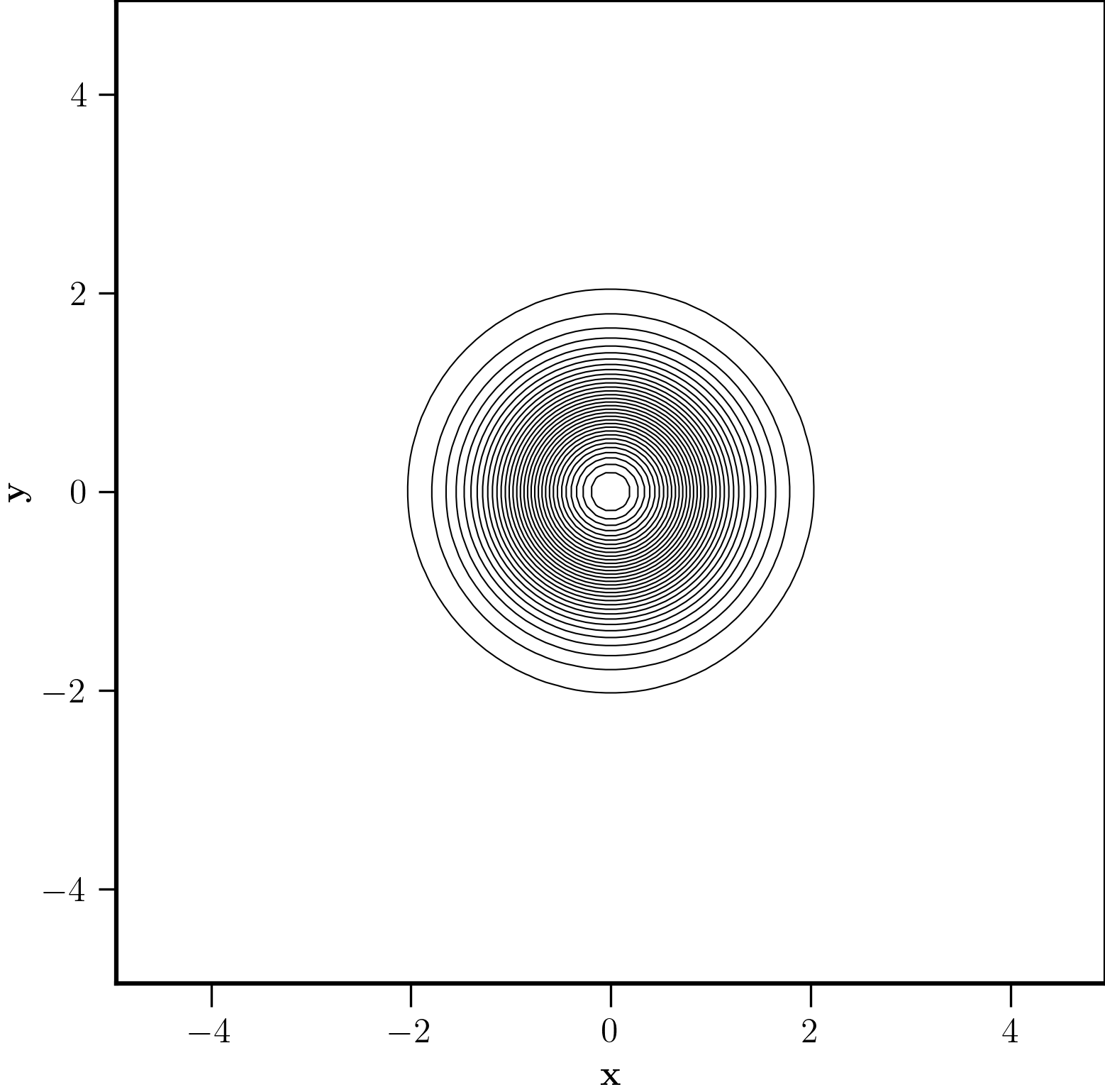}
\label{fig:isen-den}}
\subfigure[\textcolor{black}{Sensor location}]{\includegraphics[width=0.4\textwidth]{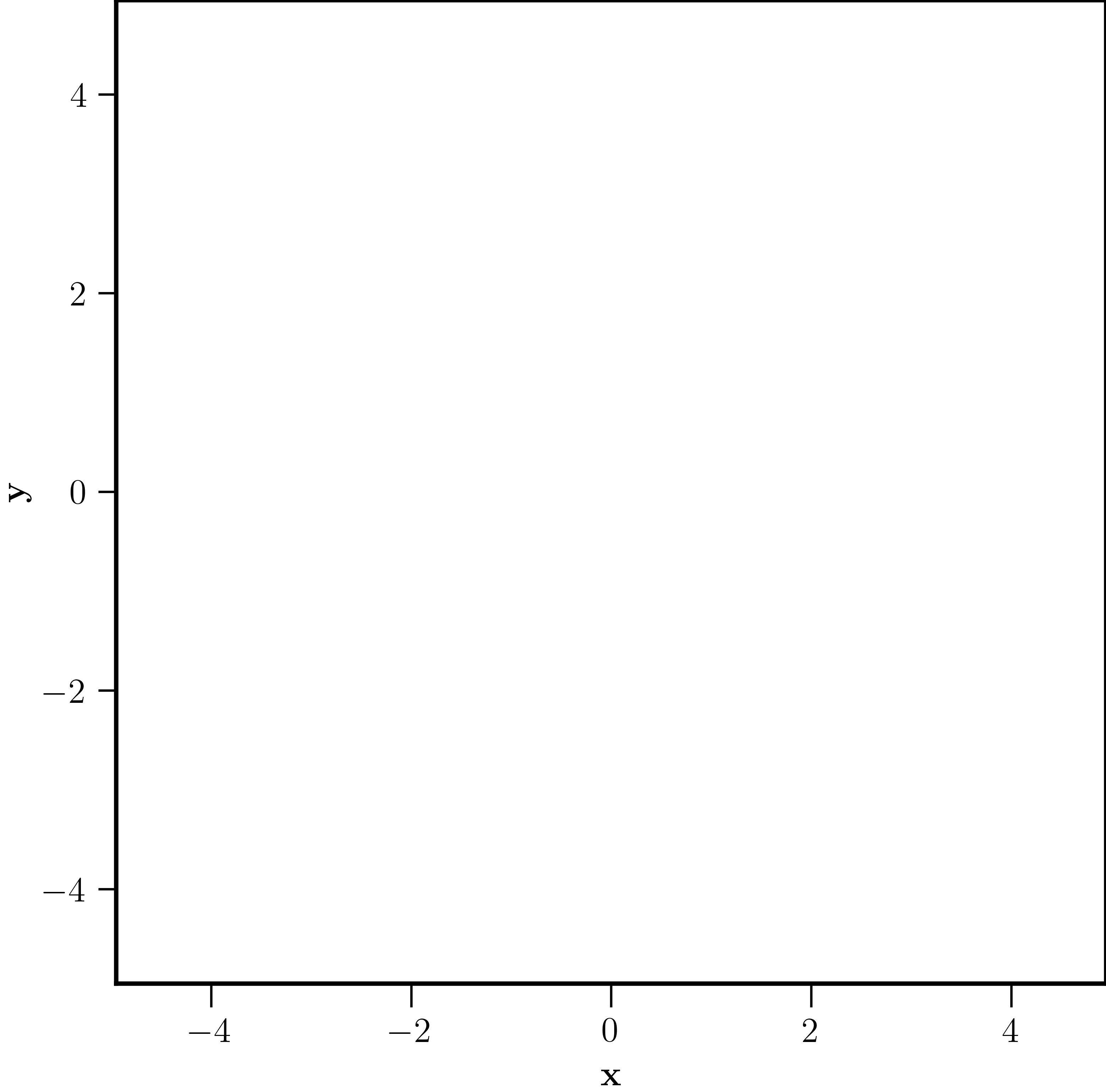}
\label{fig:isen-sensor}}
\caption{\textcolor{black}{Numerical solution (density contours) and sensor activation region for Isentropic vortex, Example \ref{ex:acc}.}}
\label{fig_isen}
\end{figure}

\textcolor{black}{In this test 2D inviscid case, the proposed numerical scheme is evaluated for the 
two-dimensional vortex evolution
problem \textcolor{black}{\cite{balsara2000monotonicity,yee1999low}}. This test case is typically considered for 
verifying the order
of accuracy of a proposed test case. Here, it is used to confirm whether the
sensor is falsely getting activated as no discontinuities are present. The computation domain is [-5, 5]
$\times$ [-5, 5] with periodic boundaries on all sides. To the mean flow, an
isentropic vortex is added, and the initial flow field is initialized as
follows:}

\begin{equation}
\textcolor{black}{p=\rho^{\gamma}, T=1-\frac{(\gamma-1) \varepsilon^{2}}{8
\gamma \pi^{2}} e^{\left(1-r^{2}\right)}, u=1-\frac{\varepsilon}{2 \pi}
e^{\frac{1}{2}\left(1-r^{2}\right)} y, v=1+\frac{\varepsilon}{2 \pi}
e^{\frac{1}{2}\left(1-r^{2}\right)} x,}
\end{equation}%
\textcolor{black}{where $r^2$ = $x^2$ + $y^2$ and the vortex strength
$\epsilon$ is taken as 5. The computations are performed to reach a final
time $t$ =10. The simulation is conducted on a grid size of 100 $\times$ 100. The density contours are shown in Fig. \ref{fig:isen-den}, and the regions of sensor activation are shown in Fig. \ref{fig:isen-sensor}. The
Fig. \ref{fig:isen-sensor} is blank as the sensor correctly detected no contact discontinuities in the flow and preserved the uniform entropy condition.}

\begin{example}\label{ex:TGV}{Inviscid Taylor-Green Vortex (Inviscid case)}
\end{example}

In this example, the performance of the contact discontinuity sensor in solving the three-dimensional inviscid Taylor-Green vortex problem, a classical benchmark problem in computational fluid dynamics, is investigated. This test case is a discontinuity-free test case. The proposed sensor should not have any effect on the solution. The initial conditions for the simulation are set on a periodic domain of size $x,y,z \in [0,2\pi)$, and the simulations are run until time $t=10$ on a grid size of $64^3$, with a specific heat ratio of $\gamma=5/3$. The flow problem is considered incompressible since the mean pressure is significantly large. The initial conditions of the test case are as follows:
\begin{equation}\label{itgv}
\begin{pmatrix}
\rho \\
u \\
v \\
w \\
p \\
\end{pmatrix}
=
\begin{pmatrix}
1 \\
\sin{x} \cos{y} \cos{z} \\
-\cos{x} \sin{y} \cos{z} \\
0 \\
100 + \frac{\left( \cos{(2z)} + 2 \right) \left( \cos{(2x)} + \cos{(2y)} \right) - 2}{16}
\end{pmatrix}.
\end{equation}

\begin{figure}[H]
\centering
\subfigure[Kinetic energy]{\includegraphics[width=0.46\textwidth]{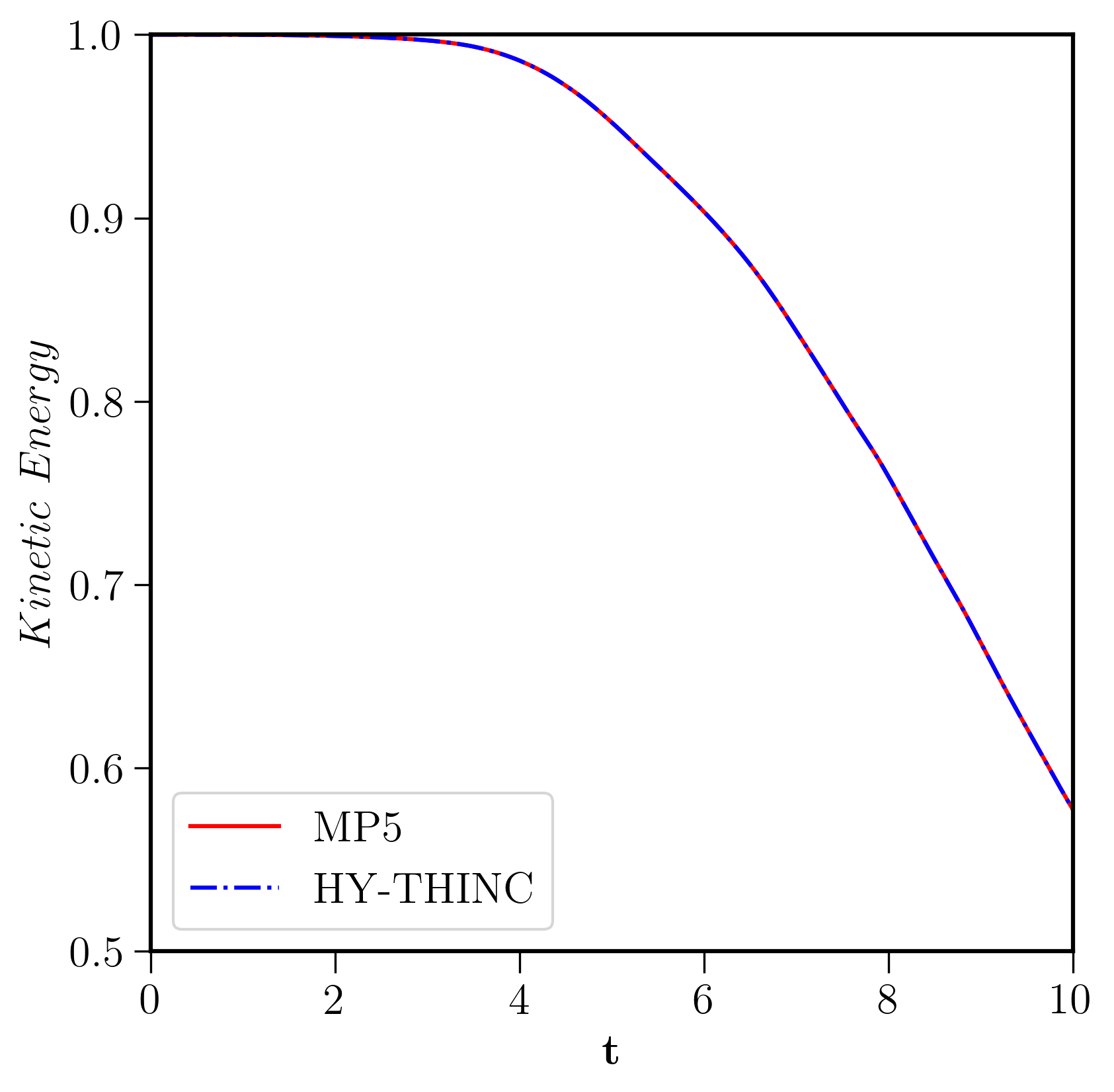}
\label{fig:TGV_KE}}
\subfigure[Enstrophy]{\includegraphics[width=0.45\textwidth]{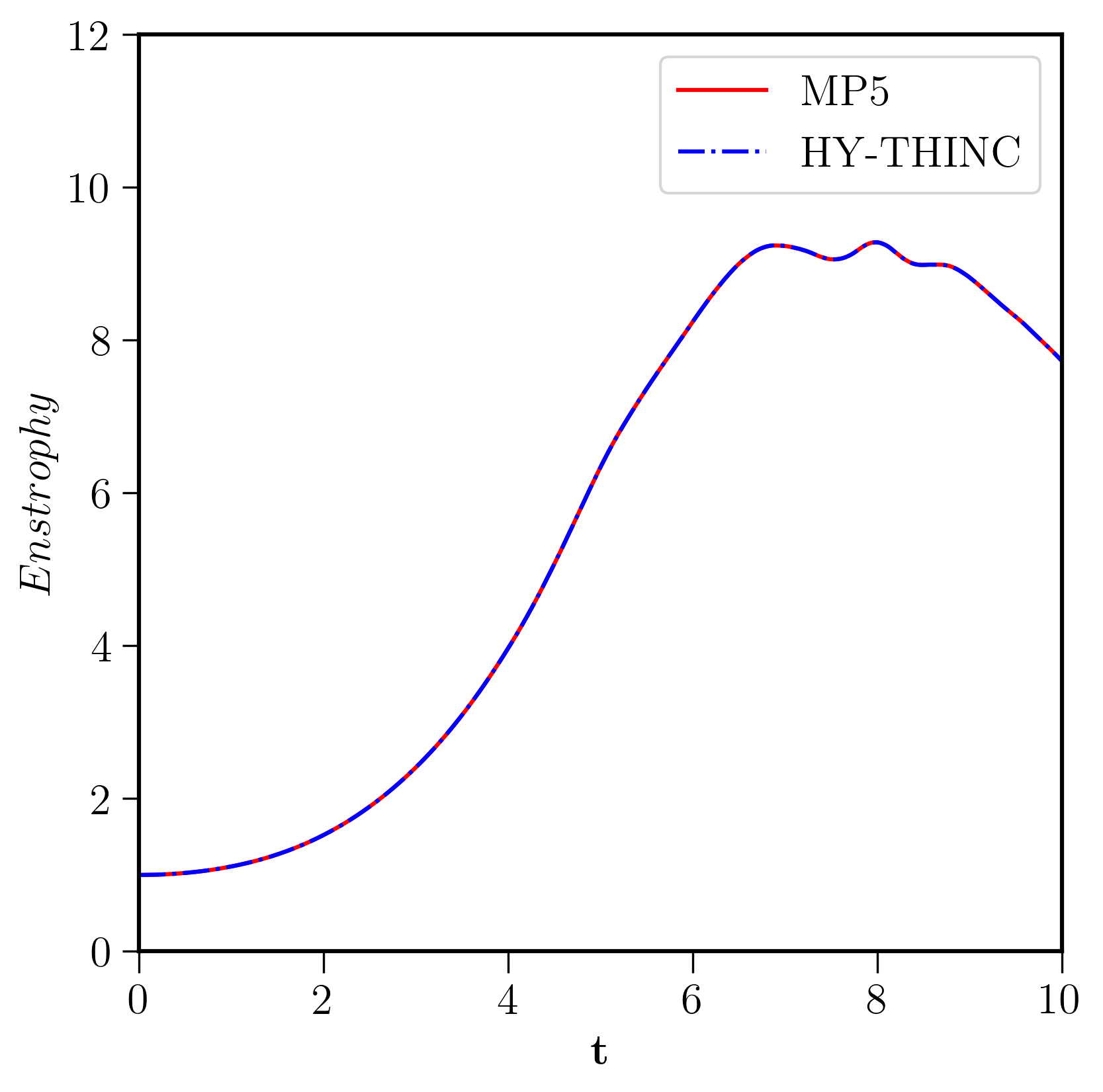}
\label{fig:TGV_ens}}
\caption{Normalised kinetic energy and enstrophy using HY-THINC and MP5 schemes for Example \ref{ex:TGV} on grid size of $64^3$. Solid red line: MP5 and dashed blue line: HY-THINC.}
\label{fig_TGV}
\end{figure}

Fig. \ref{fig_TGV} indicates that the proposed scheme (the contact discontinuity sensor) did not affect this test case as it should. The results obtained by the MP5 and HY-THINC scheme are one over the other for kinetic energy and enstrophy. Even though there are no discontinuities in this test case, the TENO-THINC scheme of Takagi et al. \cite{takagi2022novel} has improved the results (see Fig. 21 of \cite{takagi2022novel}), which indicates that either the TENO based indicator is falsely detecting smooth flow regions as discontinuities or it could be an issue of reconstructing all the variables using the THINC scheme. It is beyond the scope of the paper to analyze the TENO-THINC, but the current approach is free of such unexpected results.

\begin{example}\label{ex:dsl}{Periodic double-shear layer (Viscous case)}
\end{example}

In this 2D viscous test case, the impact of applying THINC to all the waves, as in \cite{takagi2022novel}, is investigated. The test involves two initially parallel shear layers that develop into two significant vortices at $t = 1$. All tests were run on a $ 196 \times 196 $ grid size. The non-dimensional parameters for this test case are presented in Table \ref{tab:shearLayerNondimensionalParameters}.

\begin{figure}[H]
%\begin{halfspacing}
\centering\offinterlineskip
\subfigure[Fine-grid, $800^2$.]{\includegraphics[width=0.3\textwidth]{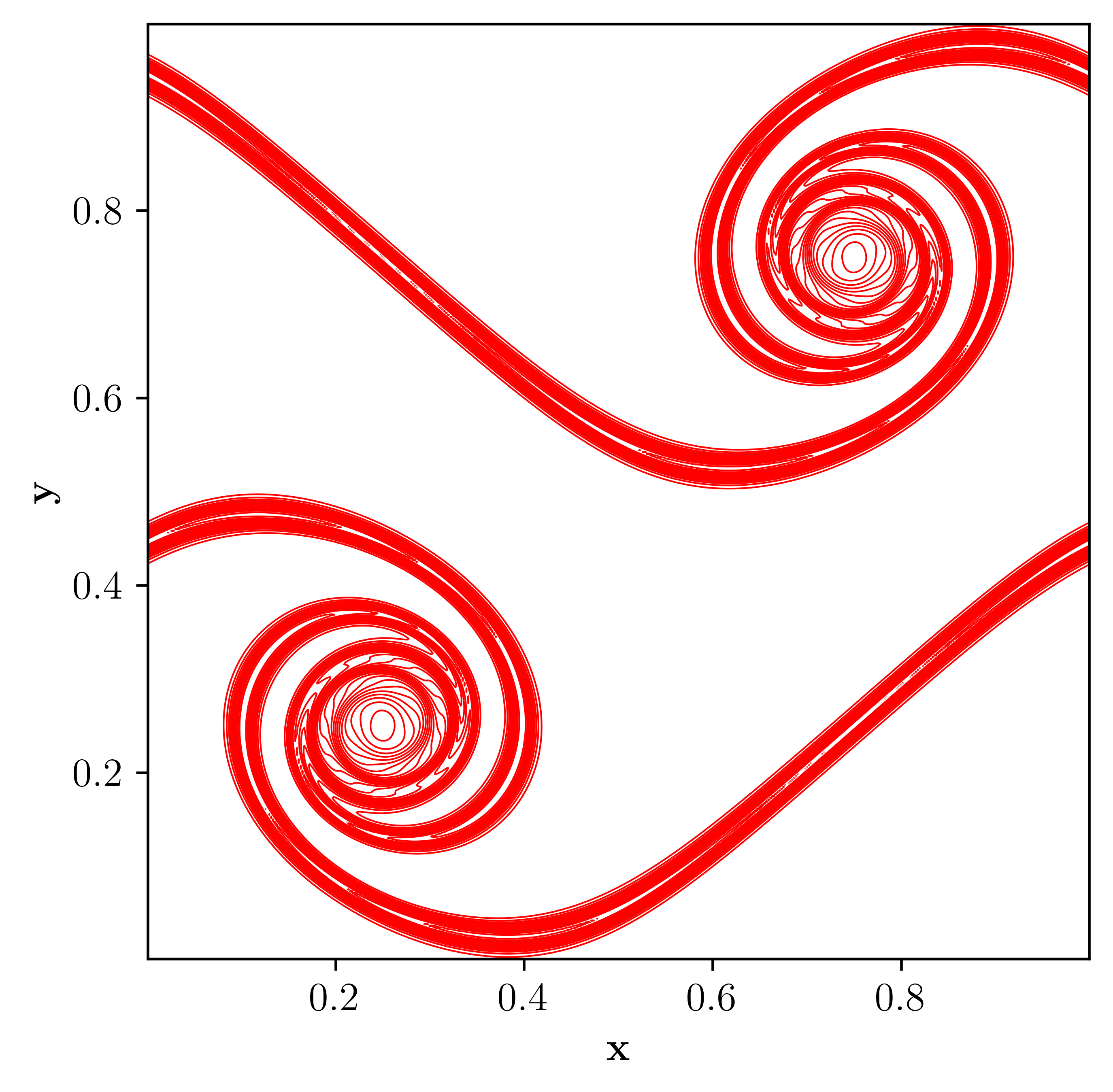}
\label{fig:fine_dsl}}
\subfigure[MP5 - Upwind.]{\includegraphics[width=0.3\textwidth]{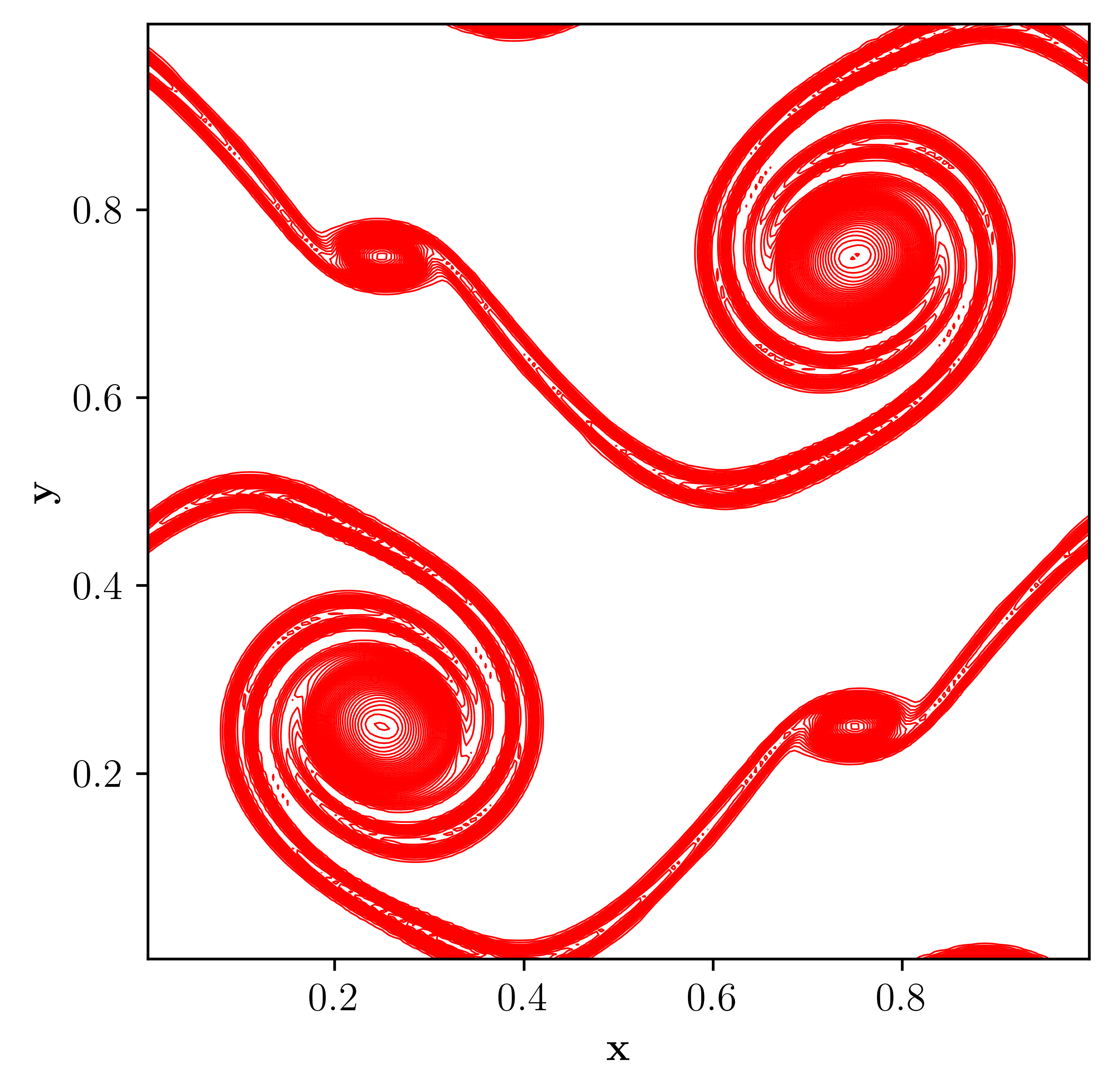}
\label{fig:mp_dsl}}
\subfigure[MP6 - Ducros.]{\includegraphics[width=0.3\textwidth]{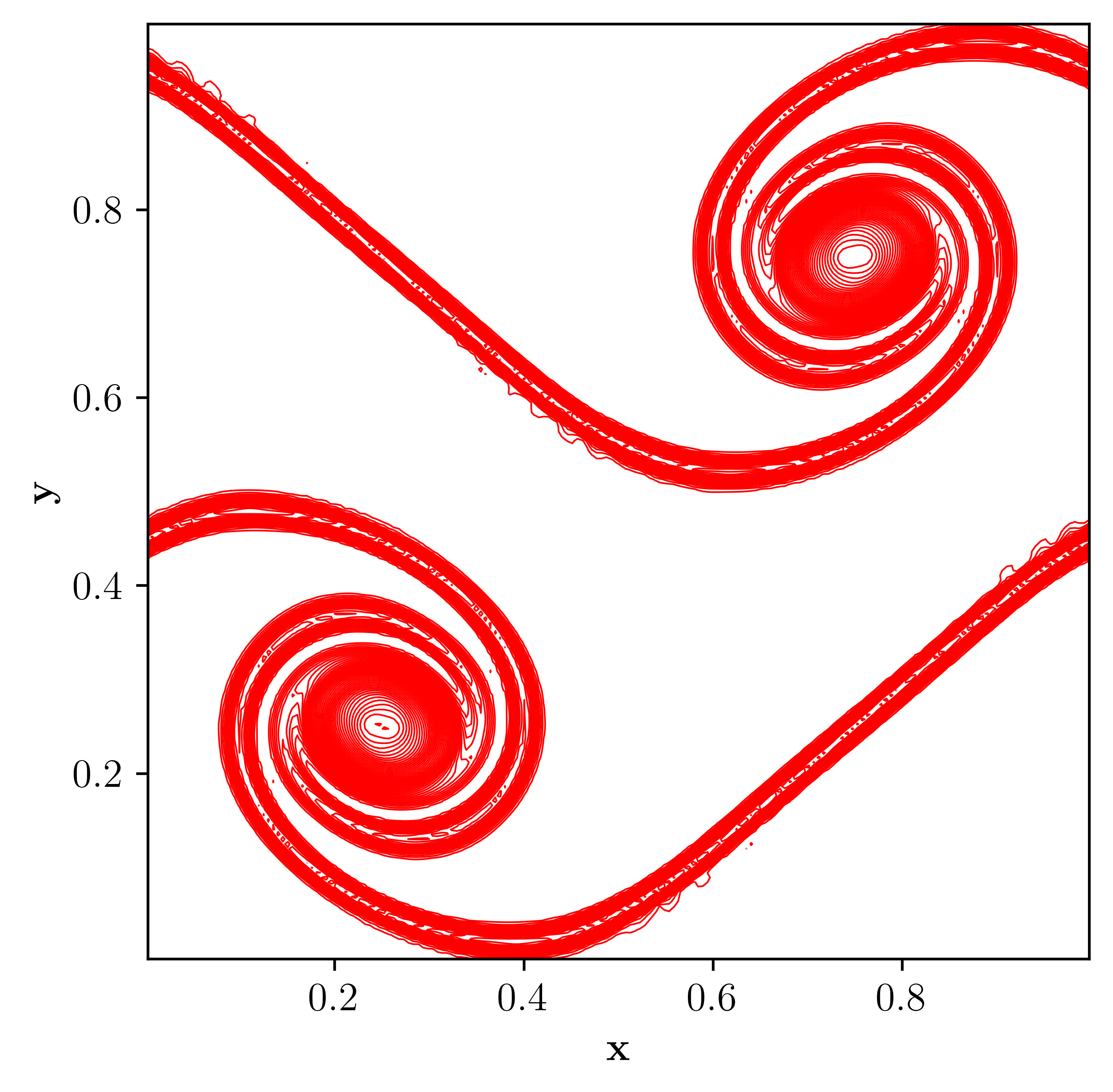}
\label{fig:mp6D_dsl}}
\subfigure[HY-THINC-D.]{\includegraphics[width=0.3\textwidth]{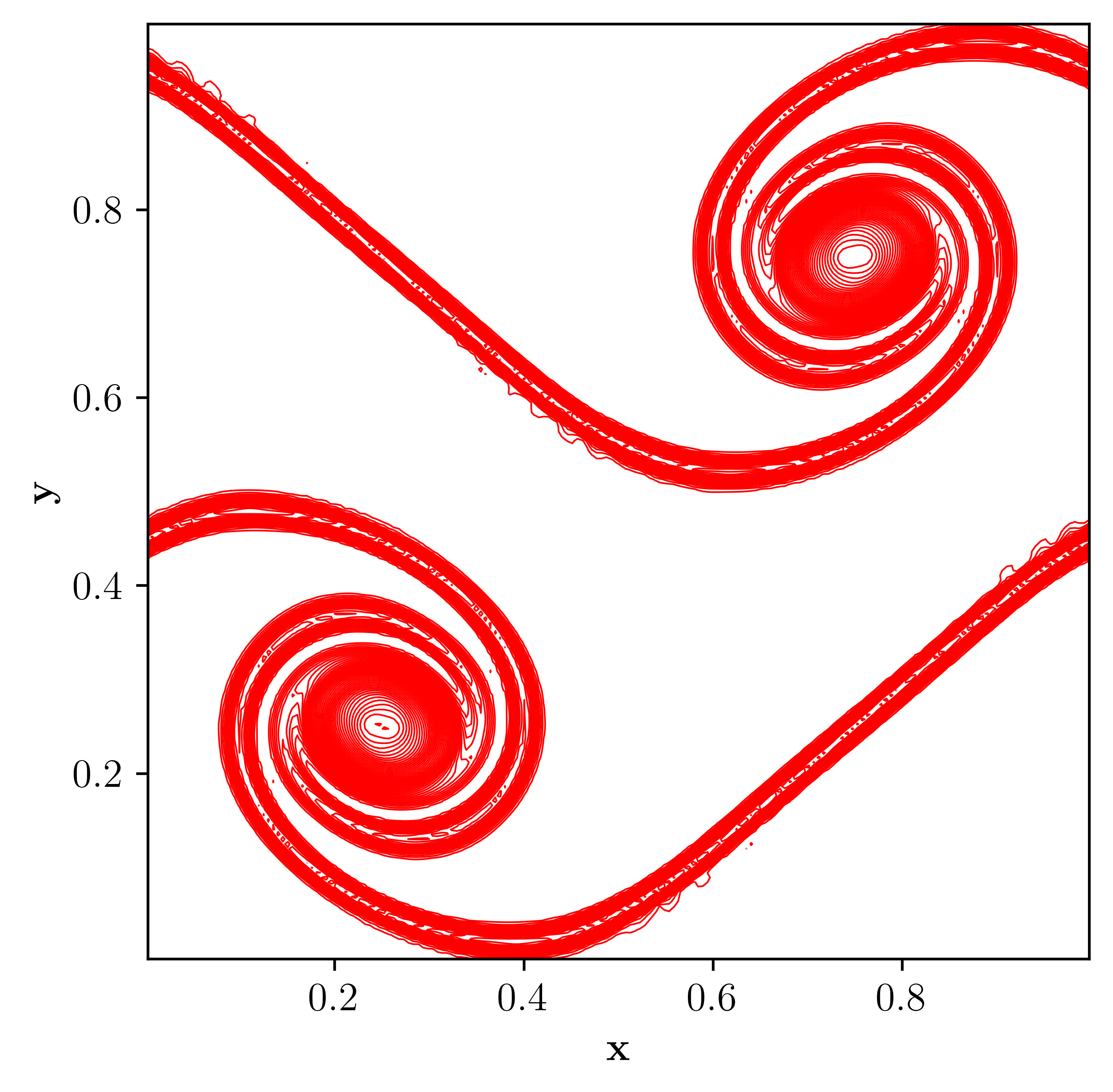}
\label{fig:hythinc_dsl}}
\subfigure[TENO5.]{\includegraphics[width=0.3\textwidth]{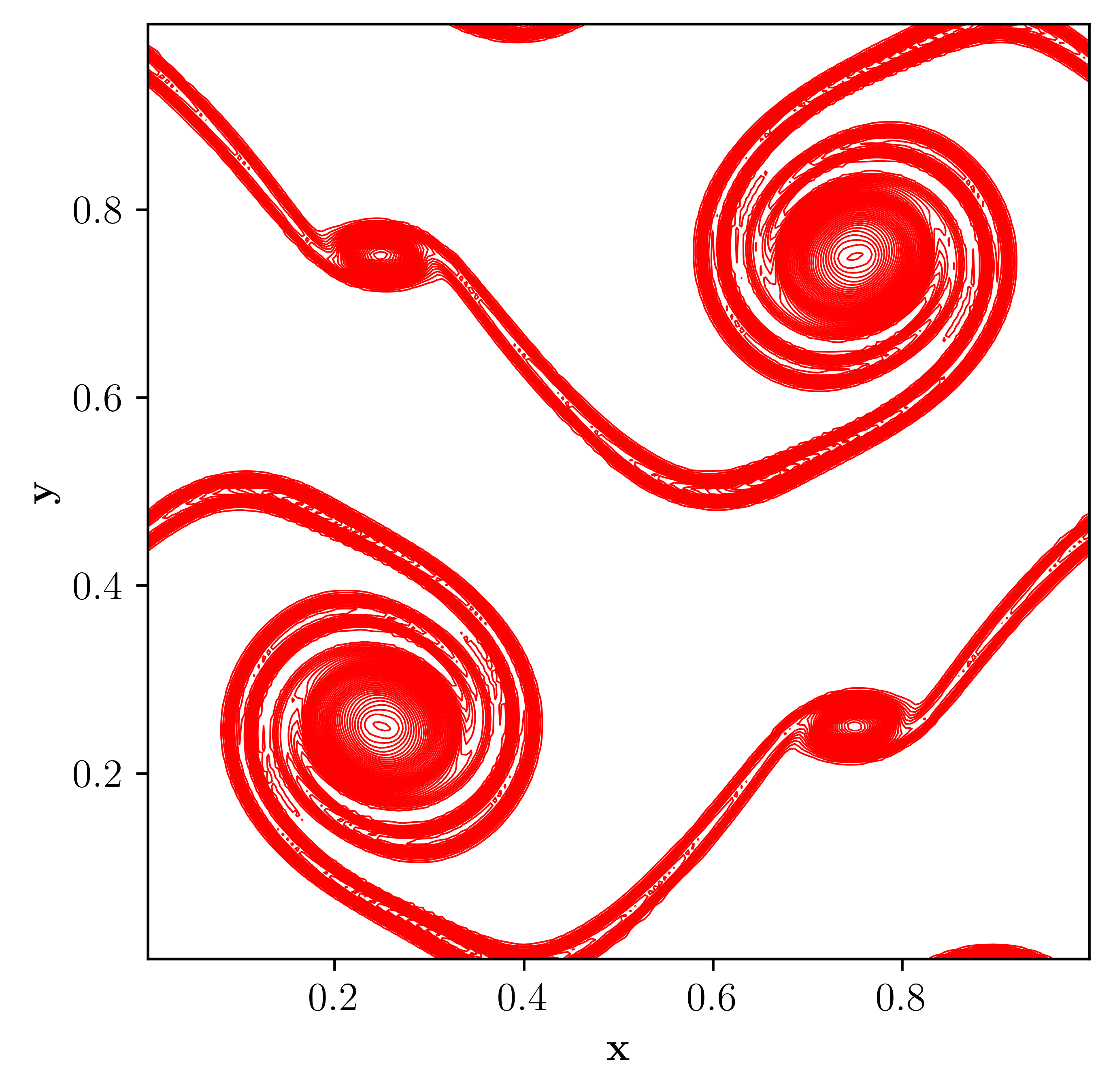}
\label{fig:teno_dsl}}
\subfigure[TENO5-THINC.]{\includegraphics[width=0.3\textwidth]{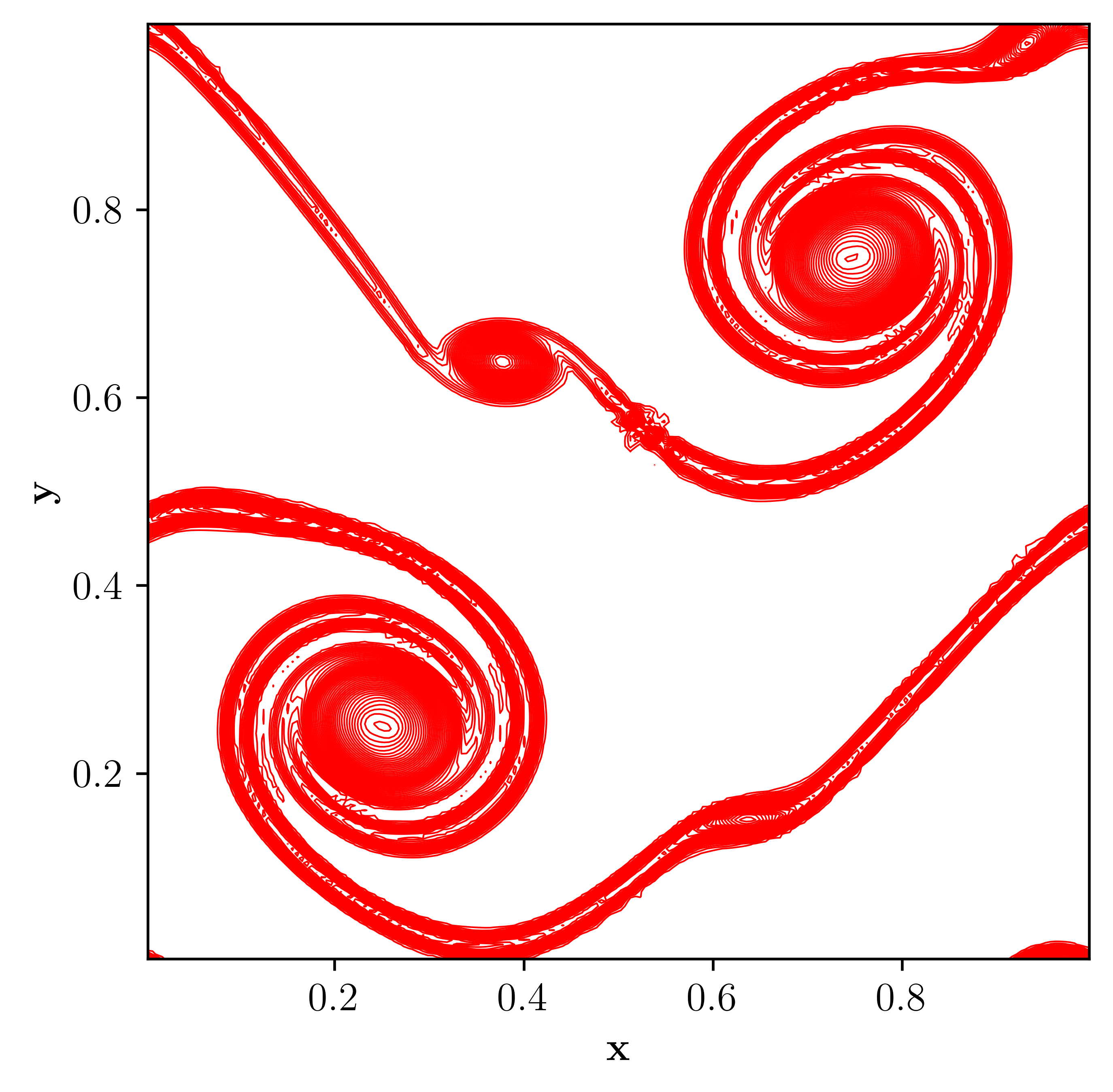}
\label{fig:tenothinc_dsl}}
\caption{Figures show $z$-vorticity contours of the considered schemes on a grid size of $196^2$ for $\mu = 1.0 \times 10^{-4}$, Example \ref{ex:dsl}.}
    \label{fig:dpsl_196}
%\end{halfspacing}
\end{figure}

\begin{table}[h!]
    \centering
    \caption{\textcolor{black}{Parameters of the periodic double shear layer test case.}}
    \begin{tabular}{c c c c}
        \hline
        \hline
        $\mathrm{Ma}$ & $\mathrm{Re}$ & Pr & $\gamma$ \\
        \hline
        0.1 & 10,000 & 0.73 & 1.4 \\
        \hline
        \hline
    \end{tabular}
    \label{tab:shearLayerNondimensionalParameters}
\end{table}
\noindent \textcolor{black}{The initial conditions were:}
\begin{subequations}
   \textcolor{black}{ \begin{align}
        p &= \frac{1}{\gamma \mathrm{Ma}^2}, \rho=1, \ u= 
        \begin{cases}
            \tanh \left[ \theta (y-0.25) \right], & \text{ if } (y \leq 0.5), \\
            \tanh \left[ \theta (0.75-y) \right], & \text{ if } (y > 0.5),
        \end{cases} 
        \\[10pt]
        v &=  0.05 \sin \left[ 2 \pi(x) \right] \text{and}\ \theta  = 120\ \text{or}\ 80, 
    \end{align}}
\end{subequations}
Simulations are conducted for $\mu = 1.0 \times 10^{-4}$ and $\mu = 3.0 \times 10^{-5}$. The reference solution $\mu = 1.0 \times 10^{-4}$ and $\theta$=120, shown in Fig. \ref{fig:fine_dsl}, was computed with the MP scheme on a $ 800 \times 800$ grid. Unphysical braid vortices and oscillations can occur on the shear layers if the grid is under-resolved for this test case. Figs. \ref{fig:mp_dsl}, \ref{fig:mp6D_dsl}, and  \ref{fig:hythinc_dsl} displays the $z$-vorticity computed by MP, MP6 - Ducros and HY-THINC-D schemes on a grid size of  196 $\times$ 196. As expected, the upwind scheme, MP5, gave unphysical braid vortices and the MP6 - Ducros and HY-THINC-D schemes, where the central scheme computes the tangential velocities, are similar to the fine grid results. In this test case, the THINC scheme is not activated as there are no contact discontinuities, which indicates the proposed sensor works reliably. 

\begin{figure}[H]
%\begin{halfspacing}
\centering\offinterlineskip
\subfigure[HY-THINC-D, Primitive variables.]{\includegraphics[width=0.3\textwidth]{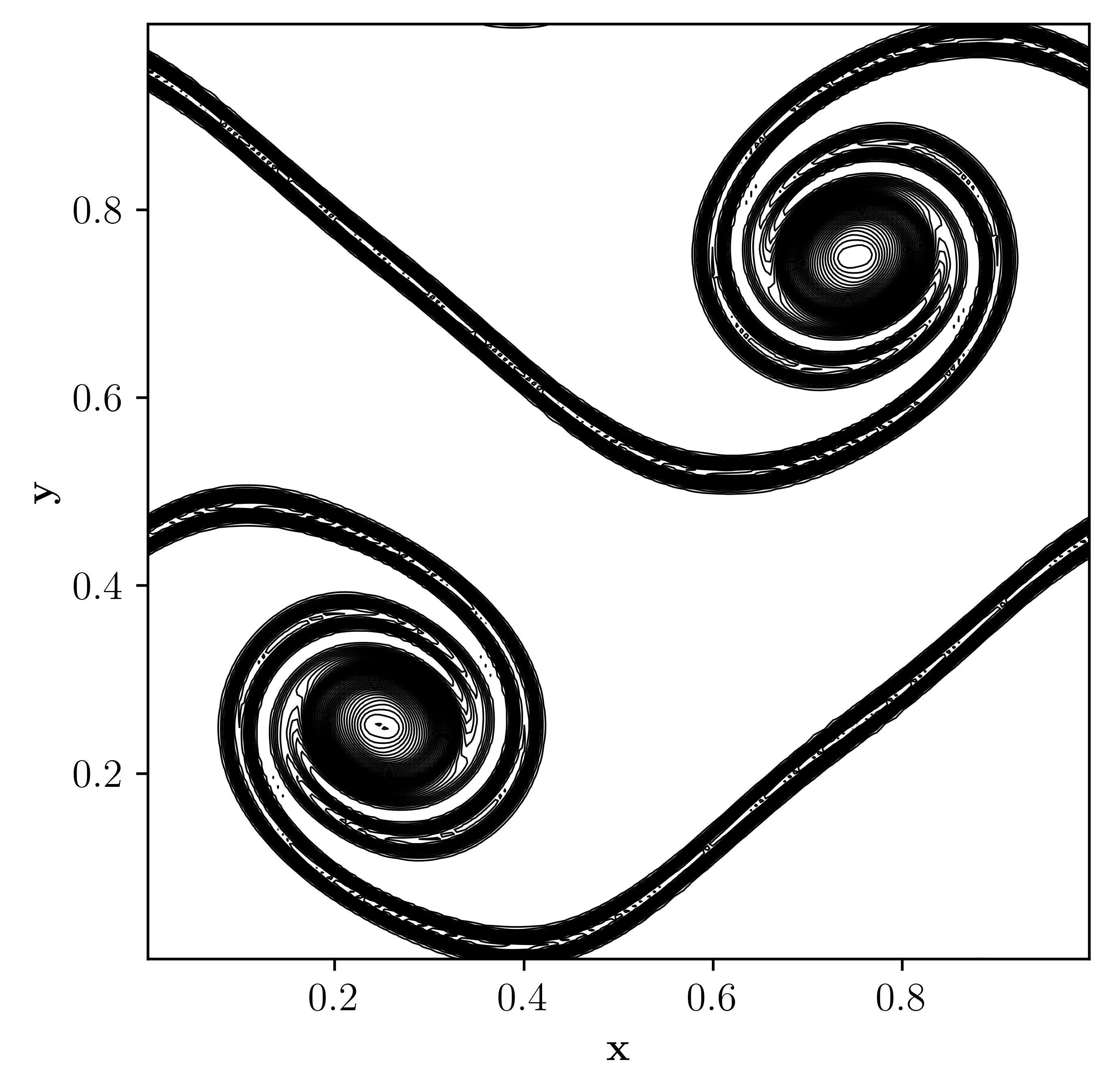}
\label{fig:primthinc_dsl}}
\caption{$z$-vorticity contours of the HY-THINC-D scheme with primitive variable reconstruction on a grid size of $196^2$ for $\mu = 1.0 \times 10^{-4}$ and $\theta$ = 120, Example \ref{ex:dsl}.}
    \label{fig:dpsl_prim}
%\end{halfspacing}
\end{figure}

Observing the Figs. \ref{fig:teno_dsl} and \ref{fig:tenothinc_dsl}, the TENO and TENO-THINC results are not identical. While the results obtained by the TENO scheme are, as expected, similar to the MP5 scheme as all the variables are computed using the upwind scheme, the TENO-THINC scheme results further deviate from the TENO5 itself. It indicates the TENO-based discontinuity sensor is falsely getting activated and is affecting the results where there are no discontinuities. Simulations with the proposed contact discontinuity sensor and Ducros sensor are also free of spurious vortices even if reconstruction is directly carried out for the primitive variables, shown in Fig. \ref{fig:primthinc_dsl}.

\begin{figure}[H]
\centering
 \includegraphics[width=0.9\textwidth]{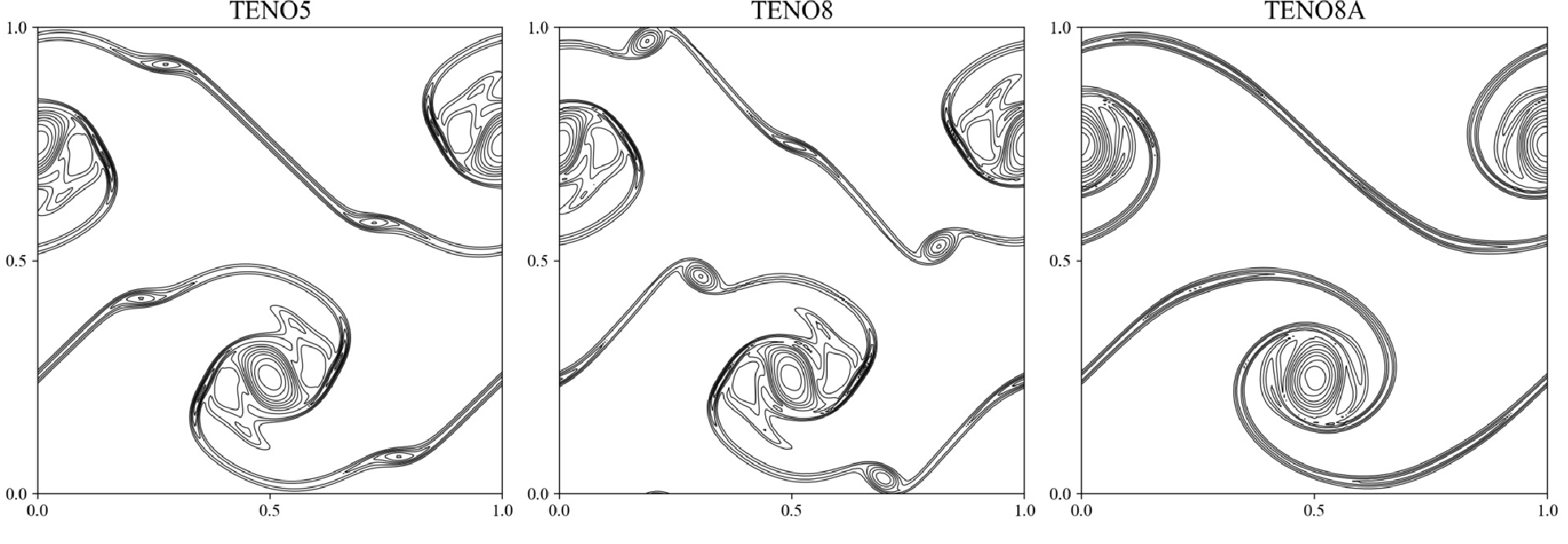}
    \caption{Figure is taken from Reference \cite{feng2024general}, where the simulations are computed on a grid size of $320^2$.}
\label{fig:teno-adams}
\end{figure}

For comparison, Figure \ref{fig:teno-adams}, adapted from \cite{feng2024general}, shows simulations performed on a grid of 320 $\times$ 320 points—2.6 times larger than the current approach. Despite this higher resolution and the optimized TENO8 scheme, braid vortices still exist in their results. These results underscore the advantages of the current approach of applying a central scheme to the tangential velocities. 

\begin{figure}[H]
\subfigure[MP5 - Upwind.]{\includegraphics[width=0.3\textwidth]{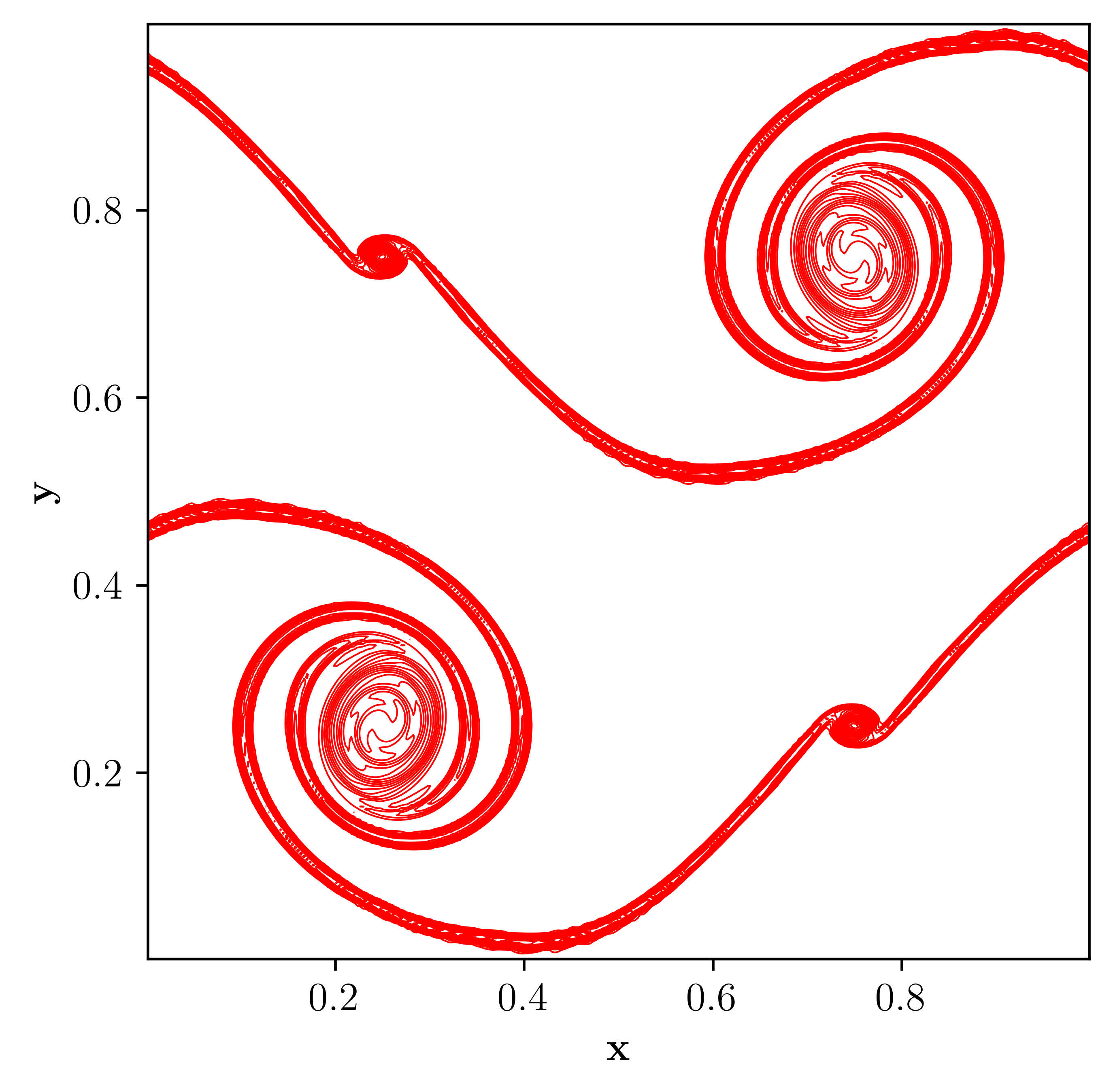}
\label{fig:mp_central_inv}}
\subfigure[MP6 - Ducros.]{\includegraphics[width=0.3\textwidth]{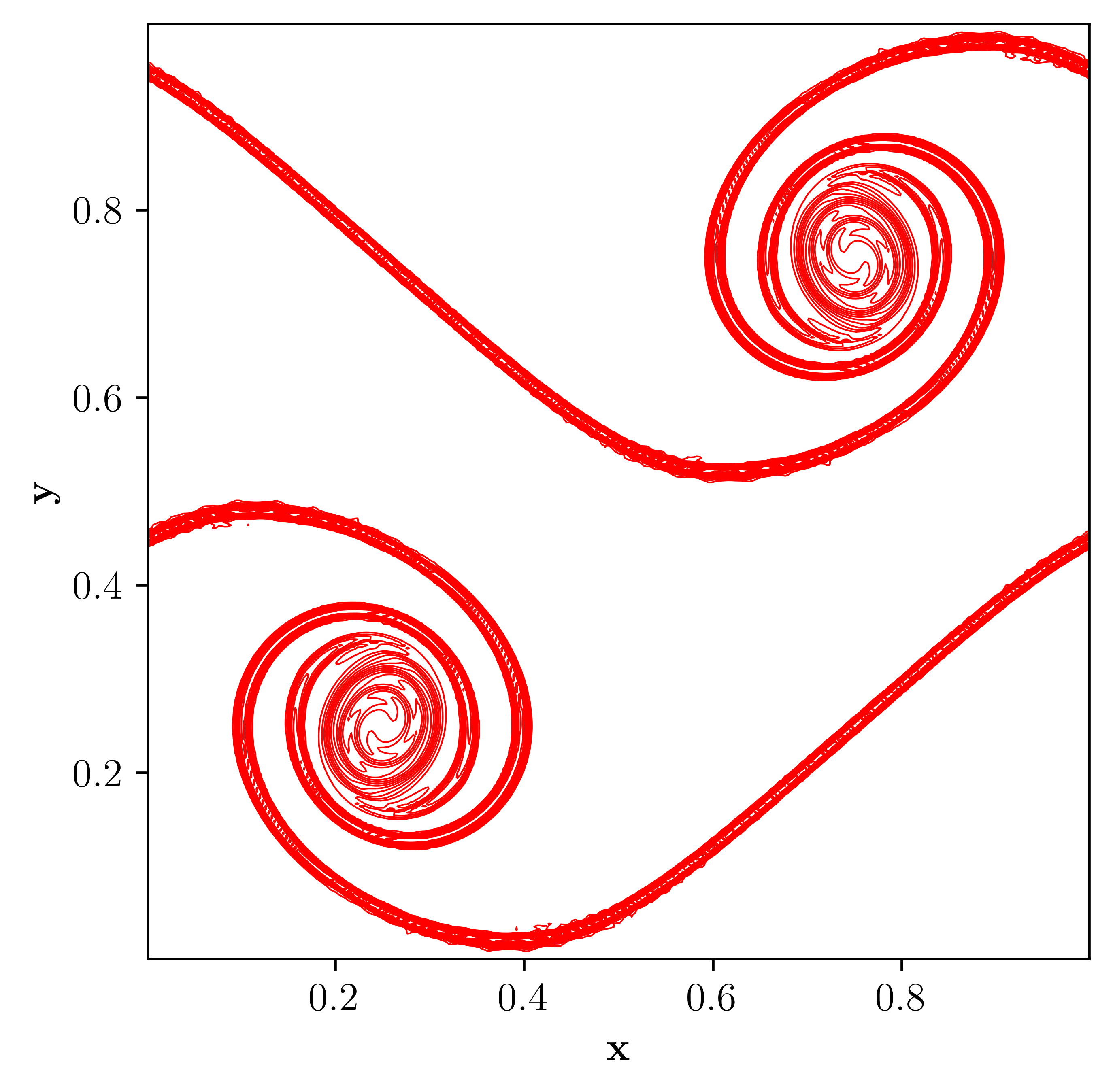}
\label{fig:mp_central_thinc_inv}}
\subfigure[HY-THINC-D.]{\includegraphics[width=0.3\textwidth]{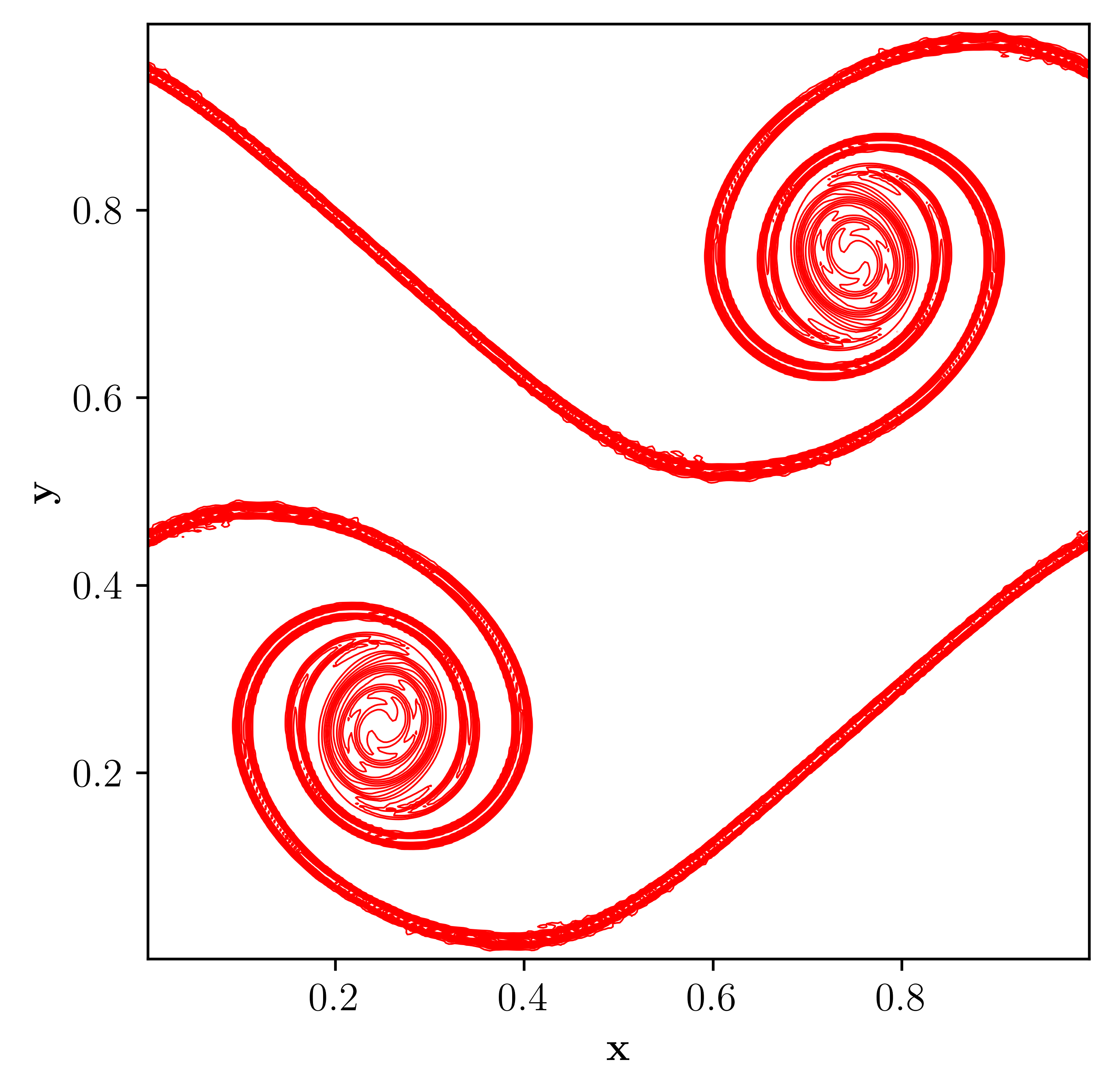}
\label{fig:mp_central_inv_tan}}
\caption{Figures show $z$-vorticity contours of the considered schemes computed on a grid size of $320^2$ for $\mu = 3.0 \times 10^{-5}$ and $\theta$ = 80, Example \ref{ex:dsl}.}
    \label{fig:dpsl_96_low}
\end{figure}

Finally, the simulation results for $\mu = 3.0 \times 10^{-5}$ and $\theta$ = 80 are shown in Fig. \ref{fig:dpsl_96_low}. Figs. \ref{fig:mp_central_inv}, \ref{fig:mp_central_thinc_inv}, and  \ref{fig:mp_central_inv_tan} displays the $z$-vorticity computed by MP, MP6 - Ducros and HY-THINC-D schemes on a grid size of  320 $\times$ 320. Once again, the upwind scheme, MP5, gave unphysical braid vortices and the MP6 - Ducros and HY-THINC-D schemes are similar to the fine grid results. For this scenario, the TENO-THINC scheme failed to pass the test (severe oscillations and eventually crashed).

It has been explained in \cite{chamarthi2024generalized} that computing tangential velocities using a central scheme will prevent unphysical vortices for this test case. In the later test cases, numerical examples will show that the tangential velocities can be computed using a central scheme, even across material interfaces.

\begin{example}\label{ex:kh}{Kelvin Helmholtz instability (Viscous case)}
\end{example}

The Kelvin-Helmholtz instability (KHI) is a hydrodynamic instability that occurs when there is a velocity shear between two fluids of different densities. This instability arises due to an unstable velocity gradient at the interface between the two fluids, leading to vortices and turbulent mixing patterns. It plays a crucial role in the evolution of the mixing layer and the transition to turbulence. This test case is typically a single-species test case but can be modified to a multi-species one. The initial conditions are set over a periodic domain of [0, 1] $\times$ [0, 1].

\begin{equation}
\begin{array}{l}
p=2.5, \quad \alpha_1\rho_1(x, y)=\left\{\begin{array}{l}
2, \text { if } 0.25<y \leq 0.75, \\
0, \text { else. }
\end{array}\right., \alpha_2\rho_2(x, y)=\left\{\begin{array}{l}
0, \text { if } 0.25<y \leq 0.75, \\
1, \text { else. }
\end{array}\right. \\
\\[5pt]
u(x, y)=\left\{\begin{array}{l}
0.5, \quad \text { If } 0.25<y \leq 0.75, \\
-0.5, \text { else }.
\end{array}\right.\alpha_1=\left\{\begin{array}{l}
1.0, \quad \text { If } 0.25<y \leq 0.75, \\
0.0, \quad \text { else }.
\end{array}\right.\\
\\[5pt]
v(x, y)=0.1 \sin (4 \pi x)\left\{\exp \left[-\frac{(y-0.75)^{2}}{2 \sigma^{2}}\right]+\exp \left[-\frac{(y-0.25)^{2}}{2 \sigma^{2}}\right]\right\}, \text{where}\ \sigma = 0.05/\sqrt {2}.
\end{array}
\end{equation}

\begin{figure}[H]
\subfigure[Initial condition.]{\includegraphics[width=0.32\textwidth]{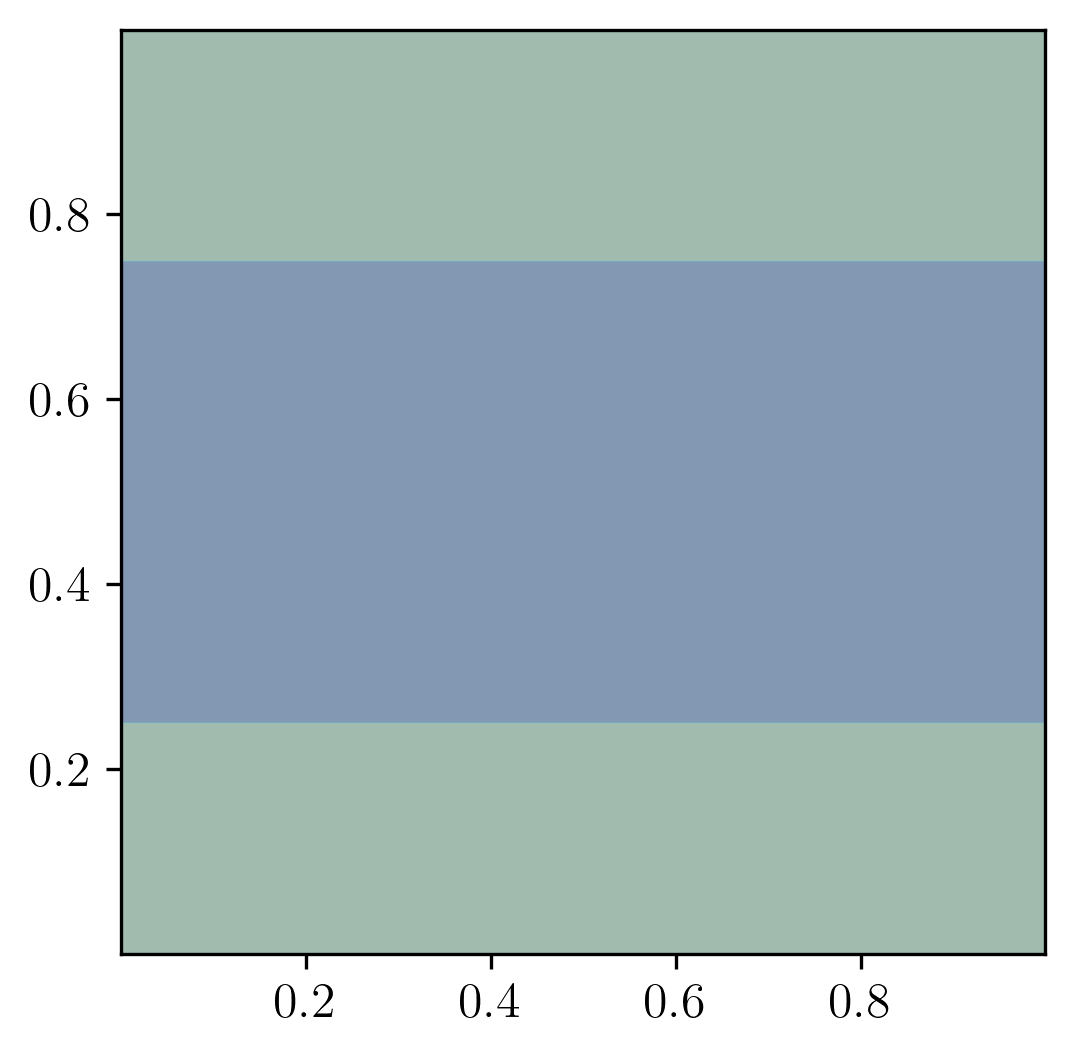}
\label{fig:kh_intial}}
\subfigure[$512^2$.]{\includegraphics[width=0.32\textwidth]{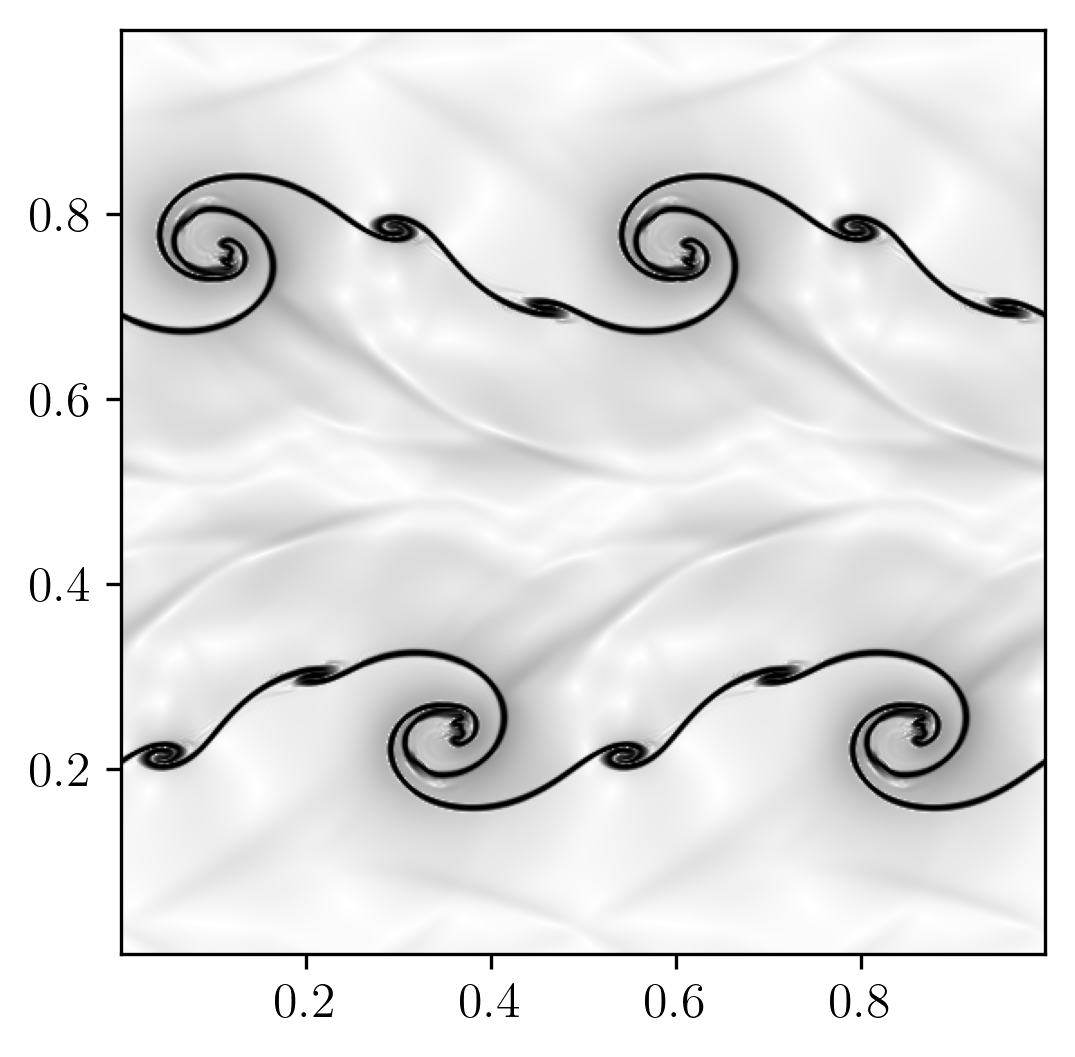}
\label{fig:kh_512_c}}
\subfigure[$2048^2$.]{\includegraphics[width=0.32\textwidth]{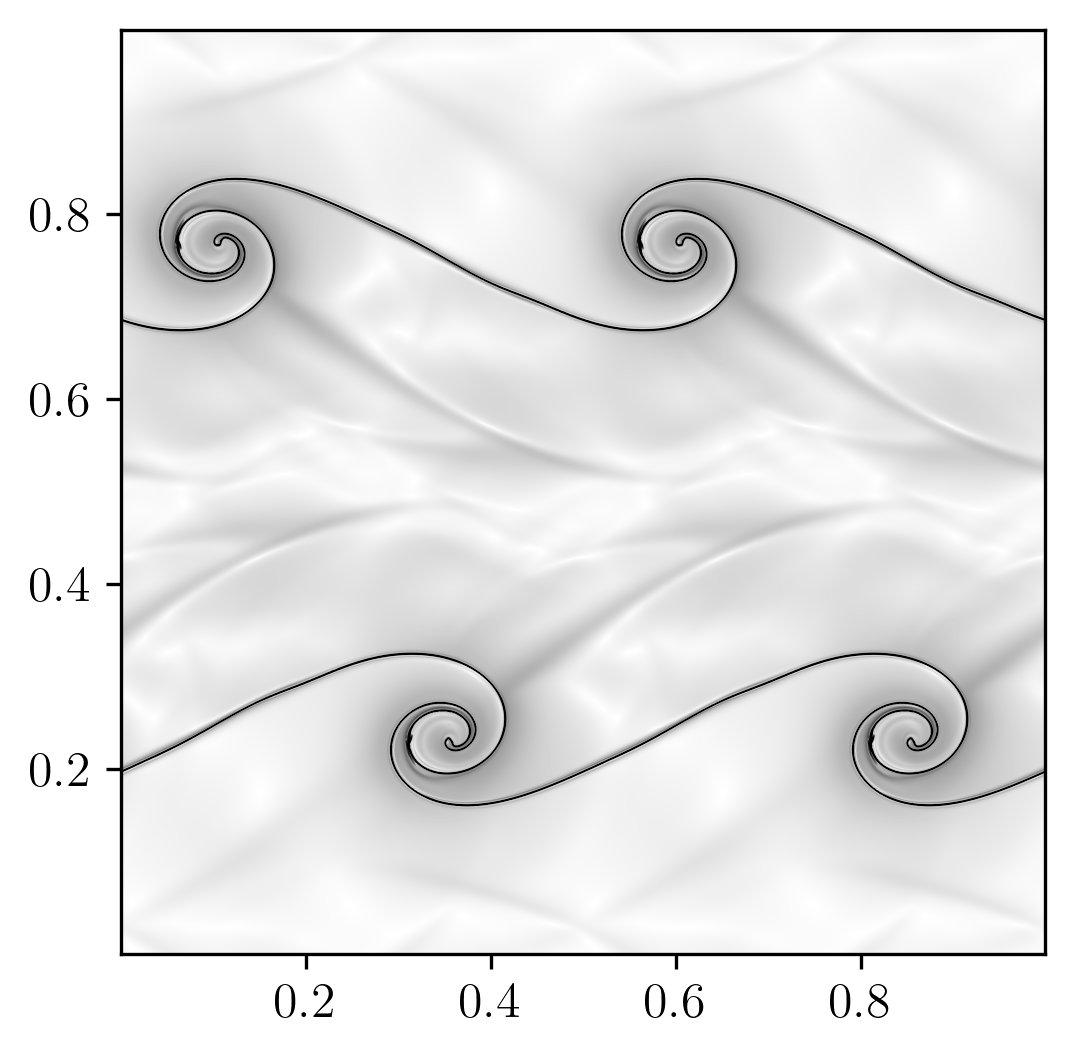}
\label{fig:kh_2048_f}}
\caption{Figures show initial condition and density gradient contours for grid sizes $512^2$ and $2048^2$, for Example \ref{ex:kh}.}
    \label{fig:kh}
\end{figure}

 Fig. \ref{fig:kh_intial} shows the initial conditions, where the green and blue colours indicate volume fractions of two different species. The specific heat ratio of the first species is taken as 1.5, and for the second species is taken as 1.4. The test case is computed using the HY-THINC-D scheme on three different grid sizes ($512^2$, $1024^2$, and $2048^2$) until a final time of $t$ = 0.8 for $\mu = 1.0 \times 10^{-4}$.  Figs. \ref{fig:kh_512_c} and \ref{fig:kh_2048_f} shows the density gradient contours on grid sizes of $512^2$ and $2048^2$, respectively. While the coarse grid simulation shows more vortical structures, they disappear with increased grid resolution.

\begin{figure}[H]
\centering\offinterlineskip
\subfigure[]{\includegraphics[width=0.32\textwidth]{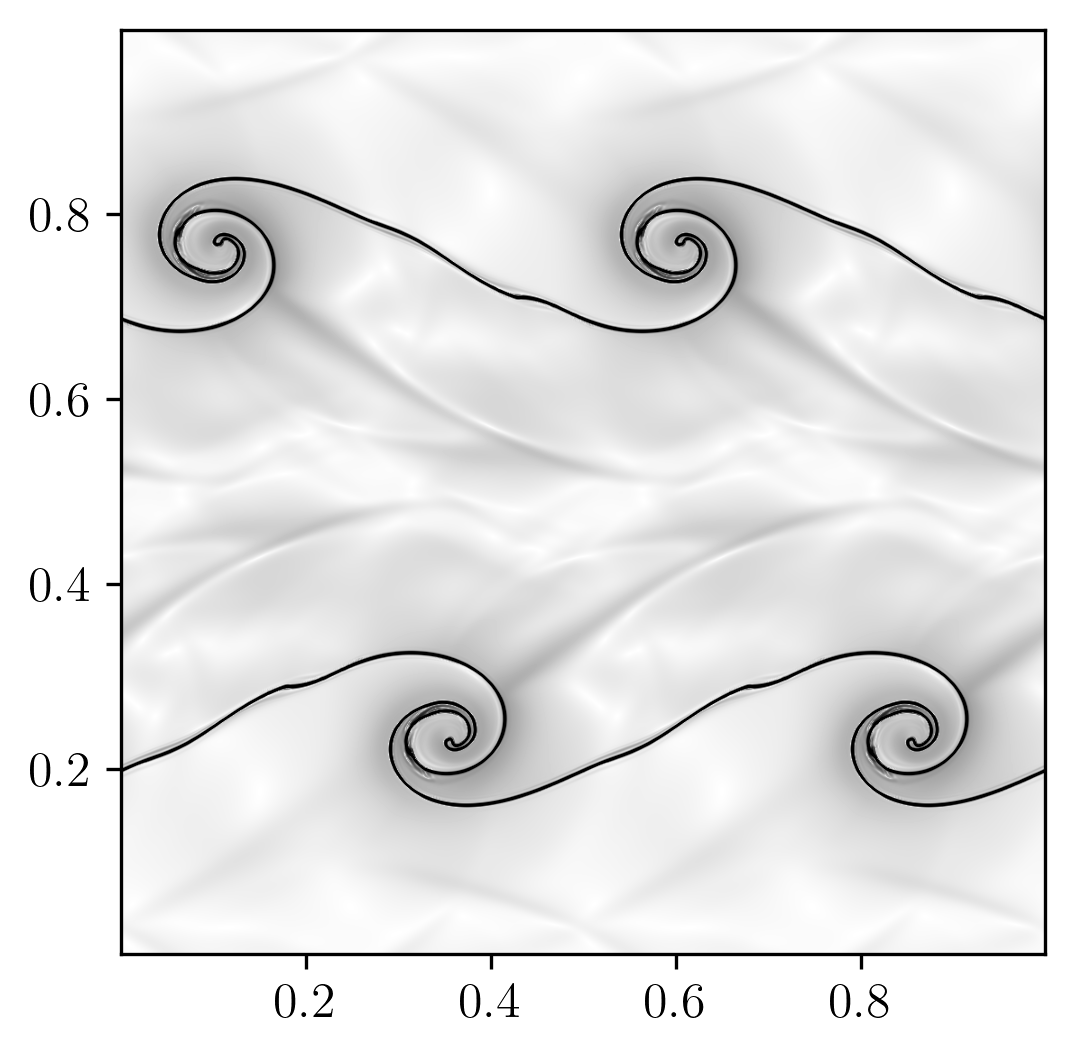}
\label{fig:kh_1024}}
\subfigure[]{\includegraphics[width=0.32\textwidth]{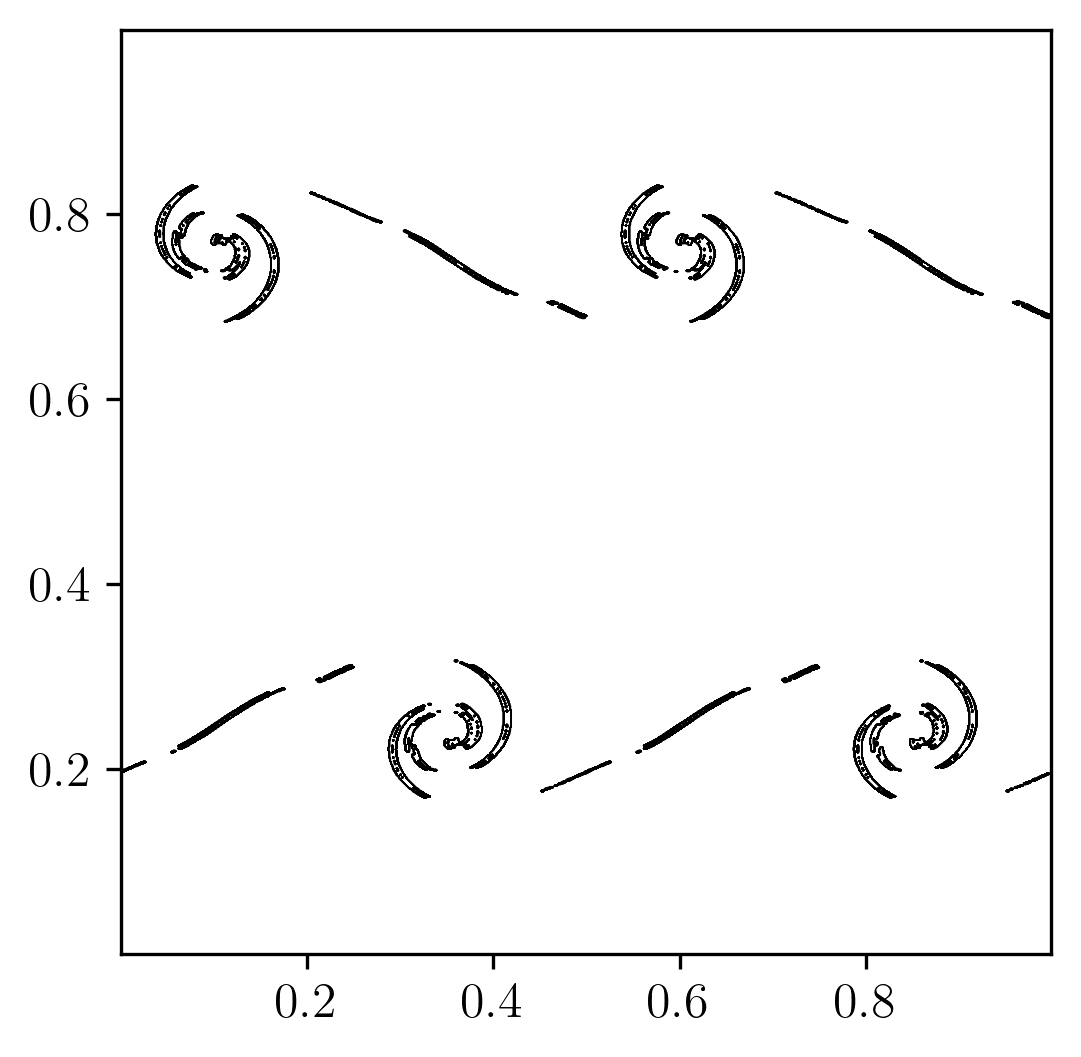}
\label{fig:kh_senx}}
\subfigure[]{\includegraphics[width=0.32\textwidth]{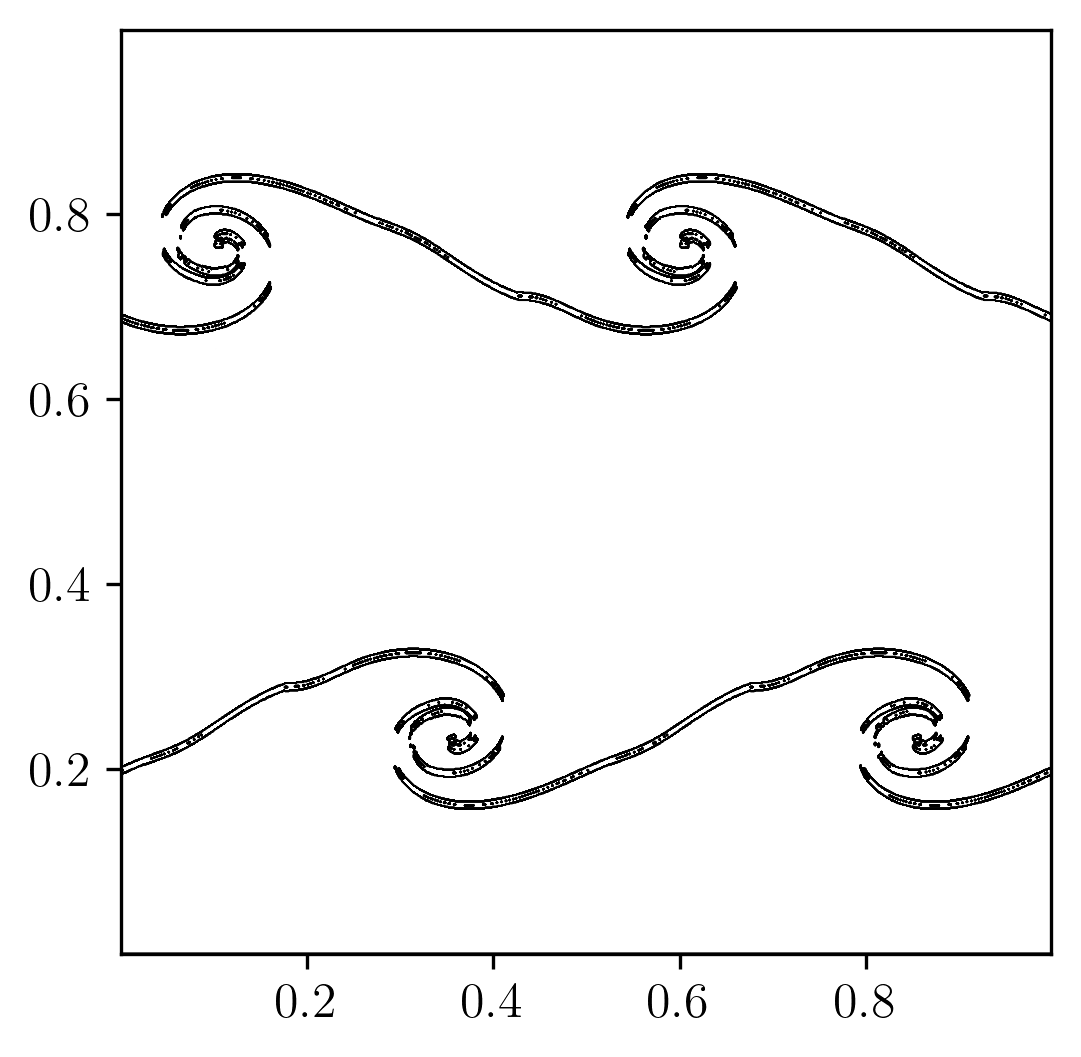}
\label{fig:kh_seny}}
\subfigure[]{\includegraphics[width=0.32\textwidth]{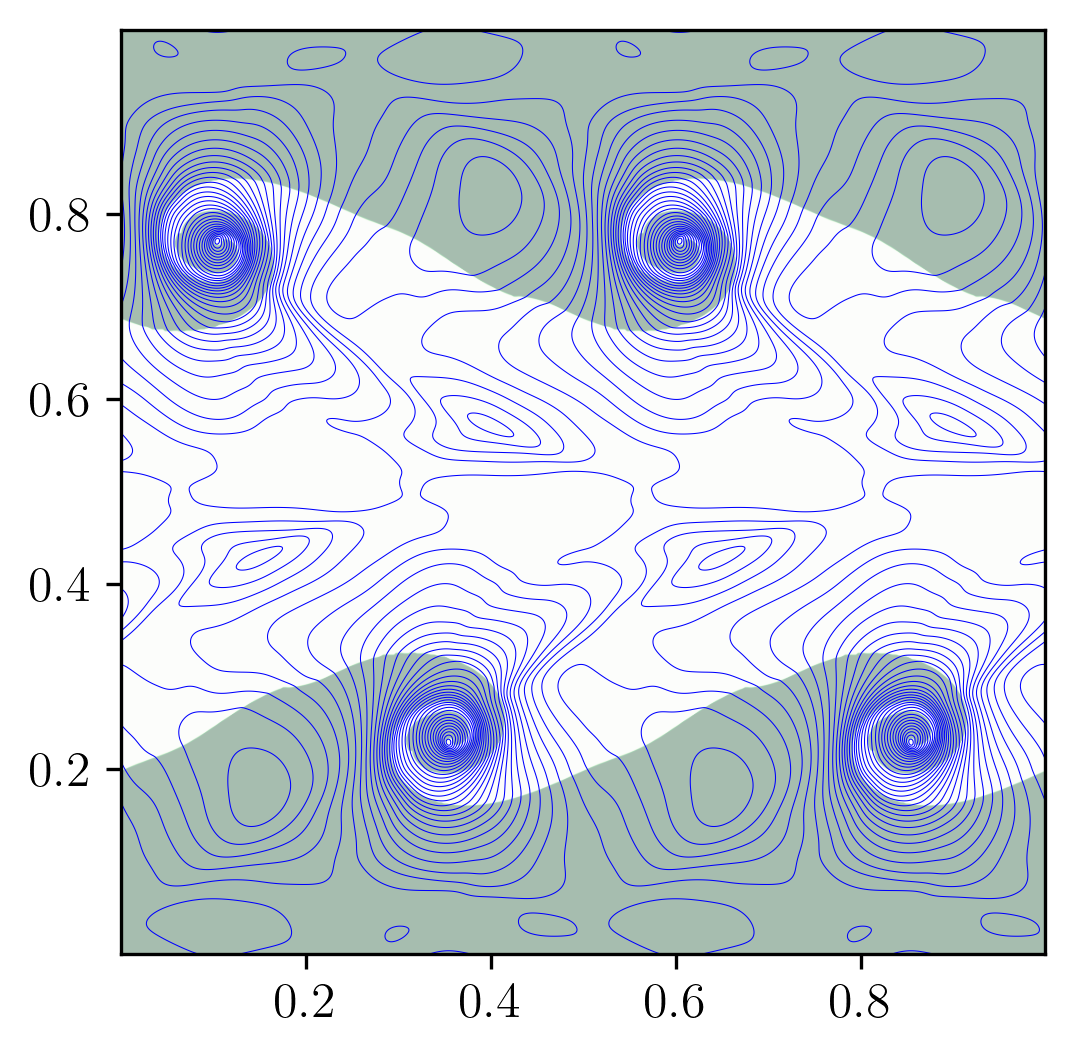}
\label{fig:kh_1024_p}}
\subfigure[]{\includegraphics[width=0.32\textwidth]{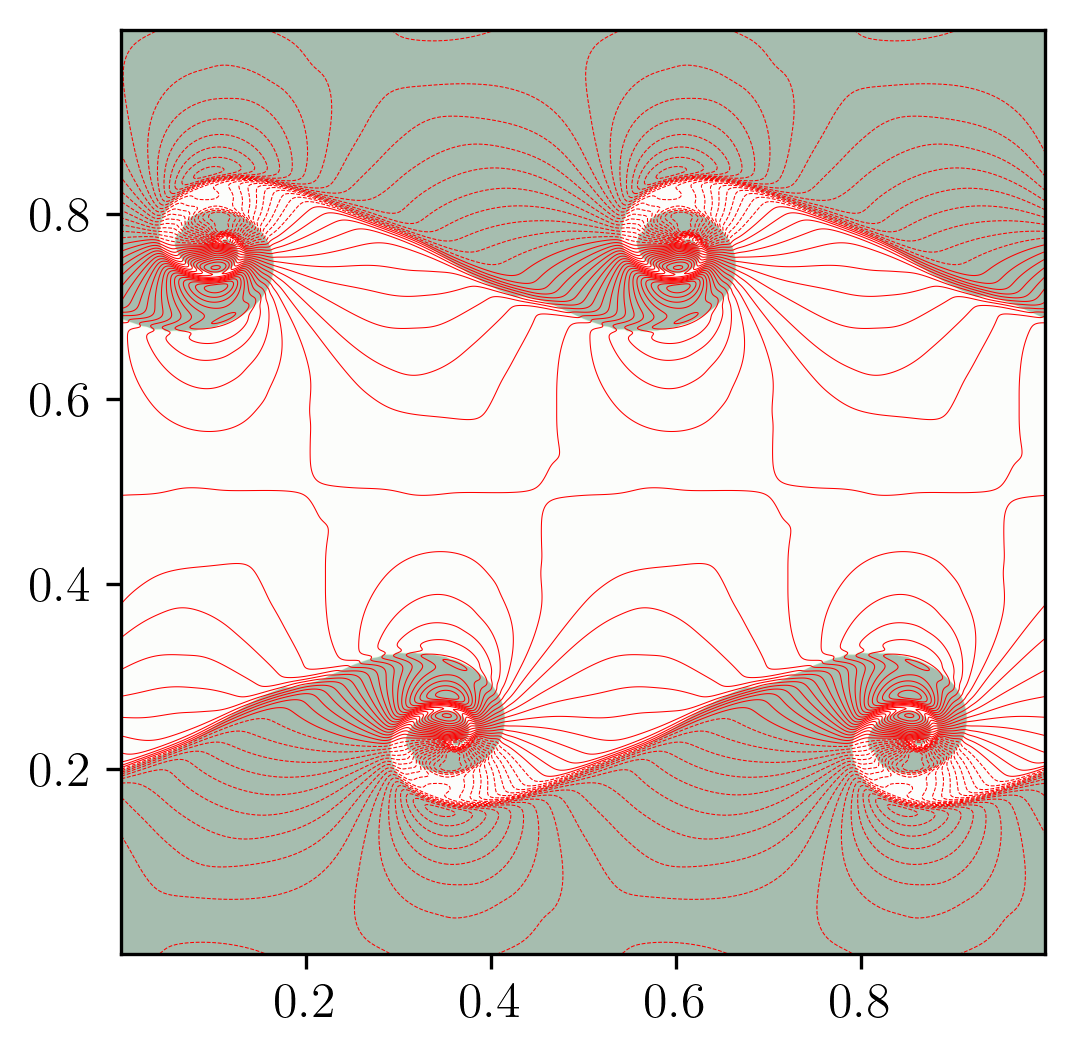}
\label{fig:kh_1024_u}}
\subfigure[]{\includegraphics[width=0.32\textwidth]{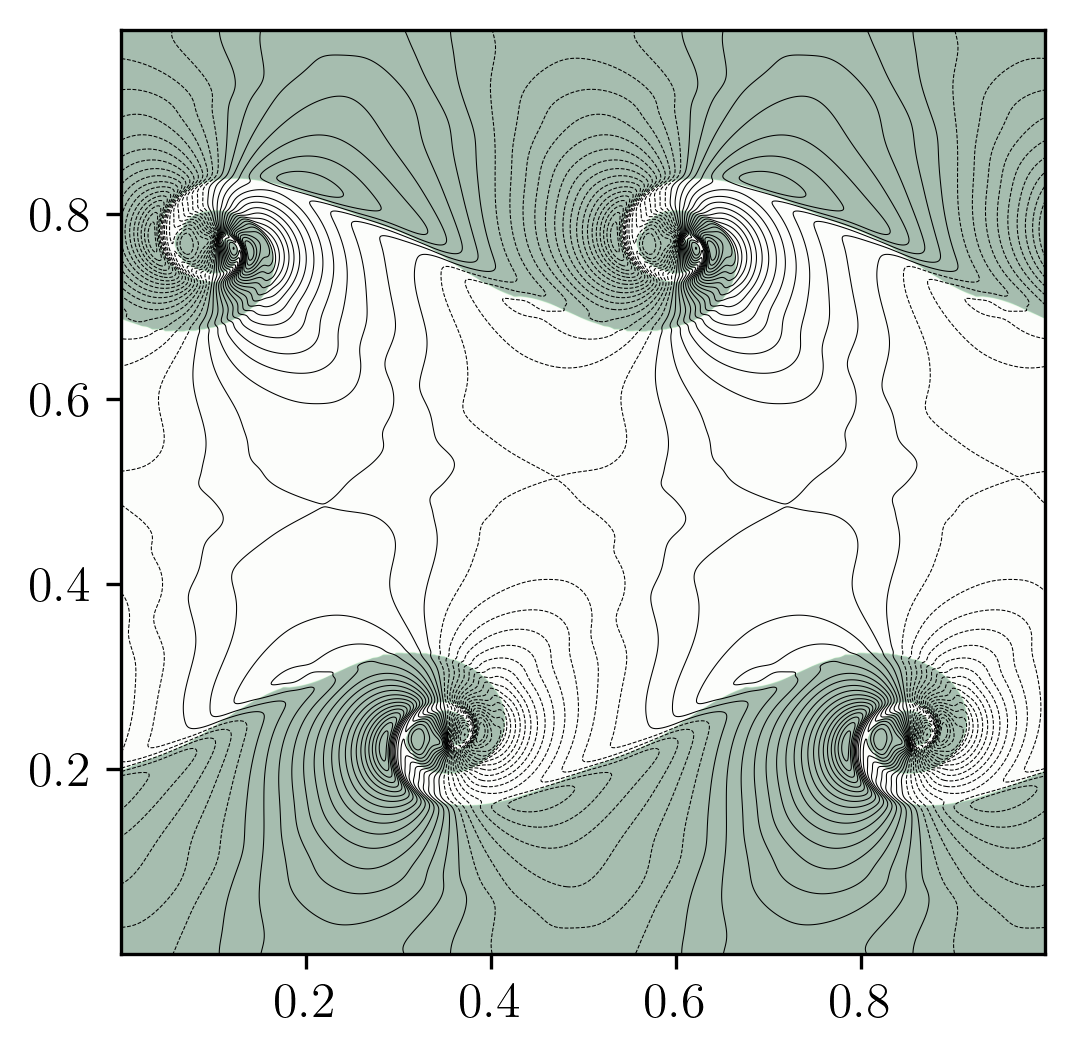}
\label{fig:kh_1024_v}}
\caption{Figures show density gradient contours, contact discontinuity sensor locations in $x-$ and $y-$ directions, pressure contours overlayed on volume fractions, $u$-velocity contours overlayed on volume fractions, and $v$-velocity contours overlayed on volume fractions, for Example \ref{ex:kh}, on a grid size of $1024^2$.}
    \label{fig:kh_well}
\end{figure}

Fig. \ref{fig:kh_1024} shows the density gradient contours on a grid size of $1024^2$. Figs. \ref{fig:kh_senx} and \ref{fig:kh_seny} show regions where the THINC scheme is used; the contours indicate the sensor locations in $x-$ and $y-$ directions. The proposed contact discontinuity sensor correctly identified the interfaces. Figs. \ref{fig:kh_1024_p}, \ref{fig:kh_1024_u} and \ref{fig:kh_1024_v} show the  pressure contours overlayed on volume fractions, $u$-velocity contours overlayed on volume fractions, and $v$-velocity contours overlayed on volume fractions, respectively. These figures plot the species volume fractions using green and white colours for better visualization. In this test case, there are no shocks, and the Ducros sensor did not detect any contact discontinuities; therefore, tangential velocities are computed using central schemes. From Figs. \ref{fig:kh_1024_u} and \ref{fig:kh_1024_v} one can see no oscillations in either velocities and the contours pass through from one material to the other.

In Figs. \ref{fig:kh} and \ref{fig:kh_well}, the characteristic variables are reconstructed, and it can assured that the tangential velocities are reconstructed using the central scheme. However, at least for the author, the characteristic variable that represents the pressure is not known. The following \textit{extreme} algorithm is considered for this test case to show that even pressure can be reconstructed using a central scheme. As explained in Section \ref{sec:eqns-gov} the primitive variable vector is $\mathbf{U}$ = $( \alpha_{1}\rho_{1}, \alpha_{2}\rho_{2}, u, v, p,\alpha_1)^T$. In the following algorithm, pressure ($b$=5) is computed using a central scheme, and the tangential velocities in each direction are computed using central schemes.

\begin{equation}
    {\bm{U}}^{L}_{i+\frac{1}{2},b} = 
    \left\{
    \begin{array}{ll}
         \text{if } b = 5\text{:} & 
{\bm{U}}^{C,Linear}_{i+\frac{1}{2},b}\\
        \\[10pt]
        \text{if } b = 1,2\text{:} & \begin{cases}
            {\bm{U}}^{L,Non-Linear}_{i+\frac{1}{2},b} & \text{if } \left( {\bm{U}}^{L,Linear}_{i+\frac{1}{2}} - {\bm{U}}_i \right) \left( {\bm{U}}^{L,Linear}_{i+\frac{1}{2}} - {\bm{U}}^{L,MP}_{i+\frac{1}{2}} \right) \geq 10^{-40},
            \\[10pt]
            {\bm{U}}^{L,Linear}_{i+\frac{1}{2},b} & \text{otherwise}.
            \\[10pt]
{\bm{U}}_{i+\frac{1}{2},b}^{L, T} & \text{if } \min \left(\psi_{i-1}, \psi_{i}, \psi_{i+1}\right)<\psi_{c}.
        \end{cases}\\ 
        \\[10pt]
        \text{if } b = 6\text{:} &  {\bm{U}}^{L,T}_{i+\frac{1}{2},b}.\\  							
    \end{array}
    \right.
    \label{eqn:prim_extreme}
\end{equation}
In $x$-direction:
\begin{equation}\label{eqn:centralScheme_x_ex}
    \mathbf{U}^{L}_{i+\frac{1}{2},b} = 
    \left\{
    \begin{array}{ll}
        \text{if } b = 3\text{:} & 
{\bm{U}}^{L,Linear}_{i+\frac{1}{2},b}
        \\[5pt]
        \text{if } b = 4\text{:} & {\bm{U}}^{C,Linear}_{i+\frac{1}{2},b}
        \end{array}
    \right.
\end{equation}
In $y$-direction:
\begin{equation}\label{eqn:centralScheme_y_ex}
    \mathbf{U}^{L}_{i+\frac{1}{2},b} = 
    \left\{
    \begin{array}{ll}
        \text{if } b = 4\text{:} & {\bm{U}}^{L,Linear}_{i+\frac{1}{2},b}
        \\[5pt]
        \text{if } b = 3\text{:} &{\bm{U}}^{C,Linear}_{i+\frac{1}{2},b}
     \end{array}
    \right.
\end{equation}

\begin{figure}[H]
\centering\offinterlineskip
\subfigure[Density gradient contours.]{\includegraphics[width=0.35\textwidth]{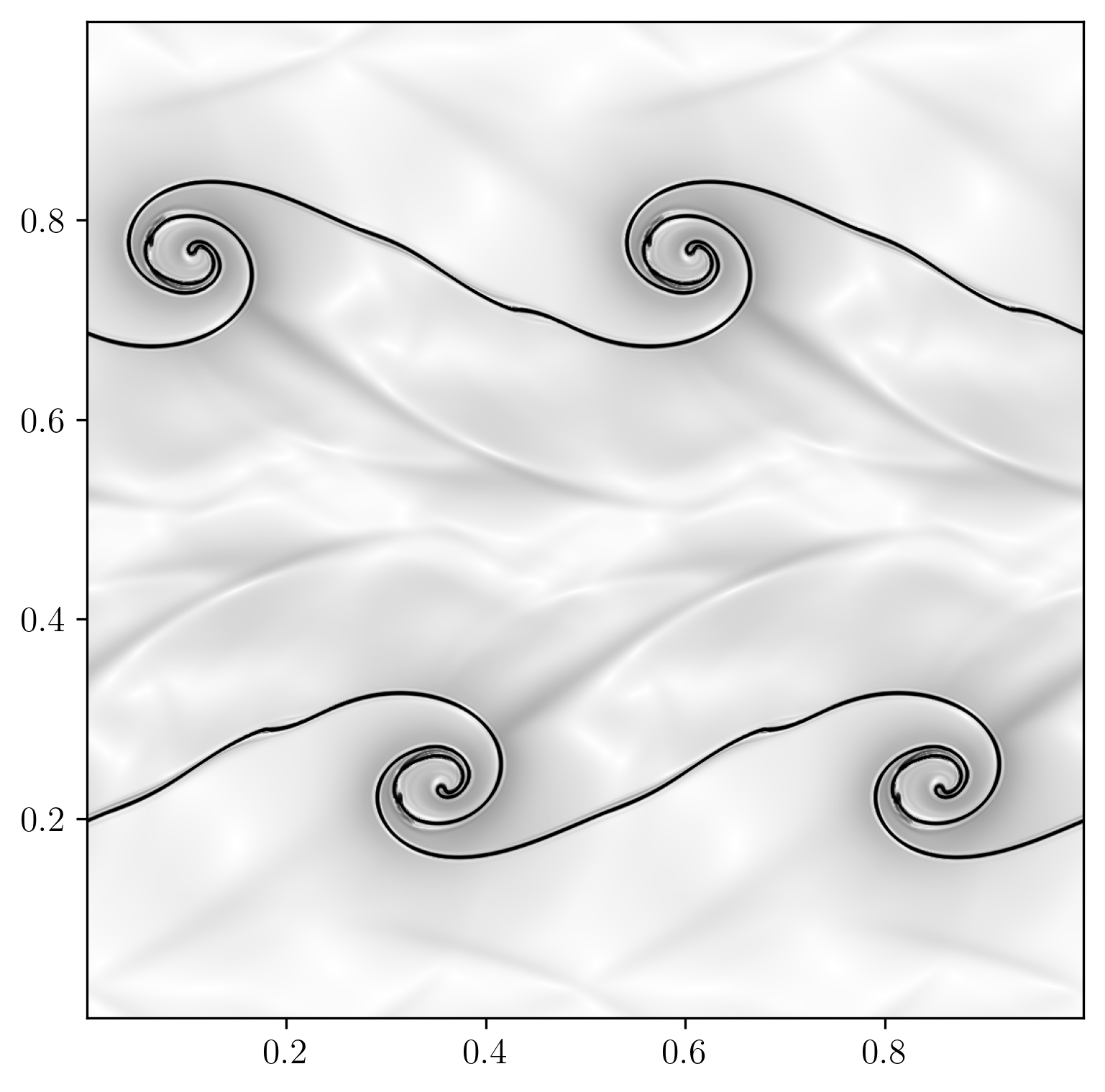}
\label{fig:primkh_den}}
\subfigure[Pressure contours.]{\includegraphics[width=0.35\textwidth]{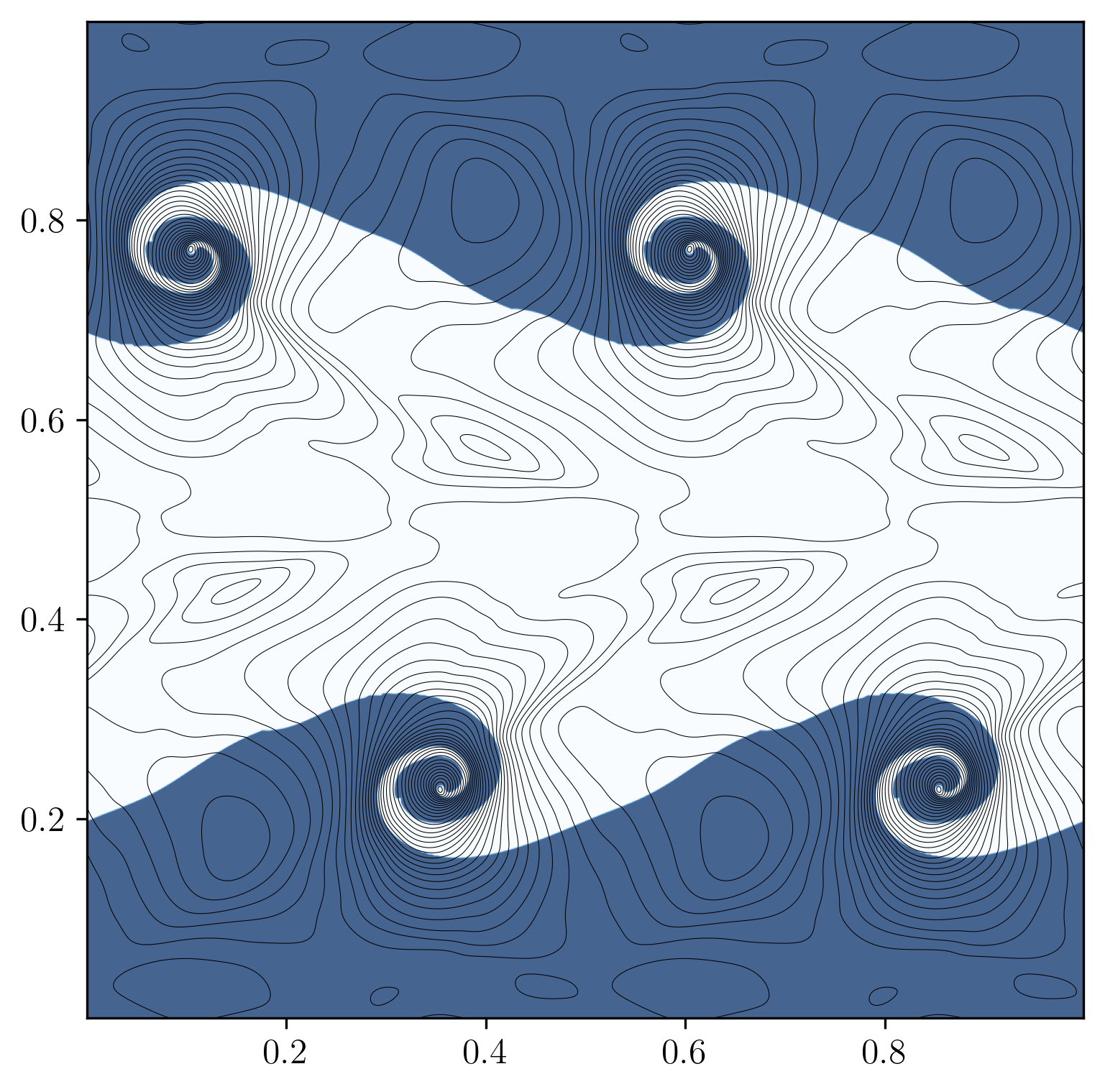}
\label{fig:prim_kh_press}}
\caption{Figures show density gradient contours and pressure contours overlayed on volume fractions, for Example \ref{ex:kh}, on a grid size of $1024^2$ using the \textit{extreme} algorithm.}
    \label{fig:kh_prim_good}
\end{figure}

Fig. \ref{fig:kh_prim_good} indicates that the algorithm mentioned above would work without any issues, and even pressure can be computed using a central scheme. These reconstructions are physically consistent as only density and volume fractions are discontinuous across the material interface, and the rest of the variables are continuous. 

 The TENO-THINC of \cite{takagi2022novel} failed for this test case, even on a coarse grid. The possible reason for failure is applying an interface sharpening approach for continuous variables across the interface (pressure and velocities). Even for the proposed sensor, applying THINC for pressure and velocities in the regions of contact discontinuities crashed.
   
\begin{example}\label{ex:triple}{Compressible triple point problem (Inviscid and viscous case)} 
\end{example}

In this test case, the multi-species compressible triple point problem is considered. This scenario poses a challenging three-state, two-dimensional Riemann problem involving two distinct materials. This benchmark test is widely used to validate the ability of interface-capturing schemes to resolve sharp interfaces. This test case emphasizes the generation of fine small-scale vortical structures along the contact discontinuities due to Kelvin-Helmholtz instabilities. The computational domain spans [0, 7] $\times$ [0, 3]. The initial conditions for this test case \cite{chamarthi2023gradient,pan2018conservative} are as follows:  

\begin{equation}
(\alpha_1 \rho_1,\alpha_2 \rho_2, u, v, p, \gamma)=\left\{\begin{array}{ll}
(1.0~~~,0.0,0,0,1.0,1.5), & \text { \text{sub-domain} [0, 1]$\times$[0, 3], } \\
(0.0~~~,1.0,0,0,0.1,1.4), & \text { \text{sub-domain} [1, 1]$\times$[0, 1.5], }\\
(0.125,0.0,0,0,0.1,1.5), & \text { \text{sub-domain} [1, 7]$\times$[1.5, 3]}.
\end{array}\right.
\end{equation}

\begin{figure}[H]
\centering
\subfigure[\textcolor{black}{Density gradient contours.}]{\includegraphics[width=0.48\textwidth]{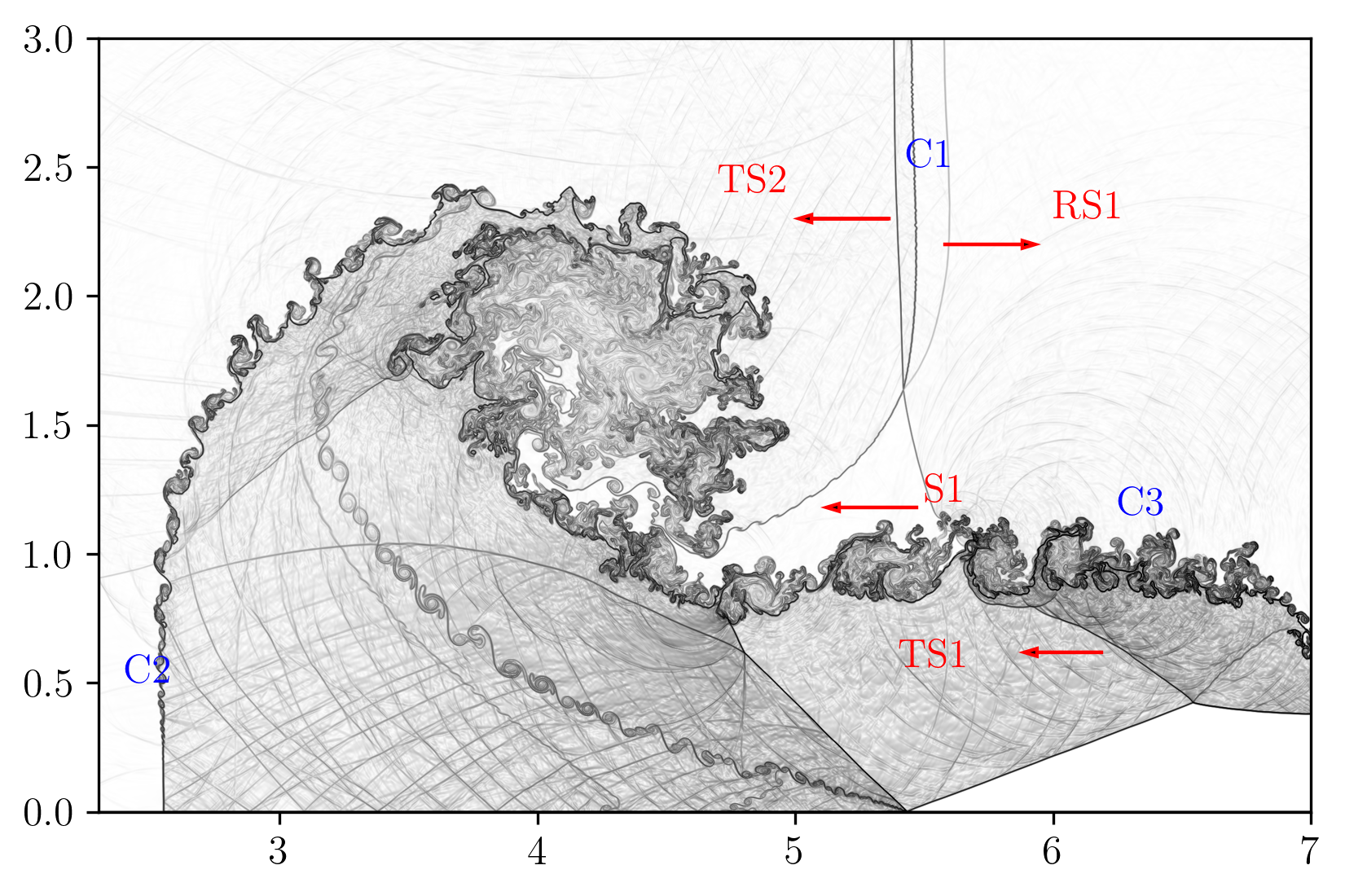}
\label{fig:hyt_f}}
\subfigure[\textcolor{black}{Vorticity contours.}]{\includegraphics[width=0.48\textwidth]{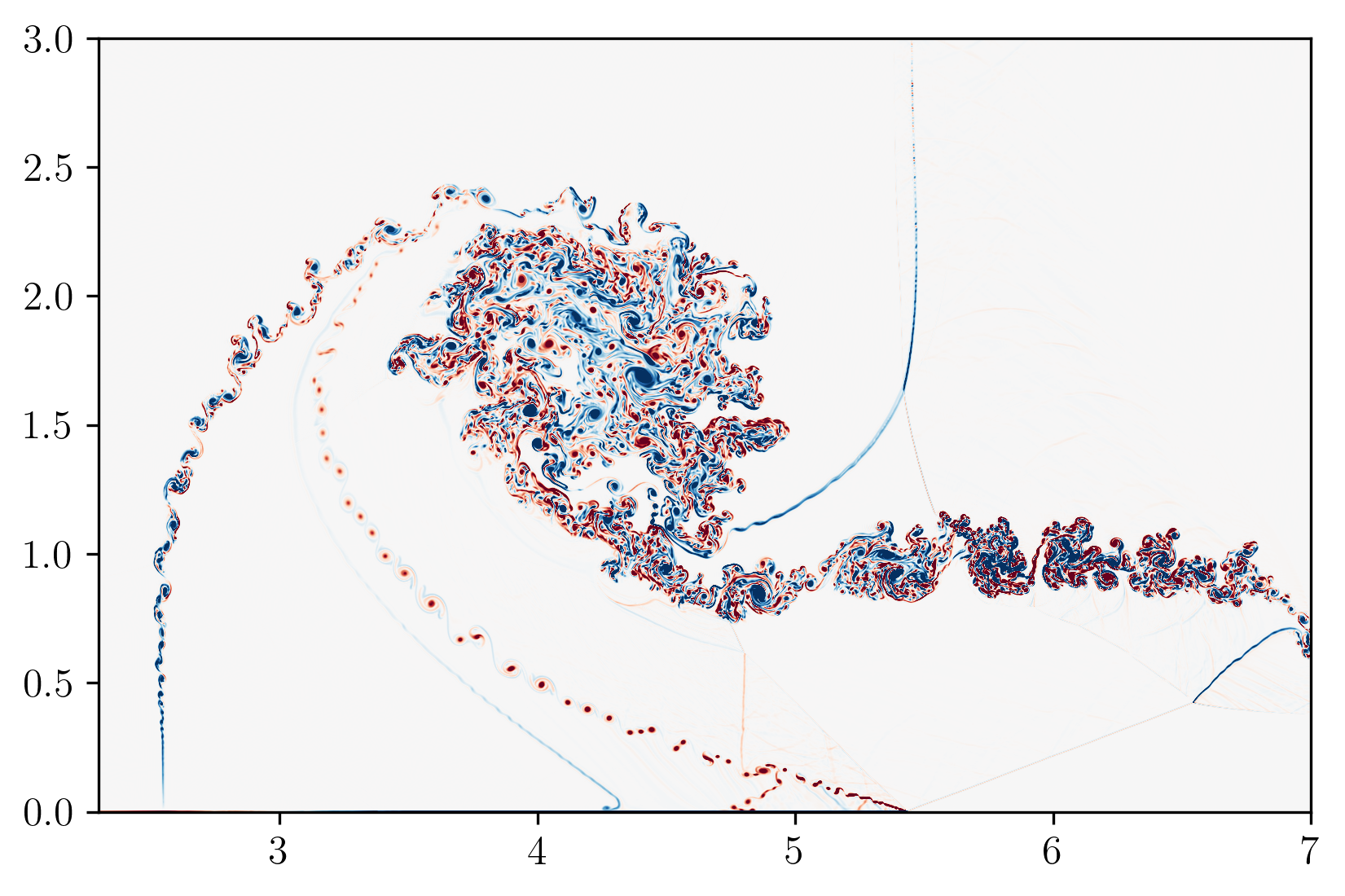}
\label{fig:hyt_f_vort}}
\subfigure[Sensor location $x-$direction.]{\includegraphics[width=0.48\textwidth]{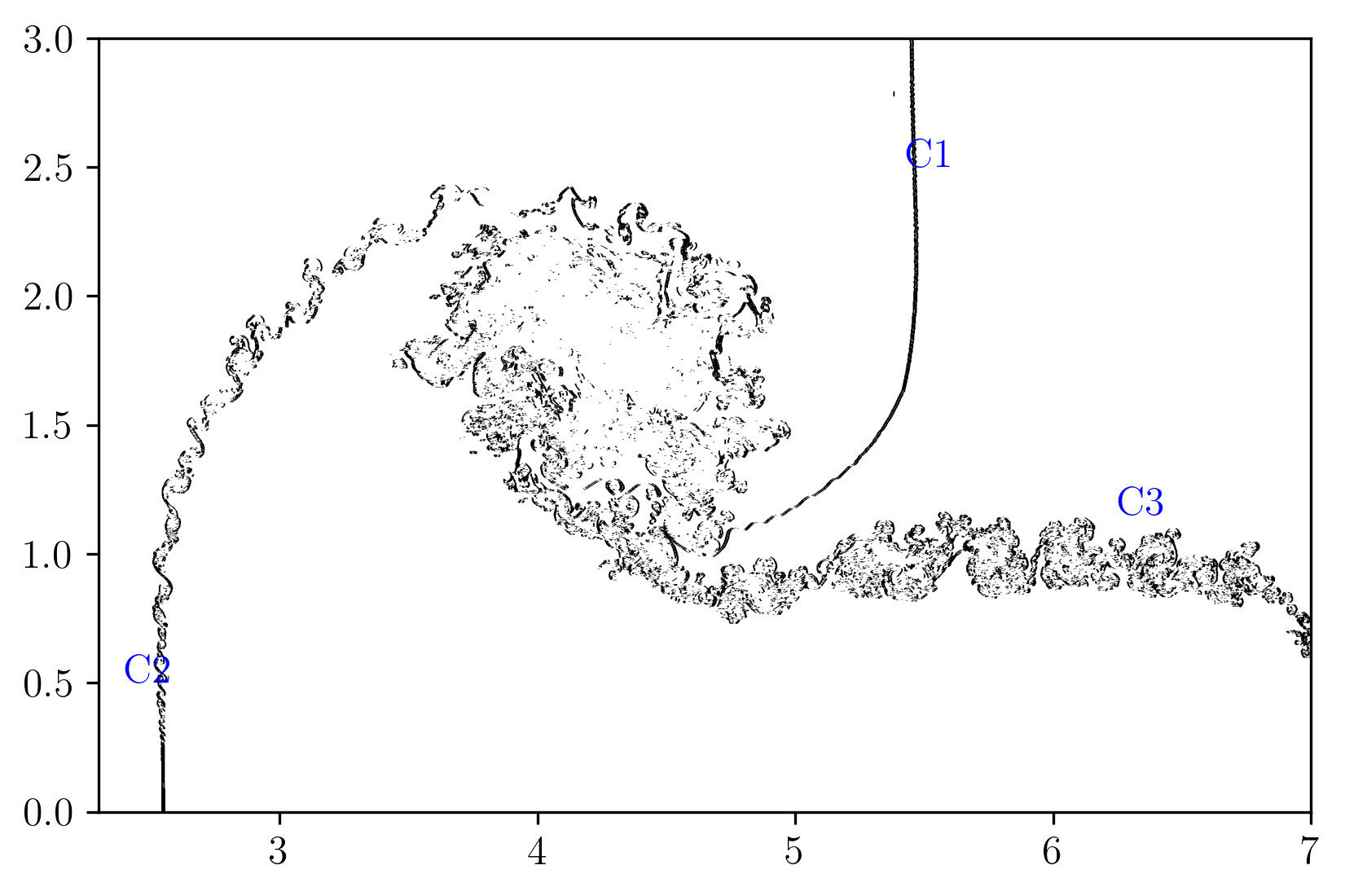}
\label{fig:hyt_f_alpha}}
\subfigure[Sensor location $y-$direction.]{\includegraphics[width=0.48\textwidth]{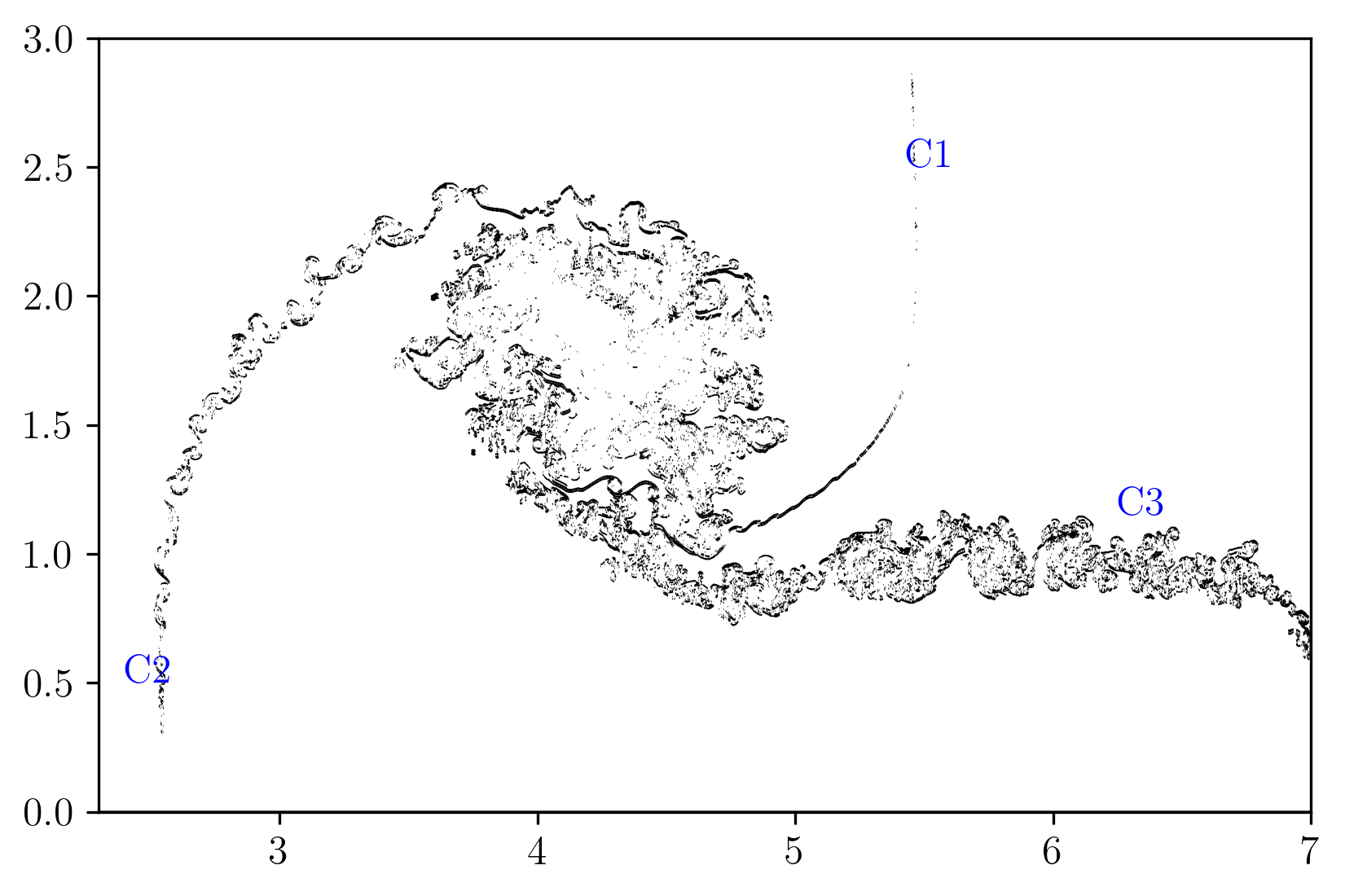}
\label{fig:hyt_sensor}}
\caption{\textcolor{black}{Density gradient contours, vorticity contours, volume fraction contours and the sensor location at time $t=5$ using the HY-THINC scheme, Example \ref{ex:triple}, on a grid resolution of 3584 $\times$ 1536 for inviscid simulation.}}
\label{fig_tapas}
\end{figure}

In the first and third subdomains, $\alpha_1$ = 1, and in the second subdomain, $\alpha_1$ = 0. Simulations are carried out in both inviscid and viscous scenarios. First, the inviscid simulation is carried out on a grid size of 3584 $\times$ 1536 (as in \cite{pan2018conservative}), and the final time considered is 5.0. Reflective boundary conditions are imposed for all the boundaries. Fig. \ref{fig_tapas} shows the results obtained by the HY-THINC scheme at $t$ = 5.0.

Figs. \ref{fig:hyt_f} and \ref{fig:hyt_f_vort} show the density gradient contours and vorticity contours, respectively. The results are similar and competitive compared to those obtained by the multi-resolution approach of Pan et al. \cite{pan2018conservative}, whose grid resolution is the same (see Fig. 12 of Ref. [69]). In Fig. \ref{fig:hyt_f}, various contact discontinuities are denoted by C1, C2 and C3. Likewise, shockwaves are denoted by S1, RS1, TS1 and TS2. Figs. \ref{fig:hyt_f_alpha} and \ref{fig:hyt_sensor} show the locations of the THINC scheme activation, which indicates that the sensor has detected all the contact discontinuities and did not detect shockwaves at all. These results justify the name \textit{contact discontinuity sensor} used for the sensor instead of being called a discontinuity detector (that detects both shocks and contacts).

\begin{figure}[H]
\centering
\subfigure[MP5.]{\includegraphics[width=0.48\textwidth]{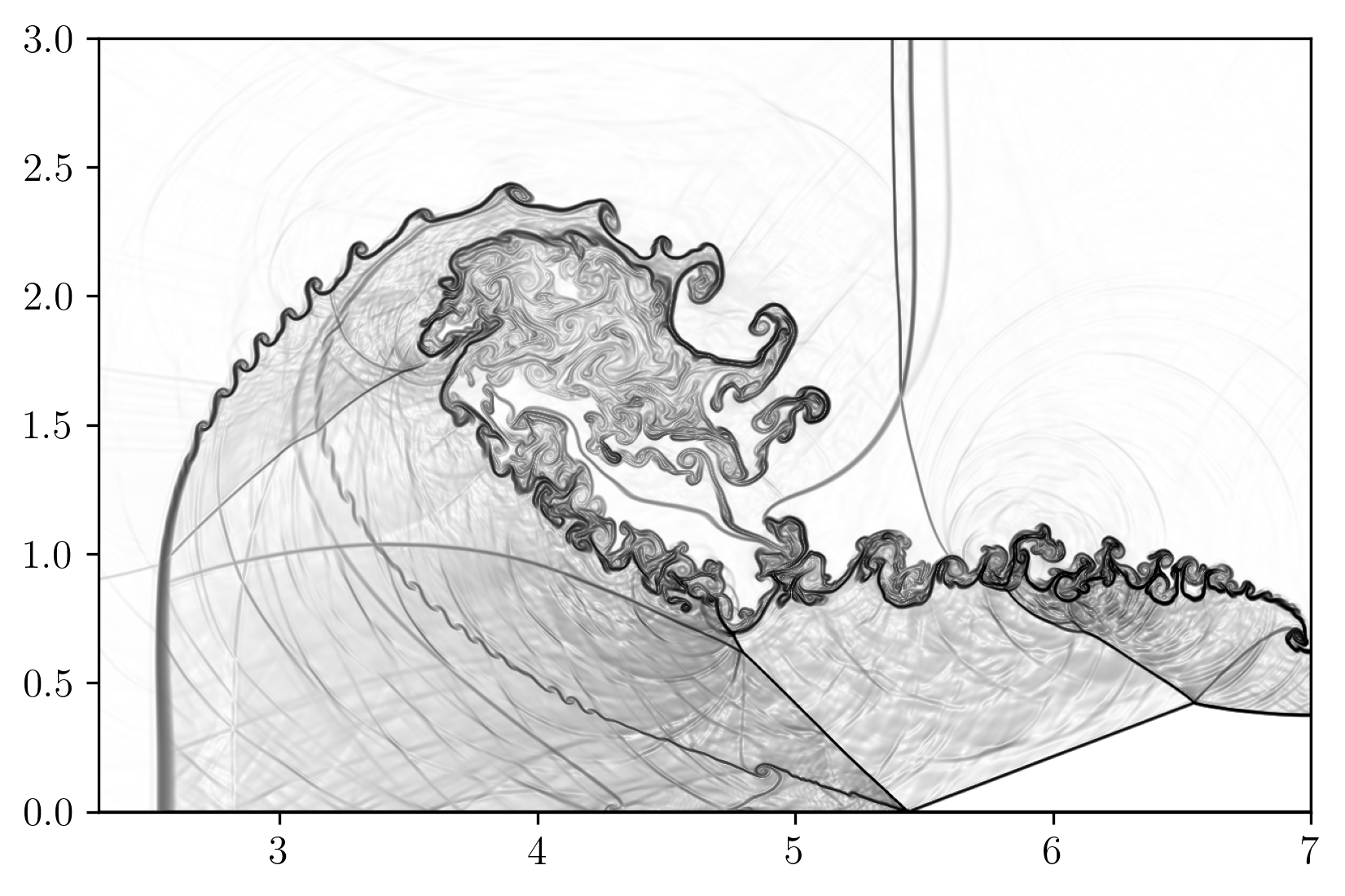}
\label{fig:mp_up}}
\subfigure[\textcolor{black}{HY-THINC.}]{\includegraphics[width=0.48\textwidth]{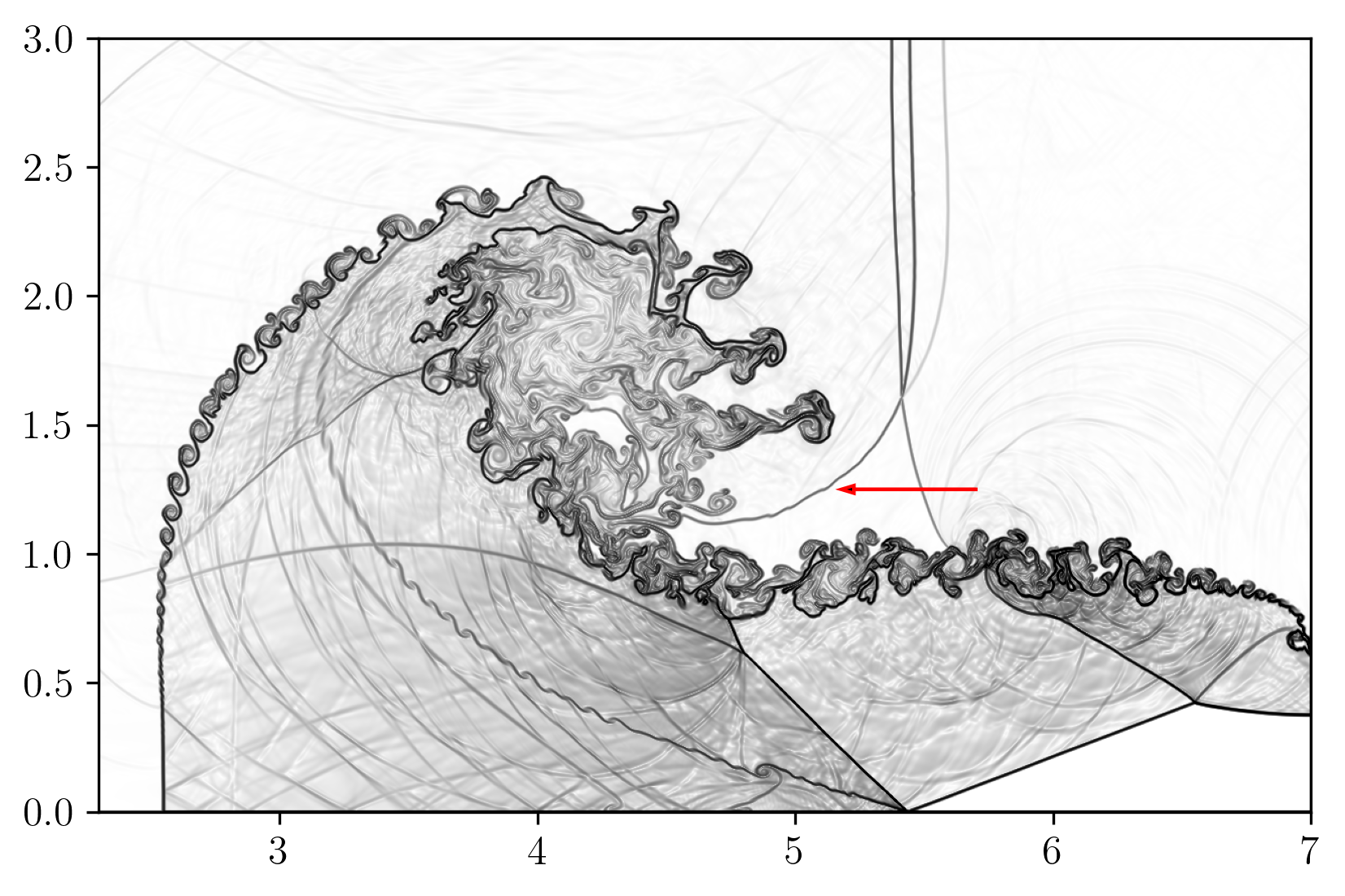}
\label{fig:mp_tp}}
\caption{\textcolor{black}{Density gradient contours at time $t=5$ using various schemes, Example \ref{ex:triple}, on a grid resolution of 1792 $\times$ 768 for inviscid simulation.}}
\label{fig_tapas_p}
\end{figure}

Fig.  \ref{fig_tapas_p} shows the density gradient contours obtained by the MP5 and HY-THINC schemes at $t=5$ on a grid size of 1792 $\times$ 768. It can be observed that the HY-THINC scheme captured the material interface within a few cells compared to the MP5 scheme based on the contact discontinuity thickness. Contact discontinuity within the material (indicated by the red arrow in Fig. \ref{fig:mp_tp}) is also detected and is computed using THINC.

\begin{figure}[H]
\centering
 \includegraphics[width=0.95\textwidth]{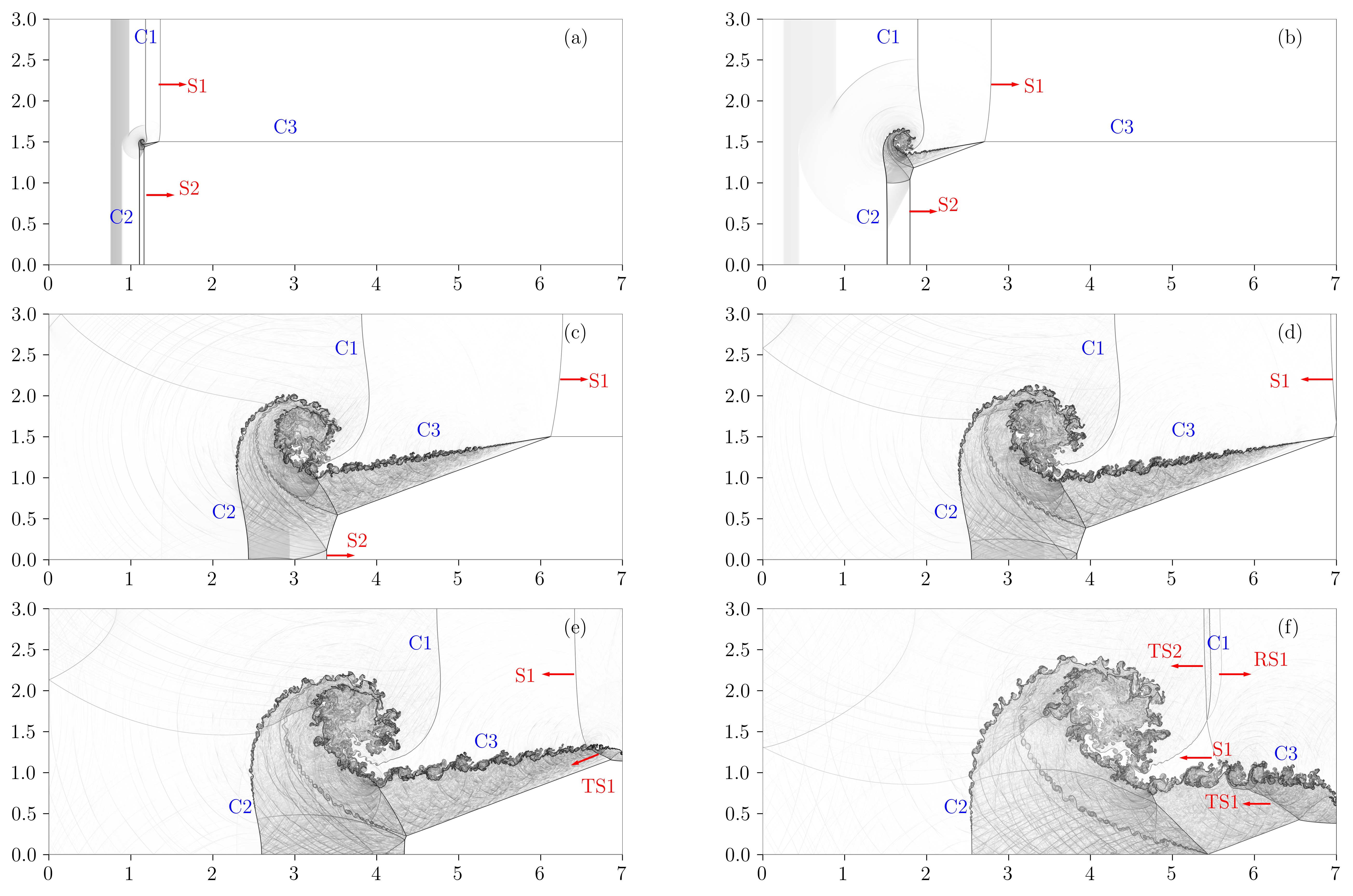}
 \caption{Numerical Schlieren images at various time instances $t$ = 0.2, 1.0, 3.0, 3.5, 4.0 and 5.0 for the compressible triple point problem using HY-THINC scheme, Example \ref{ex:triple}, on a grid resolution of 3584 $\times$ 1536 for inviscid simulation.}
 \label{fig:triple}
\end{figure}

Numerical Schlieren images at various time instances $t$ = 0.2, 1.0, 3.0, 3.5, 4.0 and 5.0 are shown in Fig. \ref{fig:triple}, which depicts the development of the shock system and are consistent with Figure 11 of \cite{pan2018conservative} and Figure 24 of \cite{chamarthi2023gradient}. The initial conditions give rise to contact discontinuities denoted as C1 and C2 in Fig. \ref{fig:triple} (a). In the vicinity of the triple point, a distinctive roll-up region takes shape due to the faster advancement of shockwave S1 compared to S2, as illustrated in Fig. \ref{fig:triple} (b). As S1 continues its trajectory, it comes into contact with the contact discontinuity labelled as C3, as depicted in Fig. \ref{fig:triple} (c), resulting in distinct vortical structures. When $t = 3.5$, shockwave S1 reaches the right boundary and reflects into the domain, Fig. \ref{fig:triple} (d). This inward movement of S1 gives rise to the formation of transmitted shock waves, denoted as TS1 and TS2, which, in turn, interact with the contact discontinuity C1, as depicted in Fig. \ref{fig:triple}(e), resulting in complex vortical structures.

\begin{figure}[H]
\centering
\subfigure[Density gradient contours.]{\includegraphics[width=0.48\textwidth]{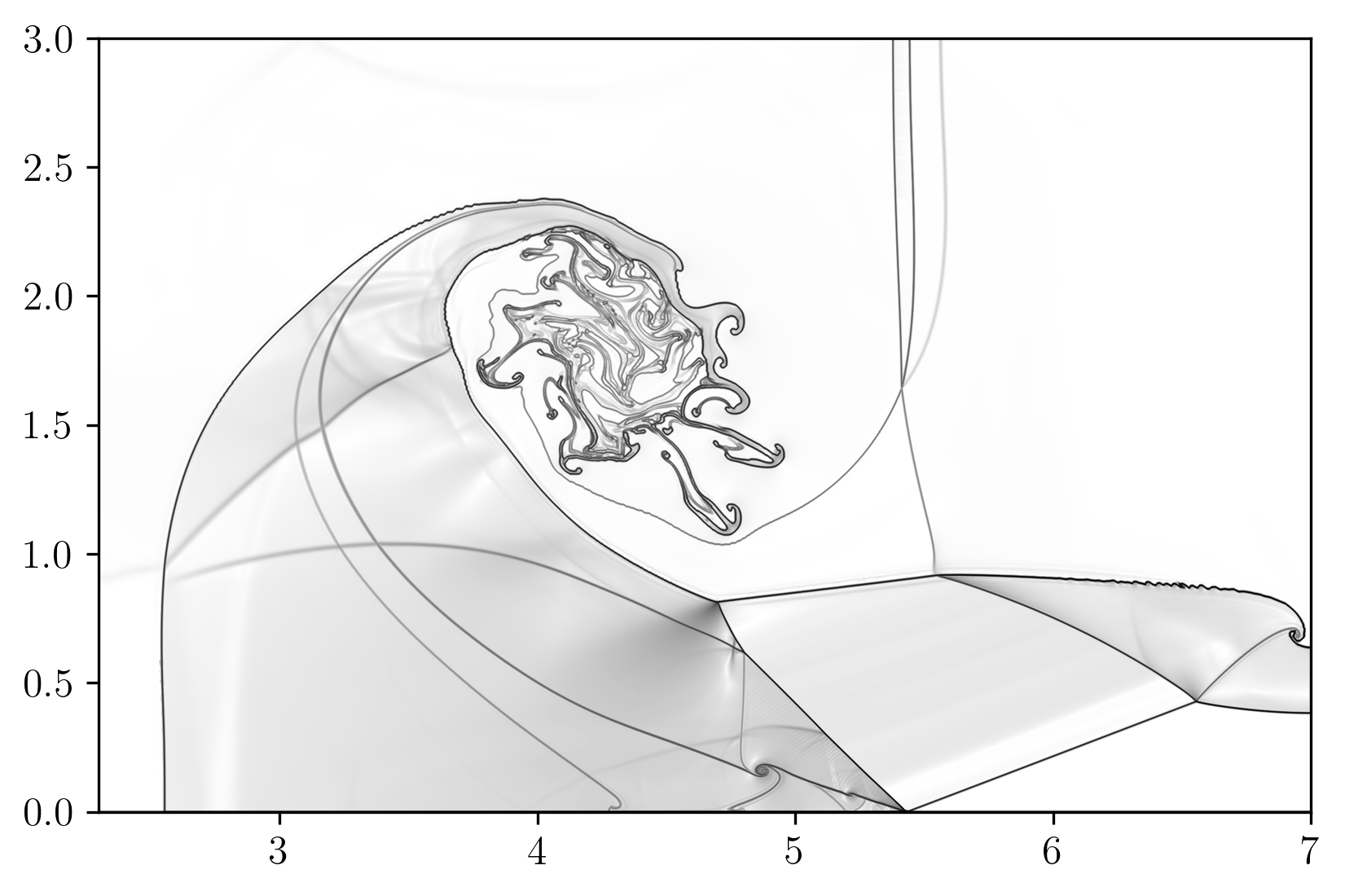}
\label{fig:hyt_tc_f}}
\subfigure[$v$ contours.]{\includegraphics[width=0.48\textwidth]{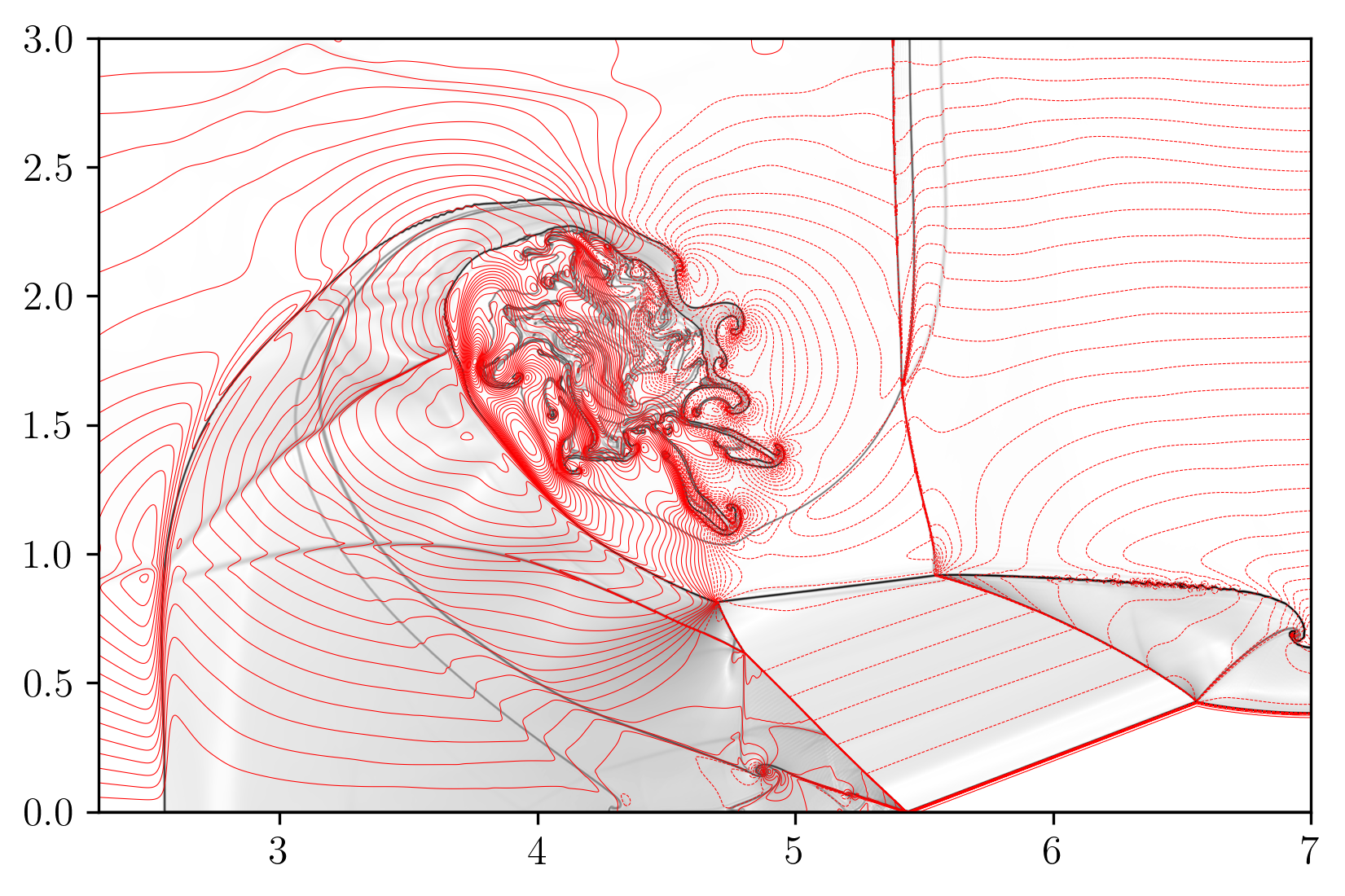}
\label{fig:hyt_tp_v}}
\caption{Density gradient contours and $v$ velocity contours at time $t=5$ using HY-THINC-D scheme, Example \ref{ex:triple}, on a grid resolution of 3584 $\times$ 1536 for viscous simulation.}
\label{fig_tapas_viscous_fine}
\end{figure}

Next, viscous simulations are carried out using the HY-THINC-D (to show that a central scheme across material interfaces can compute the tangential velocities). Fig. \ref{fig_tapas_viscous_fine} shows the fine grid solution, the grid size of 3584 $\times$ 1536, for $\mu = 1.0 \times 10^{-4}$. Fig. \ref{fig:hyt_tc_f} show the density gradient contours, and many of the vortical structures that are observed in the \textit{inviscid} solution are non-existent, yet the contact discontinuities are captured sharply. Fig. \ref{fig:hyt_tp_v} shows the $v$ velocity contours, plotted in red colour, are continuous across the material interfaces, and there are no oscillations.

\begin{figure}[H]
\centering
\subfigure[Density gradient - Characteristic variables.]{\includegraphics[width=0.48\textwidth]{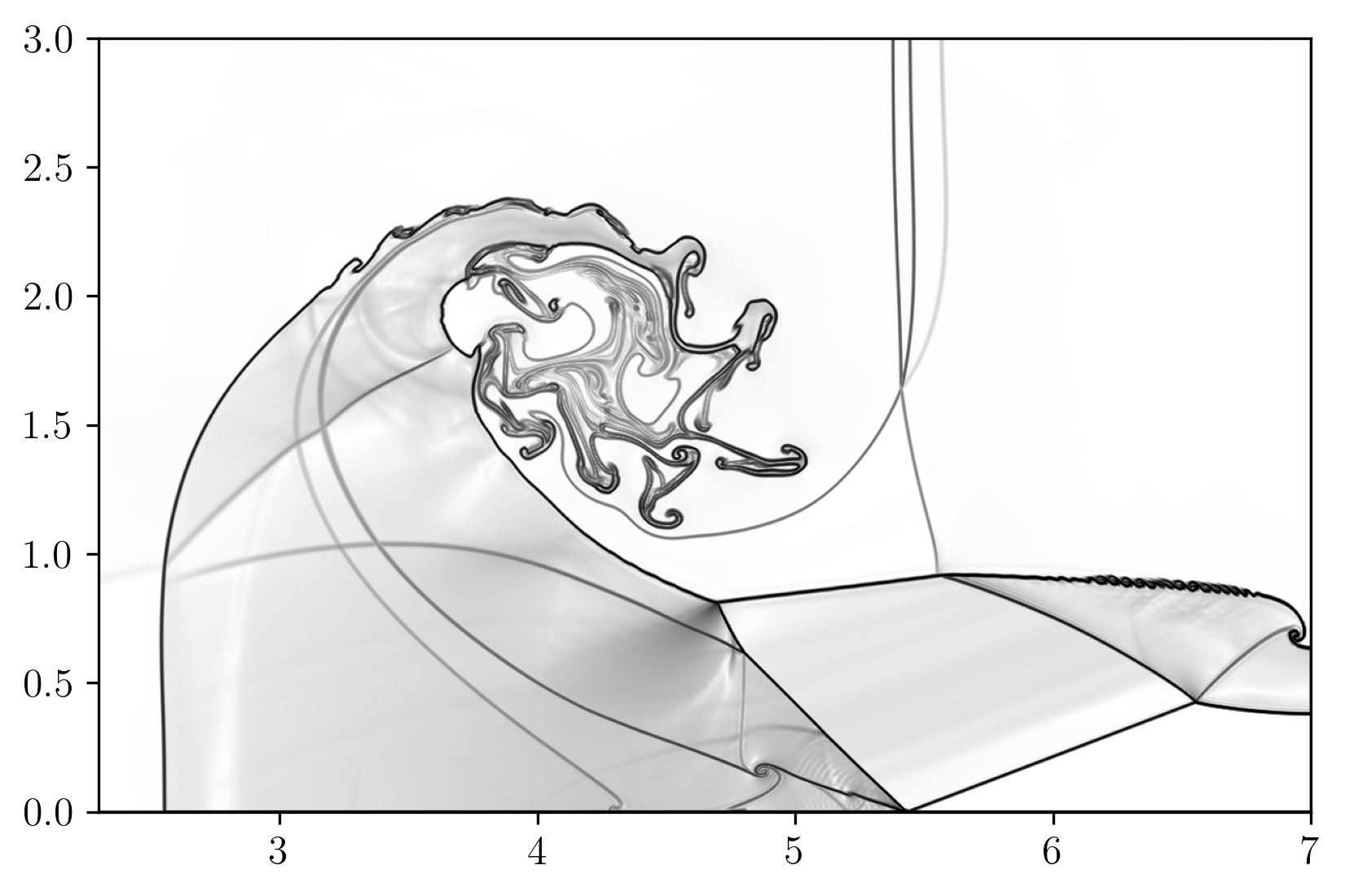}
\label{fig:hyt_tc}}
\subfigure[Density gradient - Primitive variables.]{\includegraphics[width=0.48\textwidth]{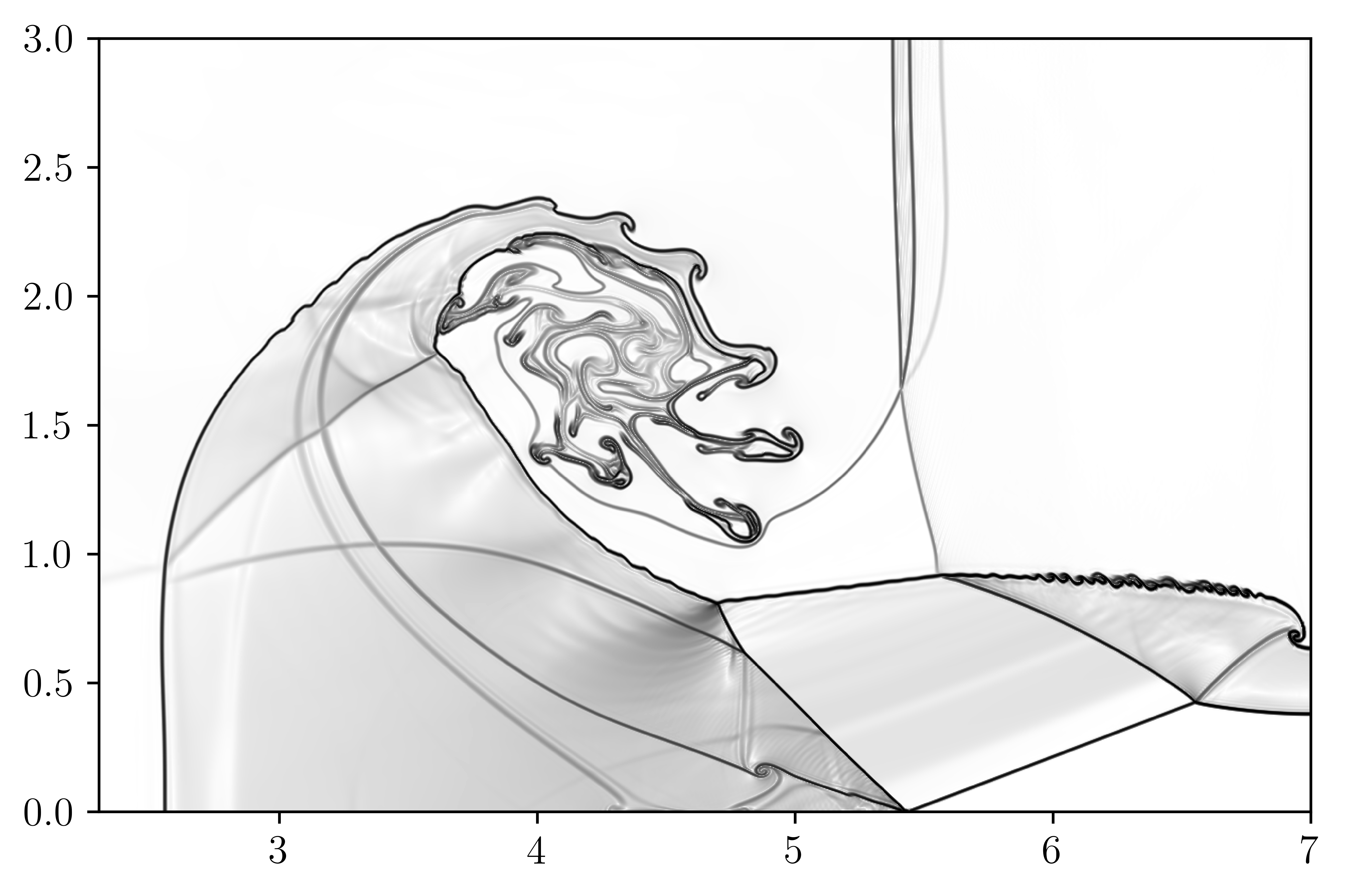}
\label{fig:hyt_tp}}
\subfigure[\textcolor{black}{Ducros sensor $x-$ direction.}]{\includegraphics[width=0.48\textwidth]{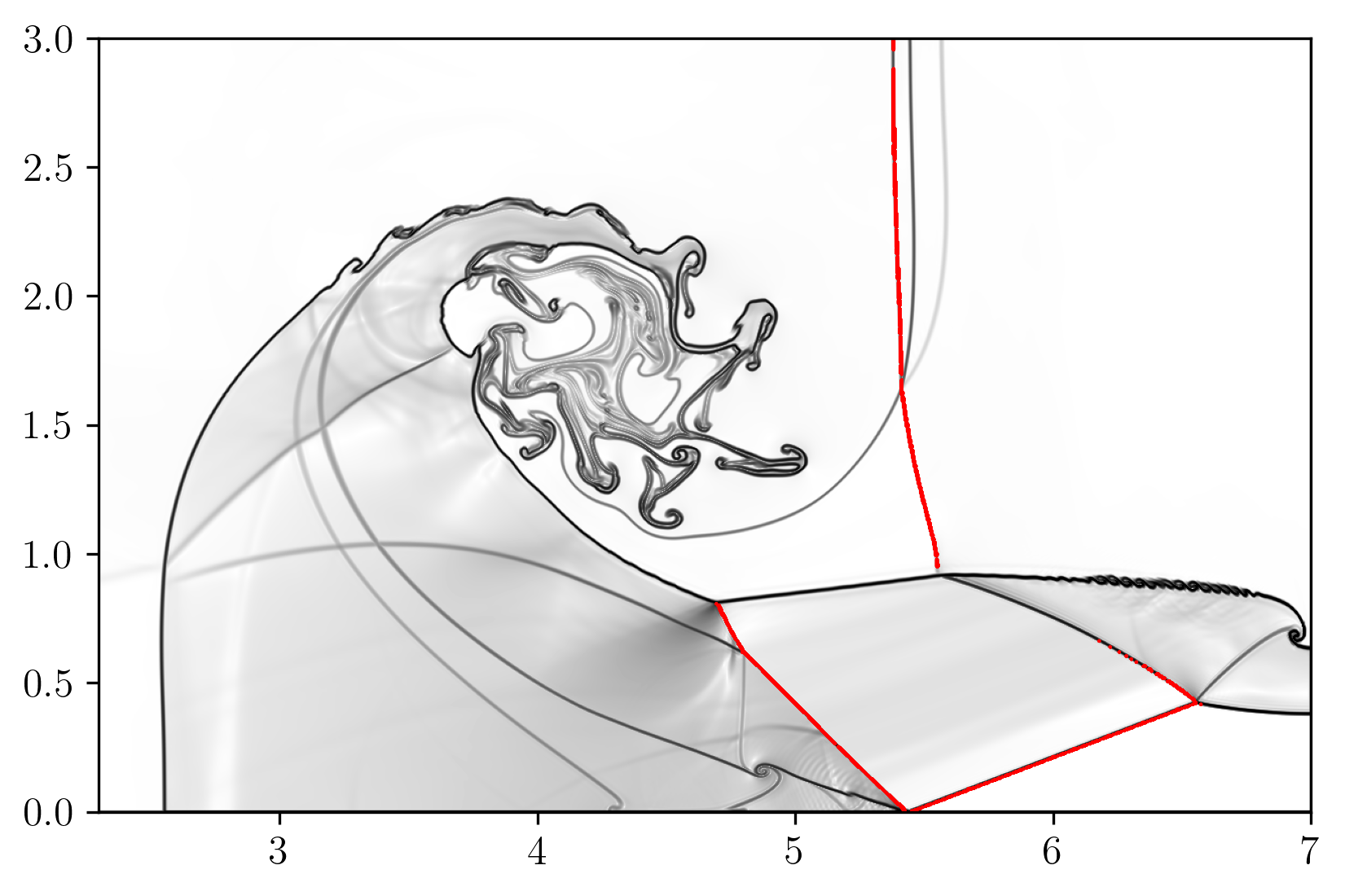}
\label{fig:hyt_ducx}}
\subfigure[\textcolor{black}{Ducros $y-$ direction.}]{\includegraphics[width=0.48\textwidth]{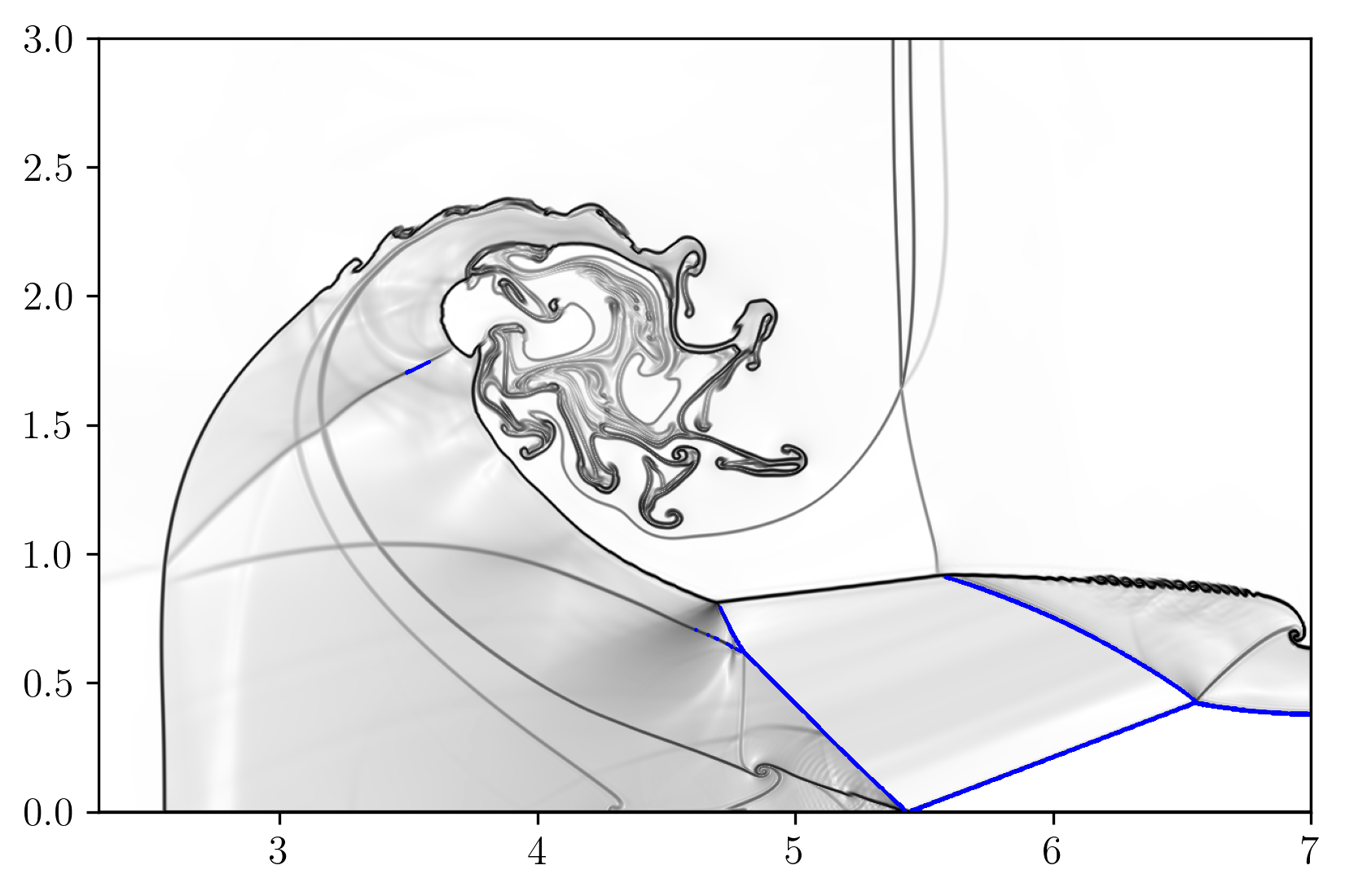}
\label{fig:hyt_ducy}}
\subfigure[\textcolor{black}{$v-$ Velocity contours.}]{\includegraphics[width=0.48\textwidth]{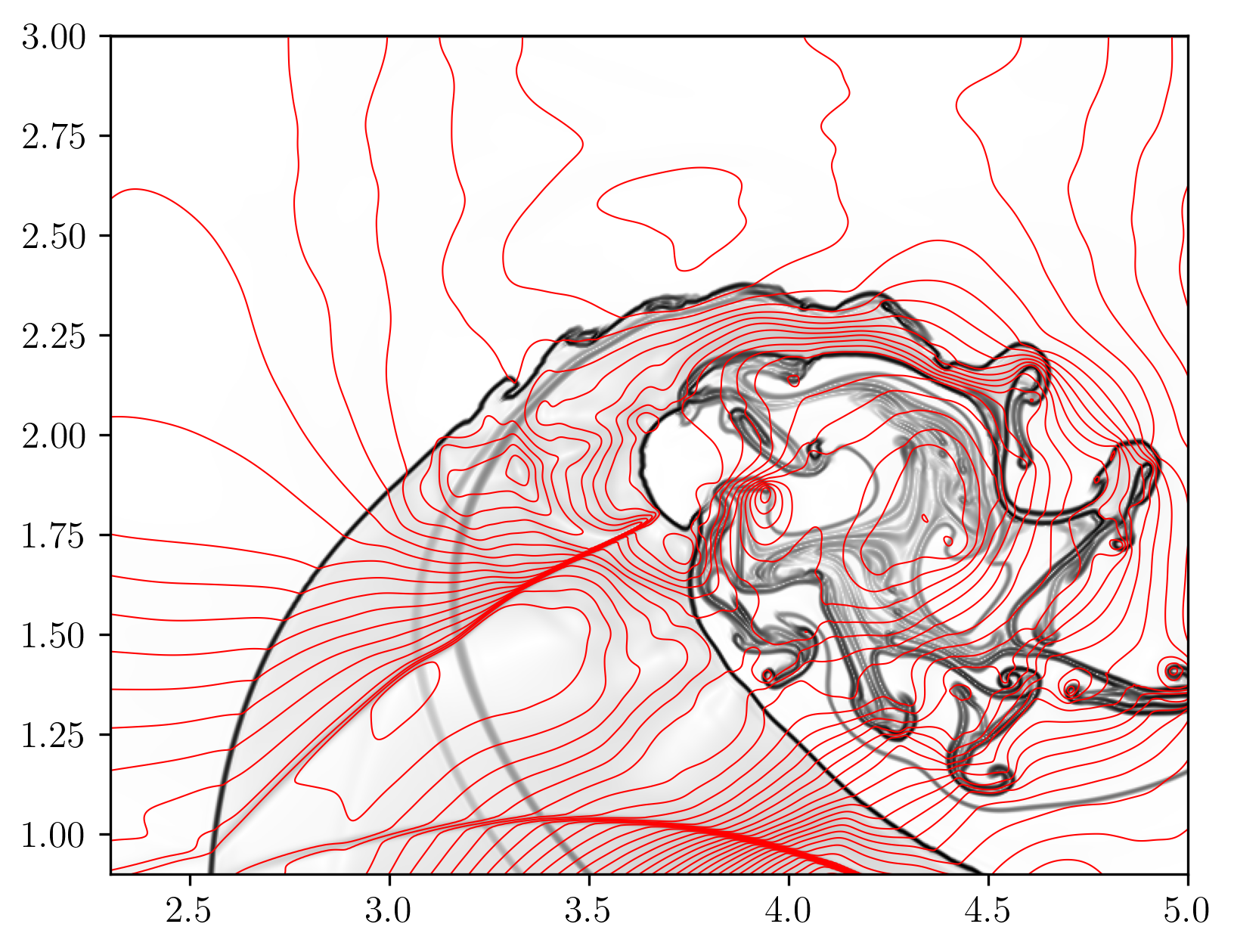}
\label{fig:hyt_v}}
\subfigure[\textcolor{black}{Pressure contours.}]{\includegraphics[width=0.48\textwidth]{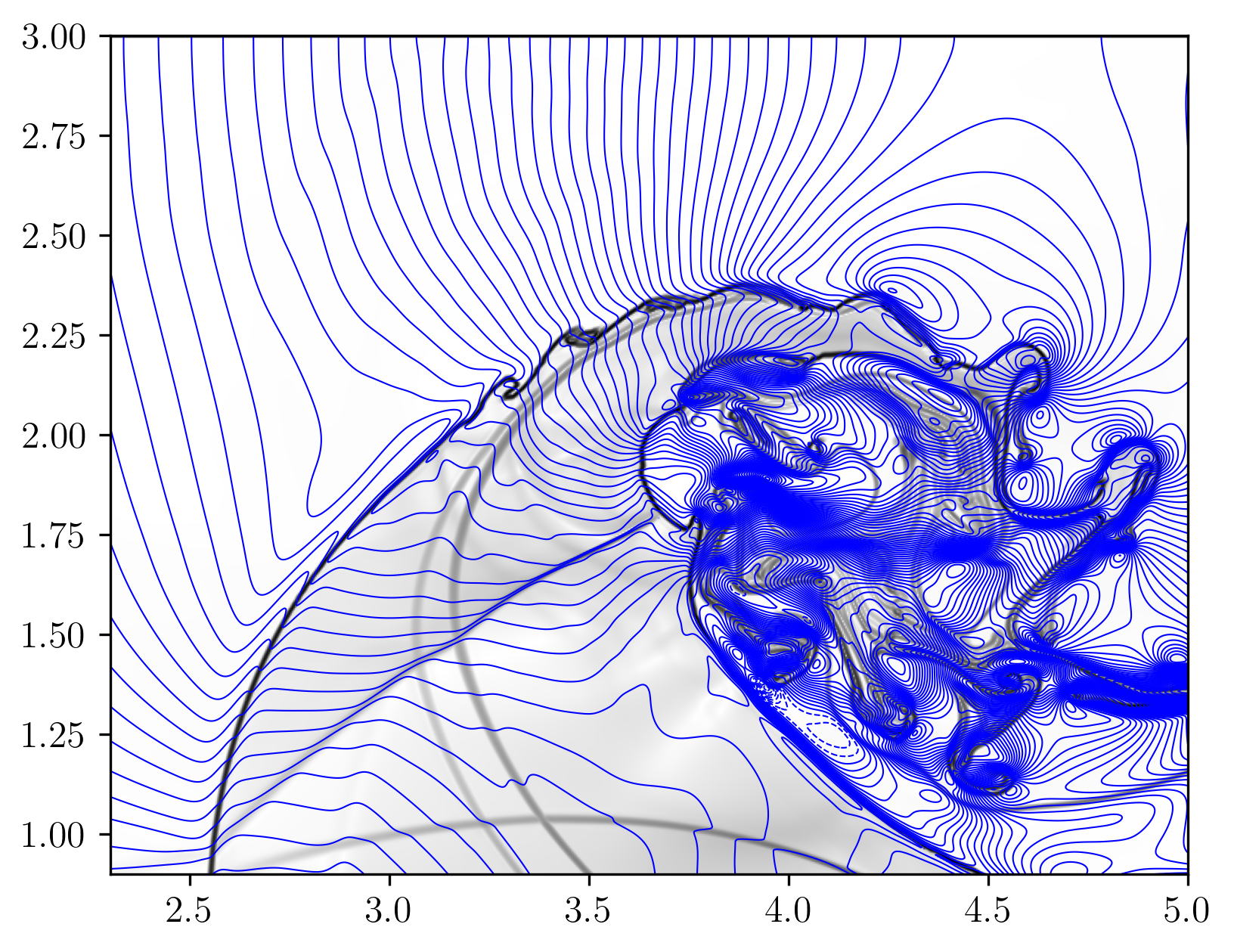}
\label{fig:hyt_p}}
\caption{Figures show density gradient contours for characteristic variable reconstruction, density gradient contours for primitive variable reconstruction, Ducros sensor locations in $x-$ and $y-$ directions, $v$-velocity contours overlayed on density gradient contours and pressure contours overlayed on density gradients, Example \ref{ex:triple}, on a grid resolution of 1792 $\times$ 768.}
\label{fig_tapas_viscous}
\end{figure}

Fig. \ref{fig:hyt_tc} shows the density gradient contours on a grid size of 1792 $\times$ 768 computed using characteristic variable reconstruction. As expected, the characteristic variable reconstruction is free of oscillations. Fig. \ref{fig:hyt_tp} shows the density gradient contours on a grid size of 1792 $\times$ 768 computed using primitive variable reconstruction, and there are mild oscillations behind the shock. Figs. \ref{fig:hyt_ducx} and \ref{fig:hyt_ducy} show regions where the Ducros sensor is activated; the contours indicate the Ducros sensor locations in $x-$ and $y-$ directions, respectively. The Ducros sensor correctly identified the shocks and did not detect the contact discontinuities, which indicates the tangential velocities computed using a central scheme across the material interfaces, and there are no oscillations. Figs. \ref{fig:hyt_v} and \ref{fig:hyt_p}  show the  $v$-velocity contours overlayed on density gradient contours and pressure contours overlayed on density gradients, respectively. The TENO-THINC scheme \cite{takagi2022novel} failed even for this test case, as the THINC scheme is used for velocity and pressure across material interfaces. These results indicate the proposed approach is physically consistent, oscillation free and robust for viscous compressible multi-species flows.

\begin{figure}[H]
\centering
 \includegraphics[width=1.0\textwidth]{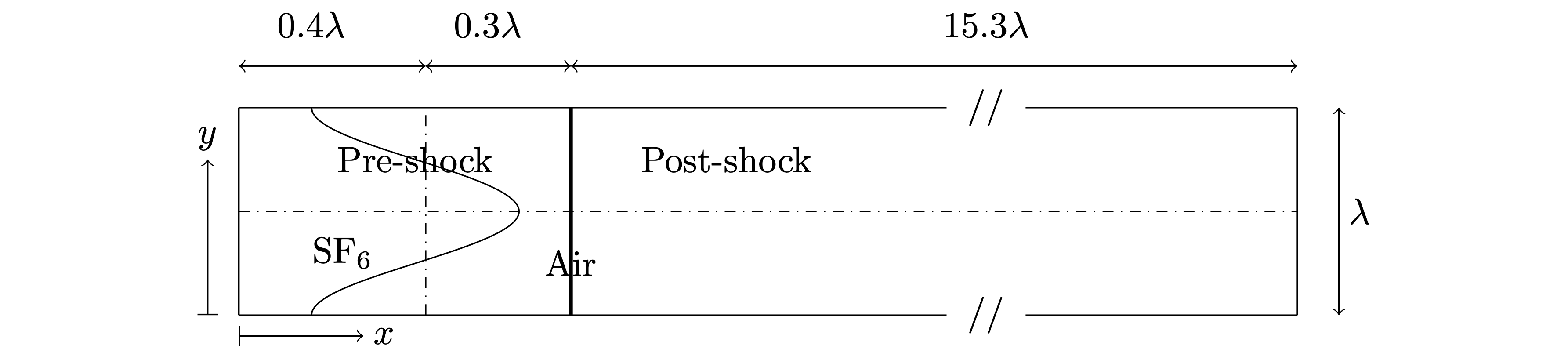}
\caption{Schematic of initial condition of Ricthmyer-Meshkov instability, Example \ref{ex:RM-viscous}.}
\label{rminstability}
\end{figure}

\begin{example}\label{ex:RM-viscous}{Two dimensional multi-species viscous Richtmeyer-Meshkov instability (Viscous case)}
\end{example}
In this test case, the two-dimensional viscous Richtmeyer-Meshkov (RMI) instability is computed \cite{chamarthi2023gradient}. RMI occurs when an incident shock accelerates an interface between two fluids of different densities.  As the shock wave hits the perturbed interface, it deforms and generates vortices due to the baroclinic effect. As time progresses, the $SF_6$, which is heavier, penetrates the air, the lighter fluid, leading to the formation of a spike. The computational domain, shown in Fig. \ref{rminstability}, for this test case, extends from $0.0 \leq x \leq 16.0\lambda$ and $0.0 \leq y \leq 1.0\lambda$ where $\lambda$ is the initial perturbation wavelength and the initial shape of the interface is given by 

\begin{equation}
\frac{x}{\lambda}=0.4-0.1 \sin \left(2 \pi\left(\frac{y}{\lambda}+0.25\right)\right),
\end{equation}
where, $\lambda$ =1.

\begin{figure}[H]
\centering\offinterlineskip
\subfigure[MP5, Coarse grid]{\includegraphics[width=0.17\textheight, angle =0]{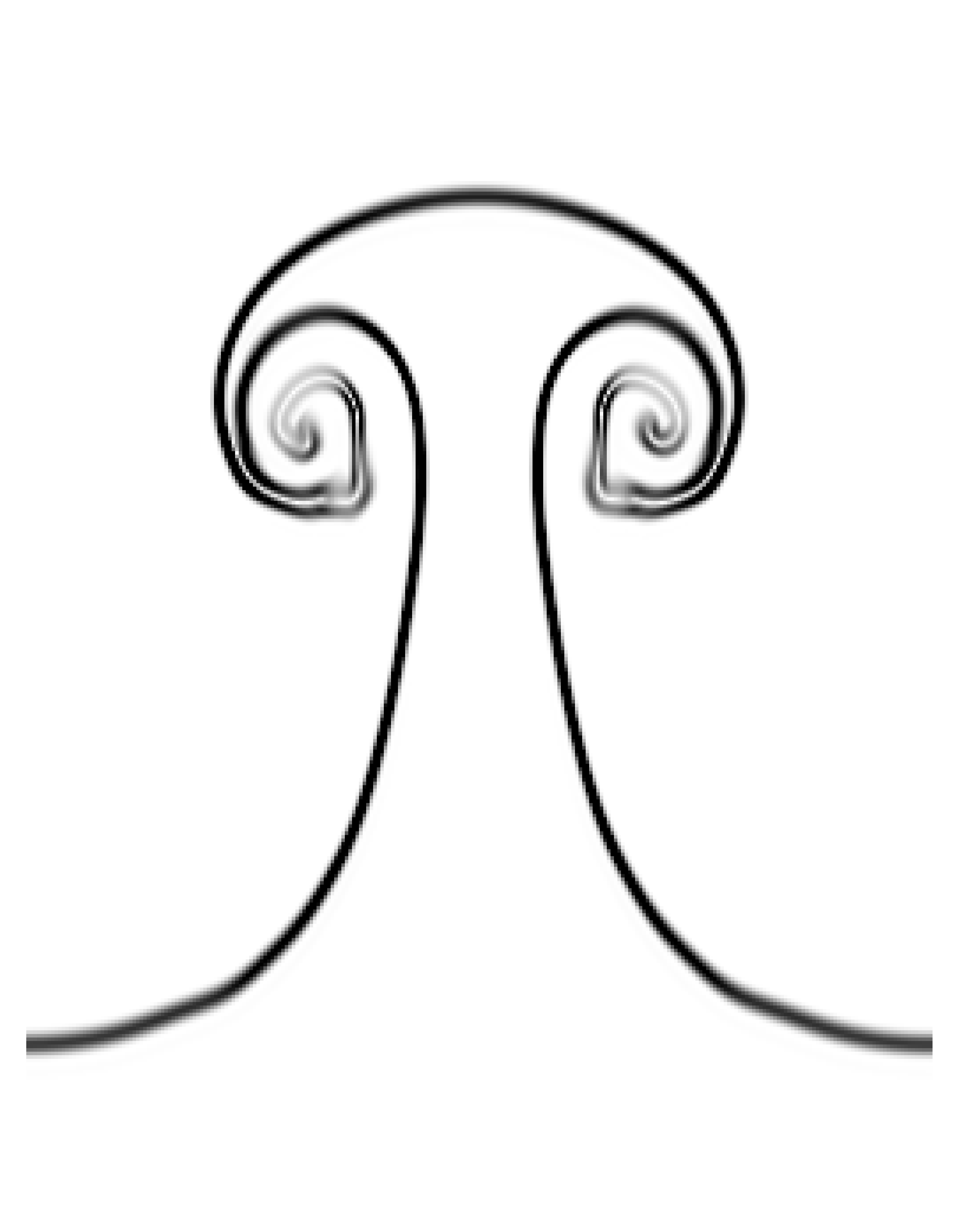}
\label{fig:MP_11_C_RMI}}
\subfigure[HY-THINC-D, Coarse grid]{\includegraphics[width=0.17\textheight, angle =0]{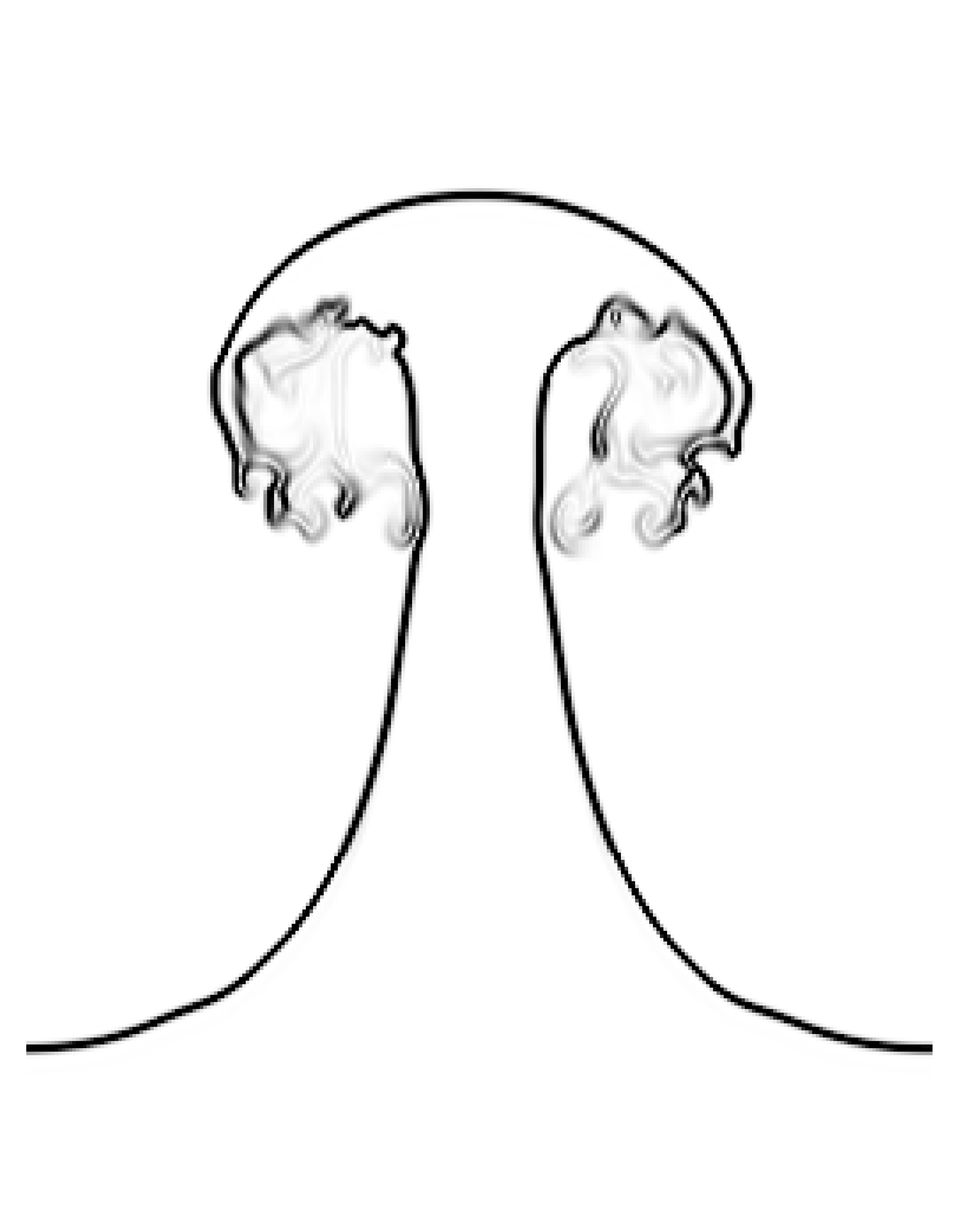}
\label{fig:HYTD_11_C_RMI}}
\subfigure[MP5, Fine grid]{\includegraphics[width=0.17\textheight, angle =0]{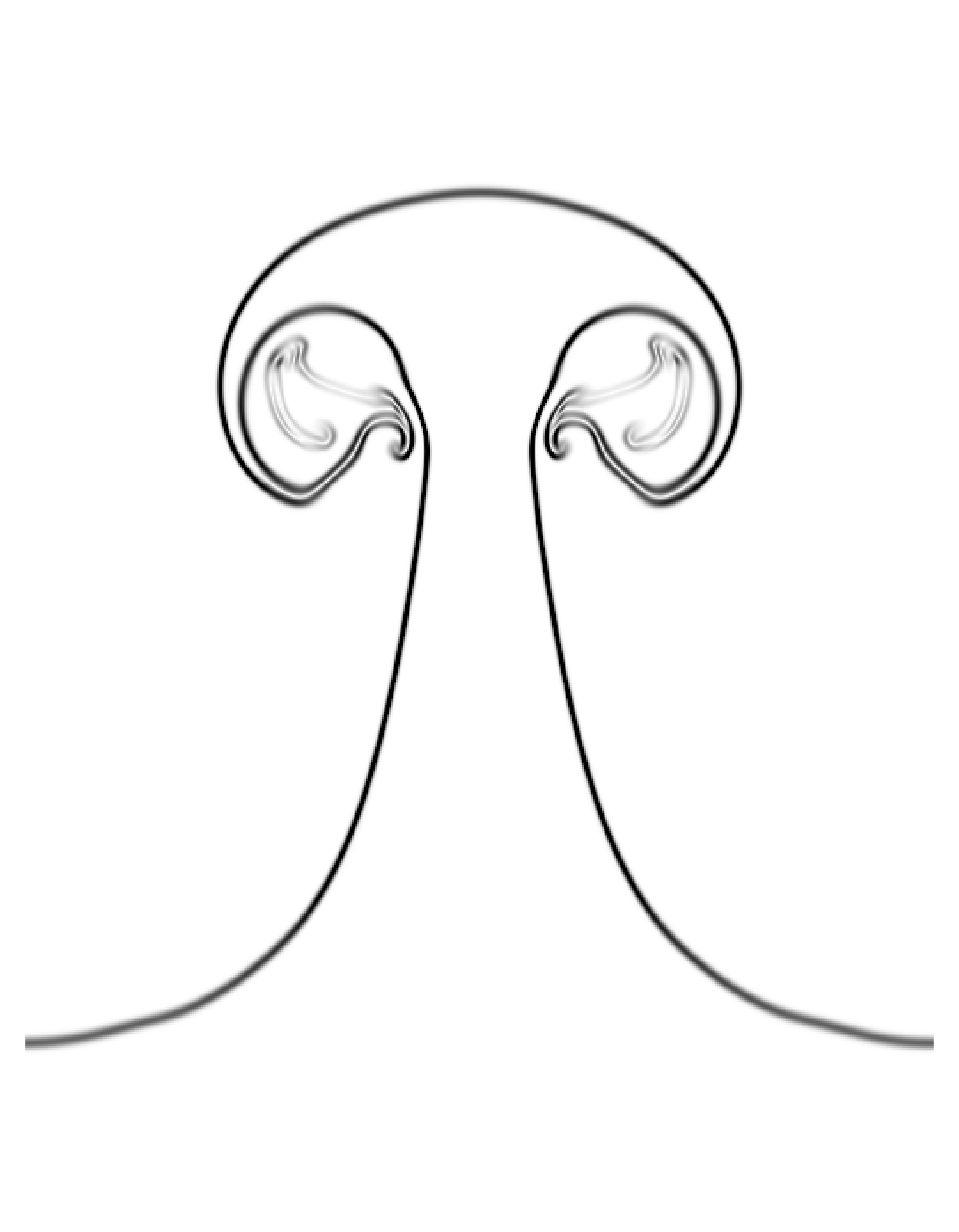}
\label{fig:MP_11_F_RMI}}
\subfigure[HY-THINC-D, Fine grid]{\includegraphics[width=0.17\textheight, angle =0]{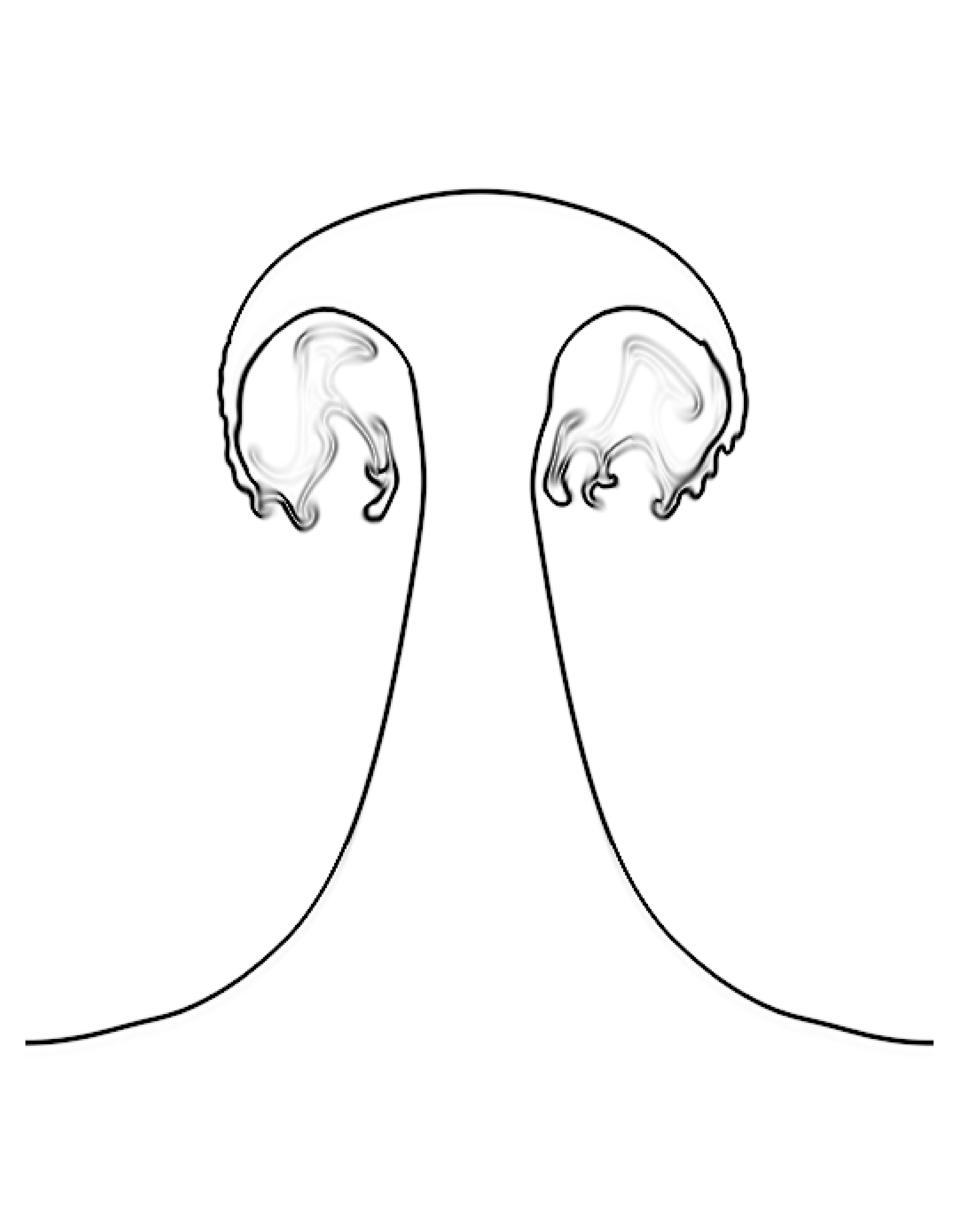}
\label{fig:HYTD_11_F_RMI}}
\caption{\textcolor{black}{Comparison of normalized density gradient magnitude, $\phi$, contours for two-dimensional viscous Richtmeyer-Meshkov instability problem in Example \ref{ex:RM-viscous}  on a grid size of 4096 $\times$ 256 and 8192 $\times$ 512. Contours are from 1 to 1.7 at time t$=$ 11.0 using the proposed scheme.}}
\label{fig_RM_viscous}
\end{figure}

The initial conditions for this test case are as follows:

\begin{equation}
(\rho, u, v, p, \gamma)=\left\{\begin{array}{ll}
(1,1.24,0,1 / 1.4,1.4), & \text { for pre-shock air } \\
(1.4112,0.8787,0,1.6272 / 1.4,1.4), & \text { for post-shock air } \\
(5.04,1.24,0,1 / 1.4,1.093), & \text { for } S F_{6}
\end{array}\right.
\end{equation}
In the current simulations, a constant dynamic viscosity of $\mu = 1.0 \times 10^{-4}$ is considered \cite{yee2007simulation, chamarthi2023gradient}. Periodic boundary conditions are applied at the top and bottom boundaries of the domain, and the initial values are set at the left and right boundaries. As a Cartesian grid is used in the present simulation, it may lead to generating secondary instabilities at the material interface, and to mitigate these secondary instabilities, the initial perturbation is smoothened by incorporating an artificial diffusion layer as proposed in Ref. \cite{Wong2017}:
\begin{equation}
\begin{array}{c}
f_{sm}=\frac{1}{2}\left(1+\operatorname{erf}\left(\frac{\Delta D}{E_{i} \sqrt{\Delta x \Delta y}}\right)\right) \\
u=u_{L}\left(1-f_{sm}\right)+u_{R} f_{sm}
\end{array}
\end{equation}
where $u$ represents the primitive variables near the initial interface, the parameter $E_i$ introduces additional thickness to the initial material interface, $\Delta D$ is the distance from the initial perturbed material interface, and subscripts $L$ and $R$ denote the left and right interface conditions. Parameter $E_i$ is chosen as 5 in this test case.
Simulations are conducted on two different grid sizes, 4096 $\times$ 256 cells and and 8192 $\times$ 512, with a constant CFL of 0.4. Computational results of normalized density gradient magnitude $\phi = $exp$(|\nabla \rho|/|\nabla \rho|_{max} )$ obtained at $t$ = 11.0 by various schemes are shown in Fig. \ref{fig_RM_viscous}.

Figs. (\ref{fig:HYTD_11_C_RMI} and \ref{fig:HYTD_11_F_RMI}) indicate no noticeable spurious oscillations for the HY-THINC-D scheme, and the interface thickness is thinner than the MP5 scheme (\ref{fig:MP_11_C_RMI} and \ref{fig:MP_11_F_RMI}) as the THINC computes the interfaces. HY-THINC-D scheme also shows improved resolution regarding the roll-up vortices, indicating the scheme's low numerical dissipation compared to the MP5 scheme. These results indicate that the proposed interface sharpening approach, with a contact discontinuity sensor, can capture material interface within a few cells, even for long-duration simulations. The simulations have no oscillations despite using a central scheme across the material interface.

\begin{example}\label{ex:multiple}{Shock wave interaction with a multi-material bubble (Inviscid case)}
\end{example}

\begin{figure}[H]
\centering\offinterlineskip
\subfigure[$t$  = 5.0 $\times$ $10^{-3}$]{\includegraphics[width=0.31\textwidth]{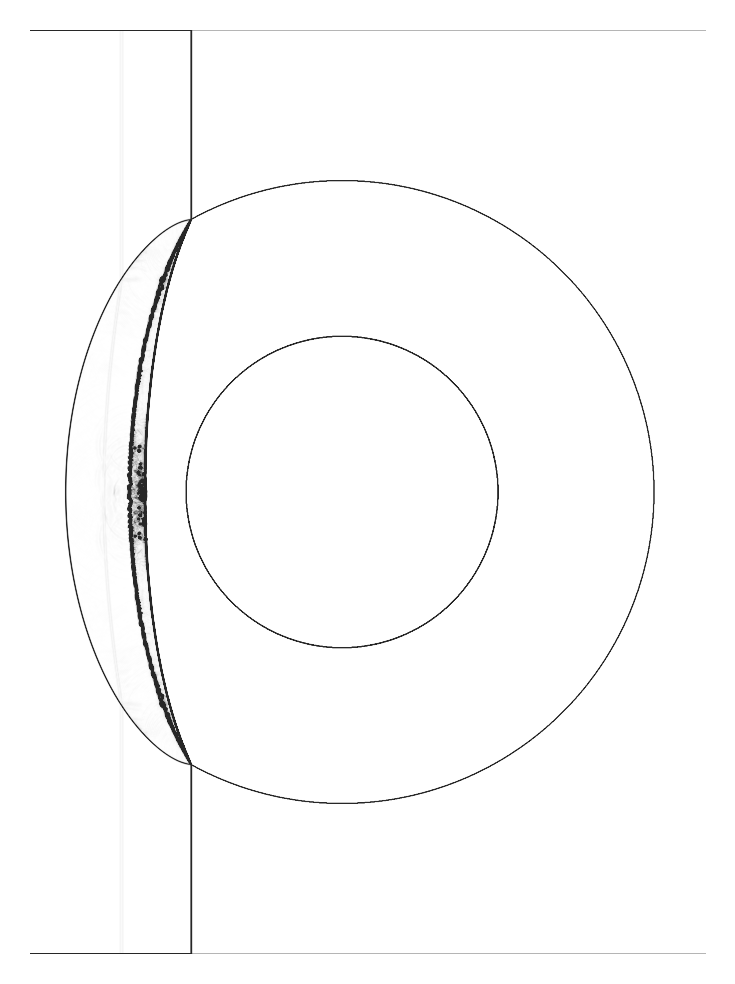}
\label{fig:hy5e-two}}
\subfigure[$t$  = 1.0 $\times$ $10^{-2}$]{\includegraphics[width=0.31\textwidth]{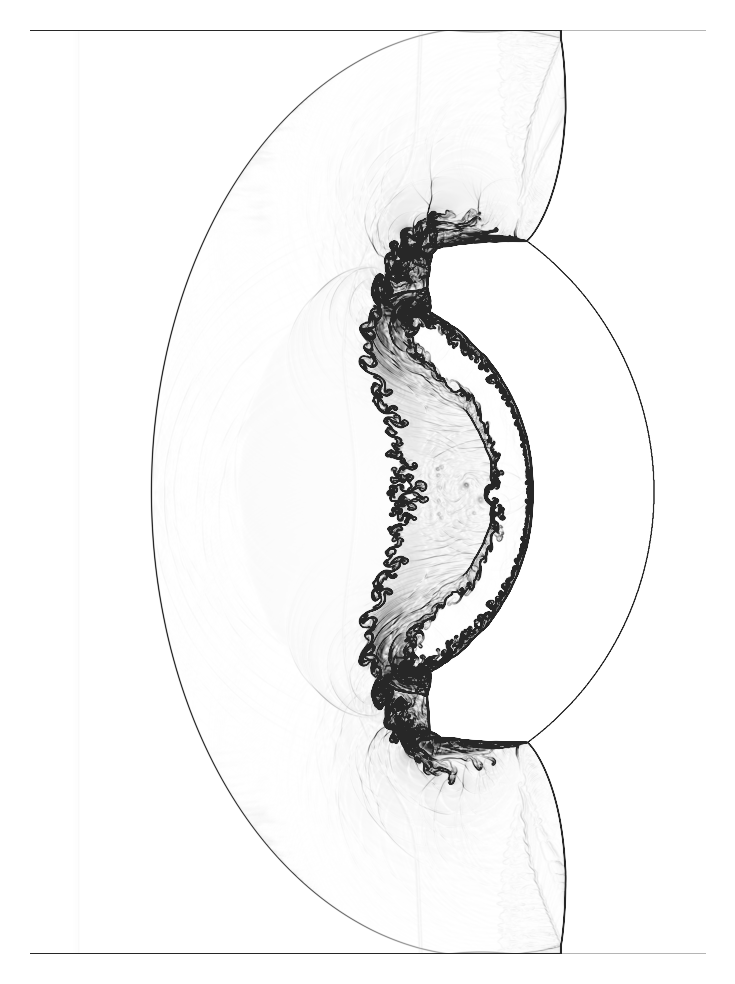}
\label{fig:hy10e-two}}
\subfigure[$t$  = 1.5 $\times$ $10^{-2}$]{\includegraphics[width=0.33\textwidth]{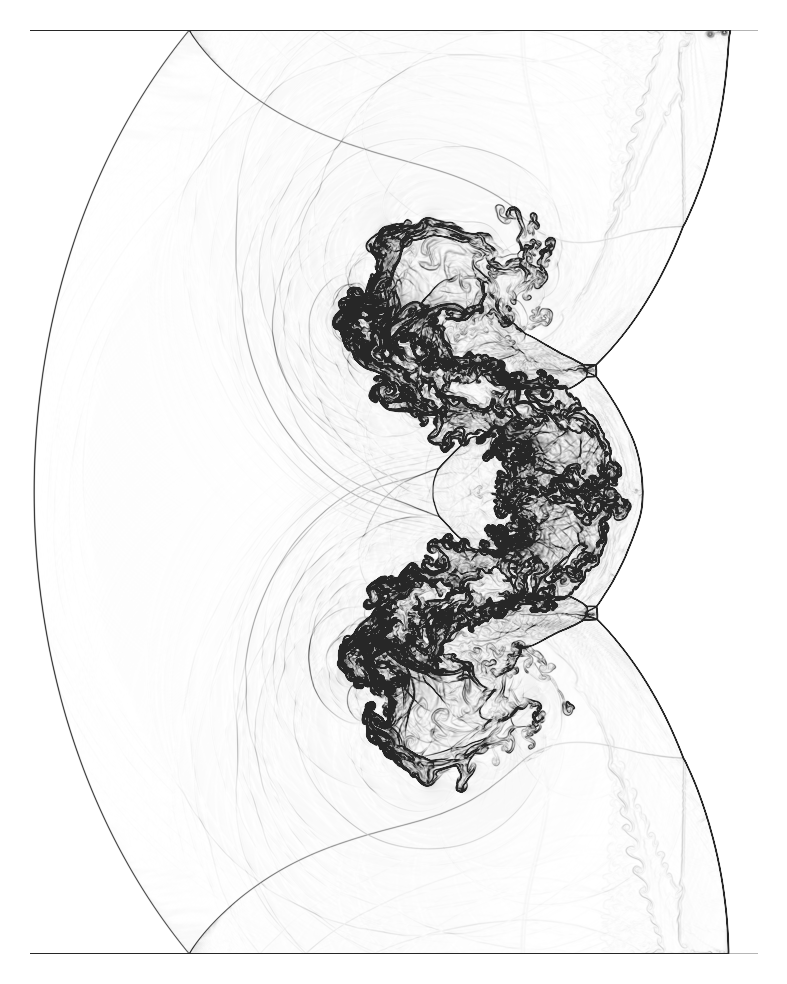}
\label{fig:hy15e-two}}
 \caption{Numerical Schlieren images \textcolor{black}{at times} $t$ =   5.0 $\times$ $10^{-3}$,  1.0 $\times$ $10^{-2}$, and 1.5 $\times$ $10^{-2}$ for the shock multiple bubble test case using HY-THINC scheme, Example \ref{ex:multiple}, on a grid resolution of 8192 $\times$ 2048, as in \cite{pan2018conservative}.}
 \label{fig:multi-1}
\end{figure}

\begin{figure}[H]
\centering\offinterlineskip
\subfigure[$t$  = 2.0 $\times$ $10^{-2}$]{\includegraphics[width=0.59\textwidth]{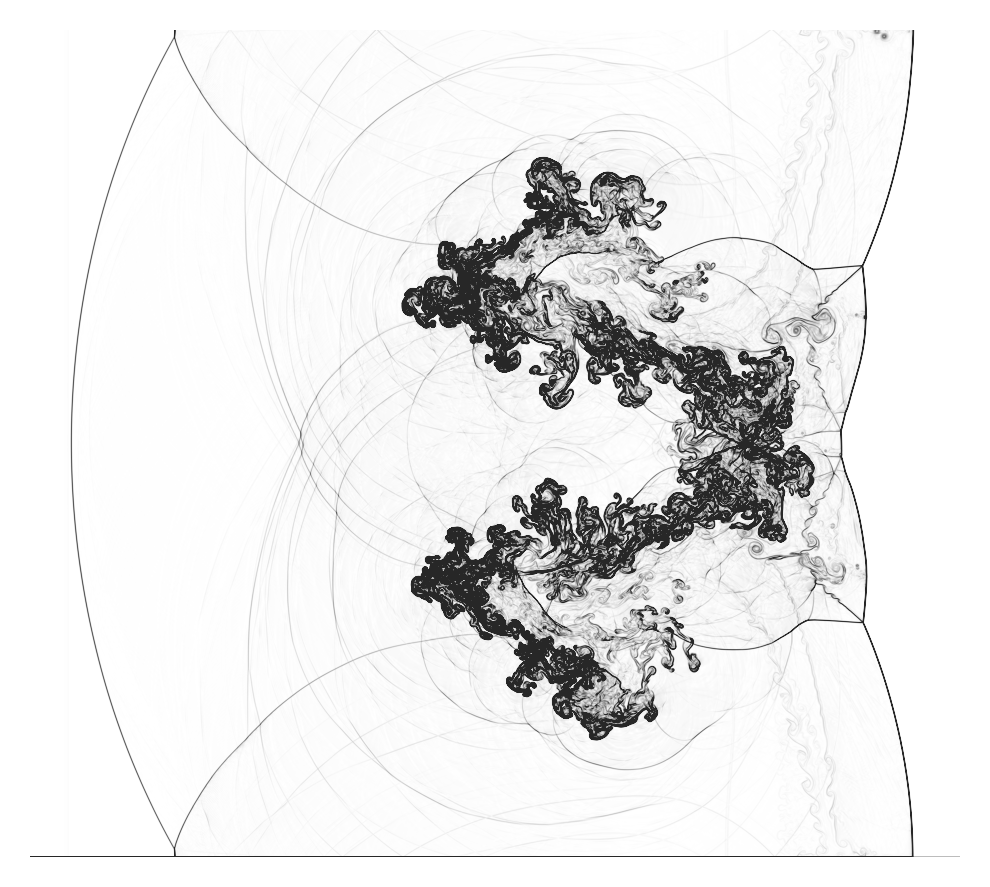}
\label{fig:hy20e-two}}
\subfigure[$t$  = 2.0 $\times$ $10^{-2}$]{\includegraphics[width=0.37\textwidth]{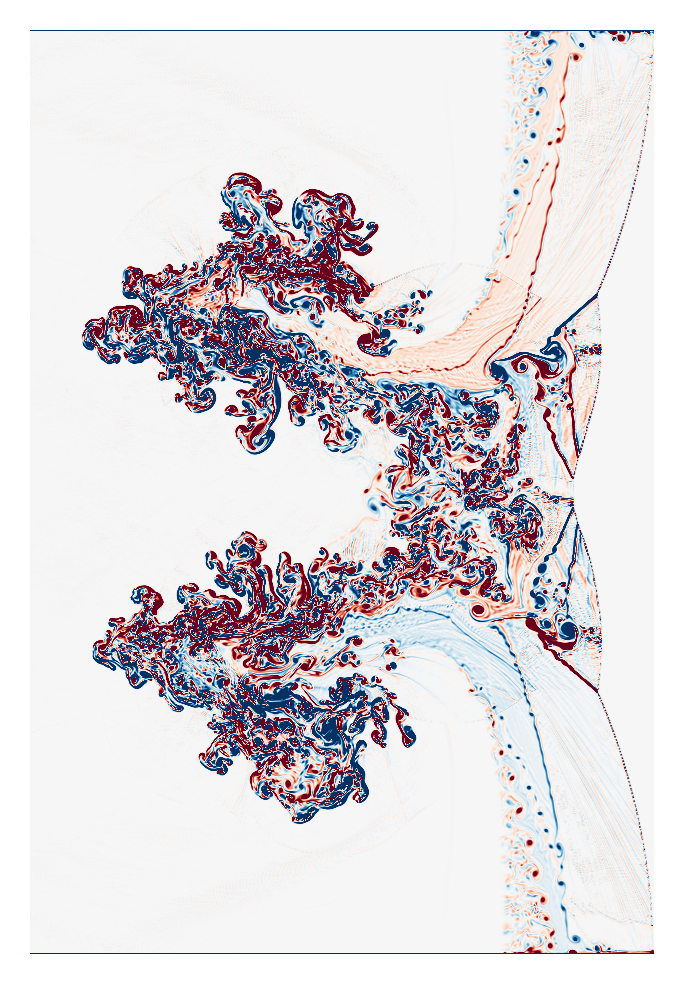}
\label{fig:hy20e-two_vort}}
 \caption{Numerical Schlieren images and corresponding Vorticity contours at $t$ = 2.0 $\times$ $10^{-2}$ $\times$ $10^{-2}$ for the shock multiple bubble test case using HY-THINC scheme, Example \ref{ex:multiple}, on a grid resolution of 8192 $\times$ 2048, as in \cite{pan2018conservative}.}
 \label{fig:multi-2}
\end{figure}

\begin{figure}[H]
\centering\offinterlineskip
\subfigure[$t$  = 3.0 $\times$ $10^{-2}$]{\includegraphics[width=0.46\textwidth]{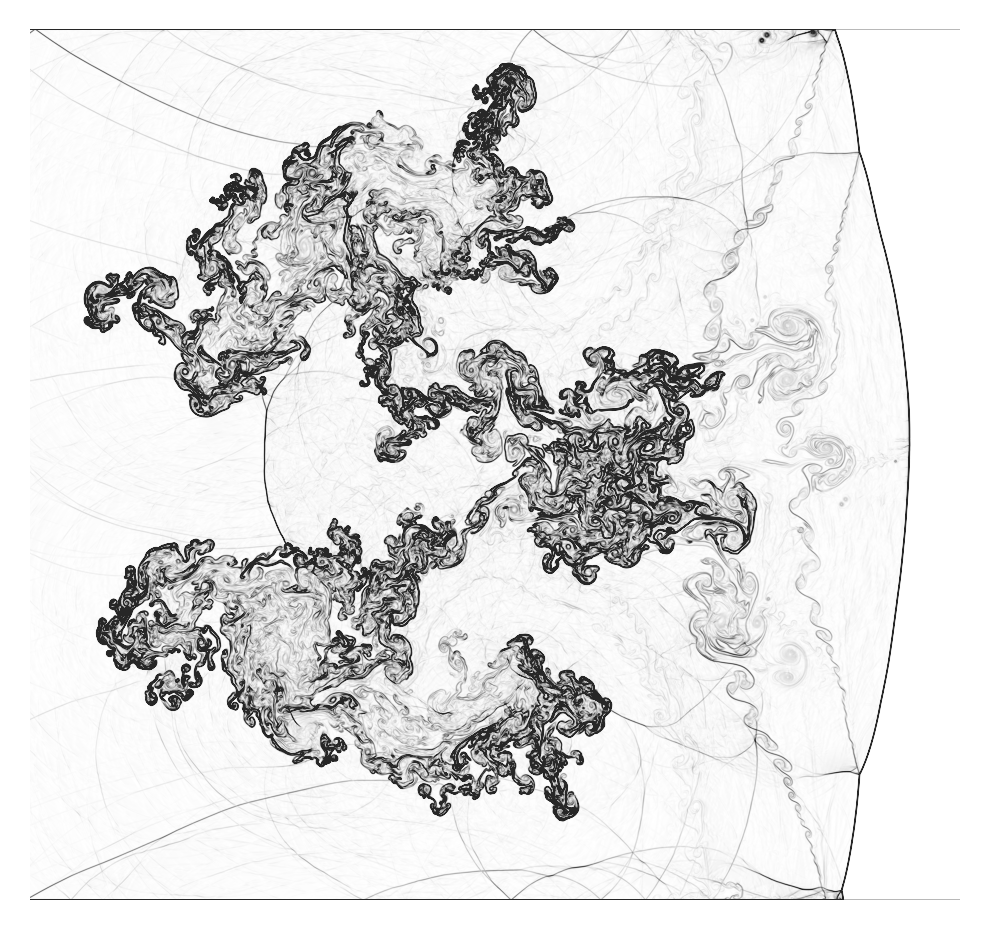}
\label{fig:hy30e-two}}
\subfigure[$t$  = 3.0 $\times$ $10^{-2}$]{\includegraphics[width=0.46\textwidth]{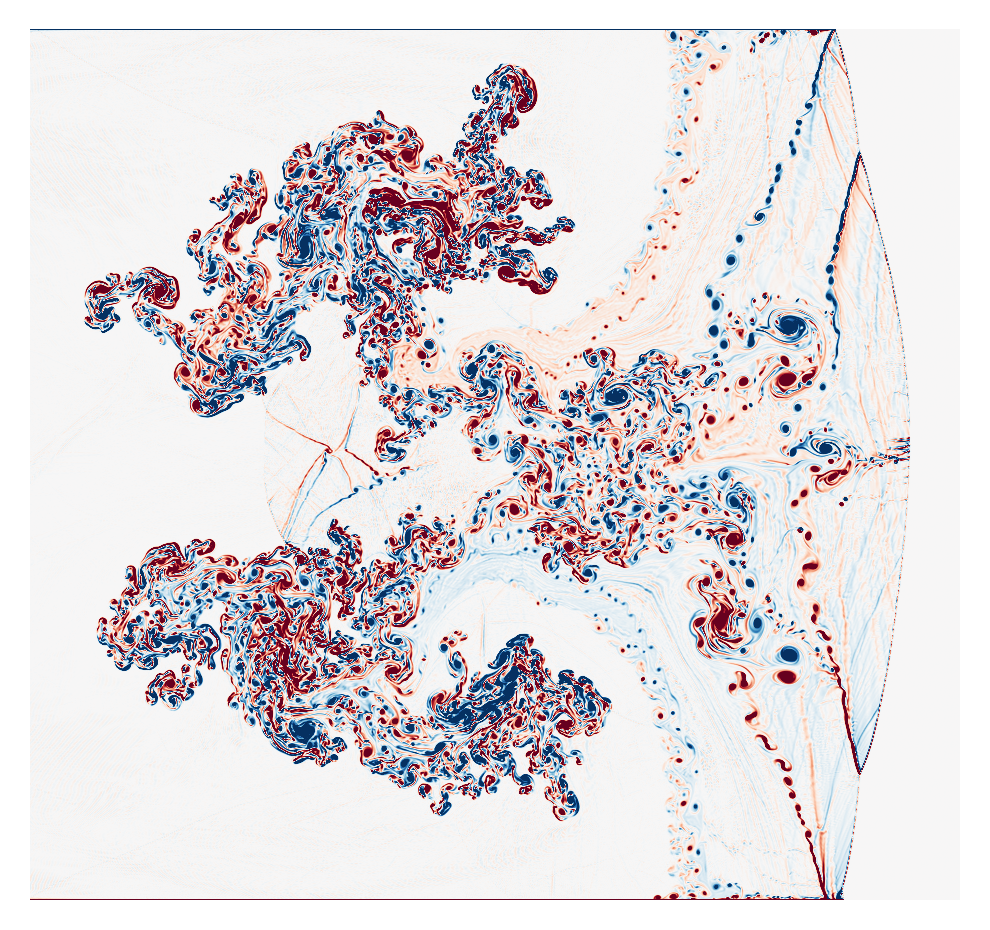}
\label{fig:hy30e-two_vort}}
 \caption{Numerical Schlieren images and corresponding Vorticity contours at $t$ =   3.0 $\times$ $10^{-2}$ for the shock multiple bubble test case using HY-THINC scheme, Example \ref{ex:multiple}, on a grid resolution of 8192 $\times$ 2048, as in \cite{pan2018conservative}.}
 \label{fig:multi-3}
\end{figure}

In this test case, a Mach 6.0 shock wave in the air meets a cylindrical helium bubble enclosed by an R22 shell considered in Ref. \cite{pan2018conservative}. It is a complex flow combining two test cases: shock interaction with a helium bubble \cite{coralic2014finite,quirk1996dynamics} and shock interaction with the R22 bubble \cite{shyue2014eulerian}. The computational domain for this test case spans [0, 0.356] $\times$ [0, 0.089]. The initial states for this test case are given by:

\begin{equation}
(\rho, u, v, p, \gamma)=\left\{\begin{array}{ll}
(5.268,5.752,0,41.83,1.400), & \text { for post-shock air, } \\
(1.000,0.000,0,\ 1.00,1.400), & \text { for pre-shock air ,}\\
(3.154,0.000,0,\ 1.00,1.249), & \text { for R22 shell ,}\\
(0.138,0.000,0,\ 1.00,1.667), & \text { for helium bubble, }
\end{array}\right.
\end{equation}
where the shock is placed initially at $x = 0.1$. A helium bubble with an initial radius of $0.15$ is placed at $x = 0.15$, $y = 0.0445$. The outer R22 shell has a radius of $0.30$ at the exact location. Symmetry boundary conditions are set for upper and lower edges, while the left and right boundaries have inflow and outflow conditions. Figs. \ref{fig:multi-1}-\ref{fig:multi-3} show the evolving density gradients and vorticity contours at five different time instances, offering insight into the progression of the shock system and the deformation of the helium and R22 bubbles obtained by the HY-THINC approach.

 At $t = 5.0 \times 10^{-3}$, shown in Fig. \ref{fig:hy5e-two}, we observe the shock wave's behaviour as it refracts upon encountering the R22 shell, creating a concave transmitted shock wave due to differences in acoustic impedance. This behaviour aligns with findings in previous research \cite{shyue2014eulerian,deng2018high}. At $t = 1.0 \times 10^{-2}$, shown in Fig. \ref{fig:hy10e-two}, we observe the shock wave's behaviour as it impacts the helium bubble. Unlike the R22 bubble, the air-helium interface produces a convex transmitted shock inside the helium bubble. As this shock propagates further downstream, it subsequently impacts the aft end of the helium bubble. From $t = 1.5 \times 10^{-2}$ to $t = 3.0 \times 10^{-2}$, R22 and helium are mixed with the ambient air resulting in complex vortical structures, as shown in Figs. \ref{fig:multi-2} and \ref{fig:multi-3}. The deformed helium bubble and R22 have a similar shape to previous numerical results of Pan et al. \cite{pan2018conservative}. Numerical simulations carried out with the standard MP failed to pass this test case due to negative density and pressure beyond $t$ =  1.0 $\times$ $10^{-2}$, indicating the proposed approach's robustness. For this test case, the ``carbuncle'' phenomenon is observed due to the use of the HLLC Riemann solver, and it \textit{may be} avoided by using a rotated HLLC-HLL or any other carbuncle-free approach. For this test case, Paula et al. \cite{paula2023robust} have conducted a three-dimensional inviscid simulation using the discrete-equations method on a grid size of 452 million grid points. It is beyond the scope of the paper and the resources available to the author to conduct a viscous simulation for this test case, as it requires hundreds of millions of grid points.

\section{Conclusions}\label{sec:conclusions}

The paper proposed a physically consistent, to the extent possible, numerical discretization approach for simulating viscous compressible multicomponent flows. The key contributions and observations of the manuscript are summarized below:

\begin{itemize}
	\item A contact discontinuity detector is proposed such that the THINC scheme can be used for the contact discontinuities and material interfaces. The detector uses the variable $s$, where $s=\frac{p}{\rho^{\gamma}}$. The detector is devised in such a way that it avoids high-frequency regions so that THINC is not applied in those regions. The study demonstrated the effectiveness of this approach through a series of benchmark tests, showcasing its ability to capture material interfaces and significantly outperform existing methods (WENO and MP).
	\item The proposed approach does not rely on volume fraction criteria to identify interfaces, which can be tedious if several species are in the domain. It can robustly identify the material interfaces (and contact discontinuity within a material) even if there are more than two species. For the hypersonic flow test case, Example \ref {ex:multiple}, where the standard shock-capturing technique failed, the proposed approach completed simulations, indicating the method's robustness.
	\item For both viscous and inviscid simulations, the THINC is applied only to the phasic densities and the volume fractions in physical space and only to entropy wave and volume fractions in characteristic space (such an approach is also consistent with the physics across a material interface \cite{batchelor1967introduction}).
\item For viscous simulations, the tangential velocities are computed using a central scheme across contact discontinuities using the Ducros sensor (a shock detector that cannot detect material interfaces) as they are continuous across the contact discontinuities. Using a central scheme did not lead to any oscillations. 
\item For shock-free viscous test cases, tangential velocities and pressure can be computed using a central scheme without any sensor, as they are continuous across the contact discontinuities. Nonlinear reconstruction techniques are required only for phasic densities and volume fractions.
\end{itemize}

\section*{Appendix}
\renewcommand{\thesubsection}{\Alph{append}}
\renewcommand{\thesubsection}{\Alph{subsection}}
\subsection{Application of contact discontinuity sensor with WENO scheme:}\label{append-a}
This appendix presents the results obtained using the WENO scheme (instead of the MP scheme) in conjunction with the proposed contact discontinuity detector and the THINC scheme . The WENO scheme is also briefly explained for clarity. In the WENO scheme, the fifth-order upwind-biased reconstruction is nonlinearly weighted from three different third-order sub-stencils. For simplicity, the reconstruction polynomials to the left side of the cell interface at $x_{i+\frac{1}{2}}$ are only presented here. The three-third order reconstruction formula of variable $U$ is given by

\begin{equation}
\begin{aligned} \label{eq:upwind_biased_stencils}
{\bar U}_{i+\frac{1}{2}}^{(0)} &= \frac{1}{6}\left(2U_{i-2} - 7U_{i-1} + 11U_{i} \right),\\
{\bar U}_{i+\frac{1}{2}}^{(1)} &= \frac{1}{6}\left(-U_{i-1} + 5U_{i} + 2U_{i+1} \right), \\
{\bar U}_{i+\frac{1}{2}}^{(2)} &= \frac{1}{6}\left(2U_{i} + 5U_{i+1} - U_{i+2} \right).
\end{aligned}
\end{equation}
The values ${\bar U}_{i+\frac{1}{2}}^{(k)}$ at cell interfaces are approximated from different sub-stencils, while $U_i$ represents the cell-averaged values at cell centers. The three third-order upwind approximation polynomials in Eqn. \ref{eq:upwind_biased_stencils} are dynamically chosen through a nonlinear convex combination. This adaptation occurs to employ a lower-order spatial discretization that avoids interpolation across discontinuities and provides the necessary numerical dissipation for shock capturing. The fifth-order WENO-Z scheme \cite{Borges2008} used in this paper is as follows:

\begin{equation}
{\bar U}_{i+\frac{1}{2}} = \sum\limits_{k=0}^{2} \omega^z_k {\bar U}_{i+\frac{1}{2}}^{(k)},
\end{equation}
where $\omega^z_k$ are the nonlinear weights which are given by,
\begin{align}\label{eqn:om_z}
\omega^z_k=\dfrac{\alpha^z_k}{\sum_{k=0}^2\alpha^z_k},\quad \alpha^z_{k}=\gamma_k\left(1+\left(\dfrac{\tau_5}{\epsilon+\beta_k}\right)^p\right), \tau_5=|\beta_0-\beta_2|, p=1,
\end{align}
where $\gamma_k$ and $\beta_k$ are ideal linear weights and smoothness indicators, respectively. $\epsilon=10^{-20}$ is a small constant to prevent division by zero. The non-linear weights of the convex combination are based on local smoothness indicators $\beta_k$. These indicators measure the sum of the normalized squares of the scaled $L^2$ norms of all derivatives of the lower-order polynomials. The goal is to assign small weights to lower-order polynomials with discontinuities in their underlying stencils, resulting in a non-oscillatory solution. Smoothness indicators $\beta_k$ are as follows:

\begin{equation}
\begin{aligned} \label{eq:smoothness}
\beta_0 &= \frac{1}{4} \left(  U_{i-2} - 4U_{i-1} +3  U_{i} \right)^2 + \frac{13}{12} \left(  U_{i-2} - 2U_{i-1} +  U_{i} \right)^2,\\
\beta_1 &= \frac{1}{4} \left(  U_{i-1} - U_{i+1} \right)^2  + \frac{13}{12} \left( U_{i-1} - 2 \bar U_{i} +  U_{i+1} \right)^2, \\
\beta_2 &= \frac{1}{4} \left( 3U_{i} - 4U_{i+1} + U_{i+2} \right)^2+ \frac{13}{12} \left( U_{i} - 2  U_{i+1} + U_{i+2} \right)^2.
\end{aligned}
\end{equation}

\begin{figure}[H]
%\begin{halfspacing}
\centering\offinterlineskip
\subfigure[WENO5-Z]{\includegraphics[width=0.38\textwidth]{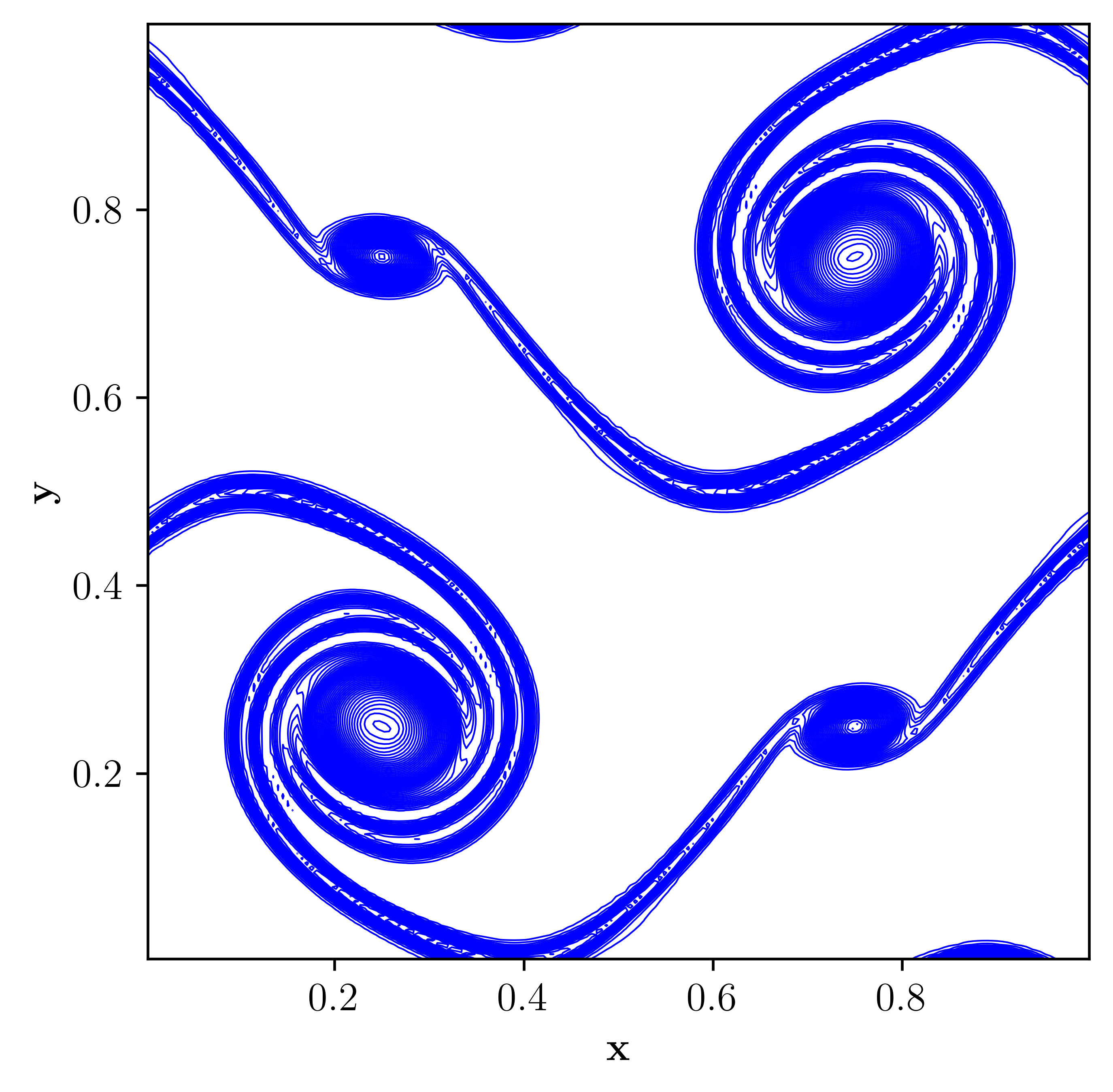}
\label{fig:weno5_dsl}}
\subfigure[WENO-Z-THINC-Ducros]{\includegraphics[width=0.38\textwidth]{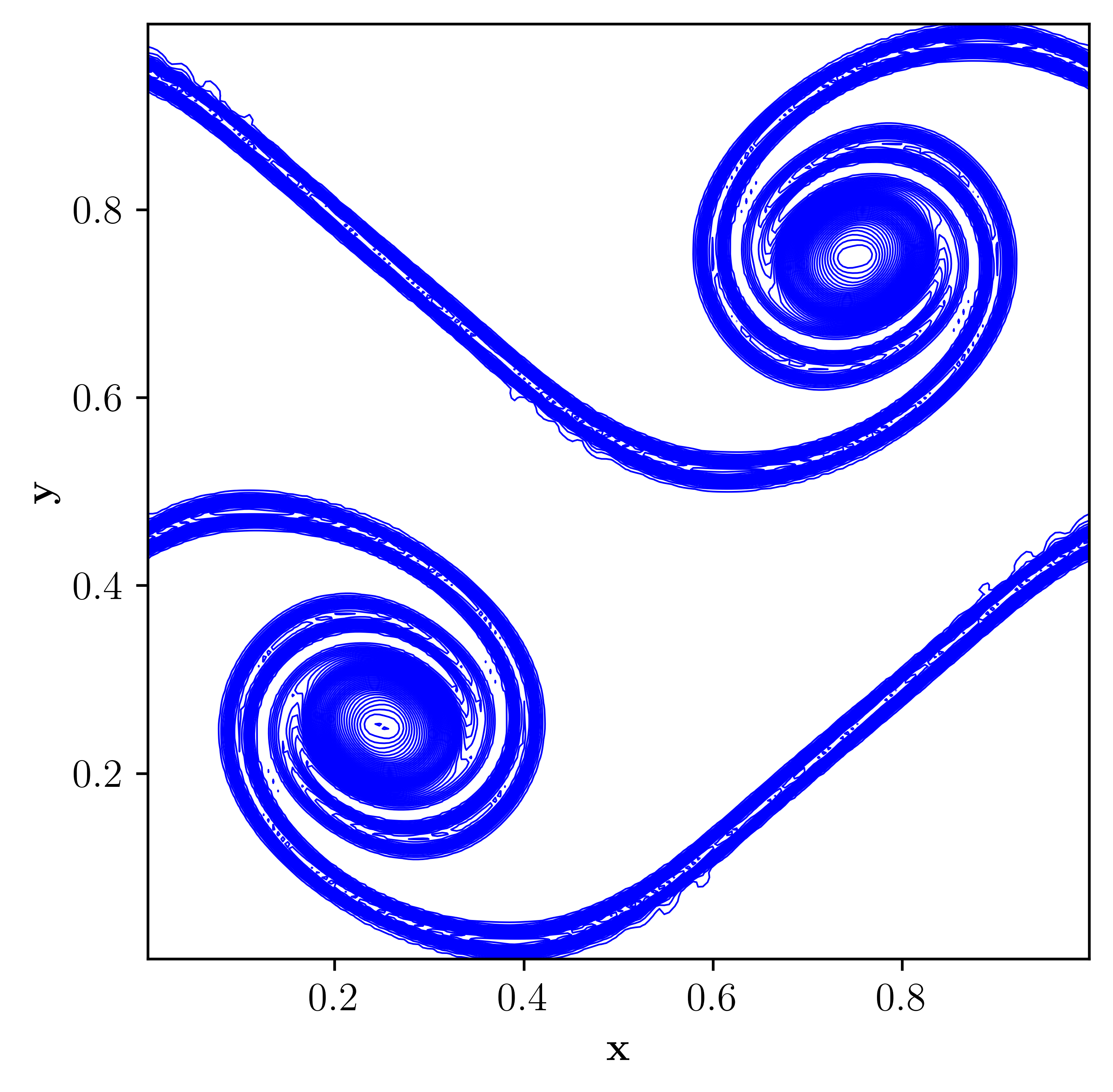}
\label{fig:weno_d_t}}
\caption{$z$-vorticity contours of the considered schemes computed on a grid size of $196^2$, Example \ref{ex:dsl}.}
    \label{fig:dpsl_196_we}
%\end{halfspacing}
\end{figure}

The parameters $a$ and $b$ in the contact discontinuity detector, Equation (\ref{detector-new}), are shown below to indicate the similarities between the smoothness indicators of the WENO and the $a$ and $b$ in the contact discontinuity detector.

\begin{equation}
\begin{aligned}
&a  = \frac{13}{12} \left|s_{i-2} - 2 s_{i-1} + s_{i}\right| + \frac{1}{4} \left|s_{i-2} - 4 s_{i-1} + 3 s_{i}\right|,\\
&b  = \frac{13}{12} \left|s_{i} - 2 s_{i+1} + s_{i+2}\right| + \frac{1}{4} \left|3 s_{i} - 4 s_{i+1} + s_{i+2}\right|, \ \text{where} \ s=\frac{p}{\rho^{\gamma}}, \ \text{and}\ \rho=\rho_{1} \alpha_{1}+\rho_{2} \alpha_{2}.
\end{aligned}
\end{equation}
 $a$ and $b$ are infact $\beta_0$ and $\beta_2$ without the squares and different choice of variable. To the author's knowledge, these smoothness indicators are not used to detect contact discontinuities in this manner, so the THINC scheme can be applied robustly to improve the resolution of contact discontinuities. Fig.  \ref{fig:dpsl_196_we} shows the $z$-vorticity contours of the WENO5-Z and the WENO-Z-THINC-Ducros schemes for the periodic shear layer test case, Example \ref{ex:dsl}. As expected, the WENO-Z-THINC-Ducros approach is free of spurious vortices, unlike that of WENO5-Z, as the tangential velocities are computed using the central scheme.

\begin{figure}[H]
\centering
\subfigure[\textcolor{black}{WENOZ.}]{\includegraphics[width=0.48\textwidth]{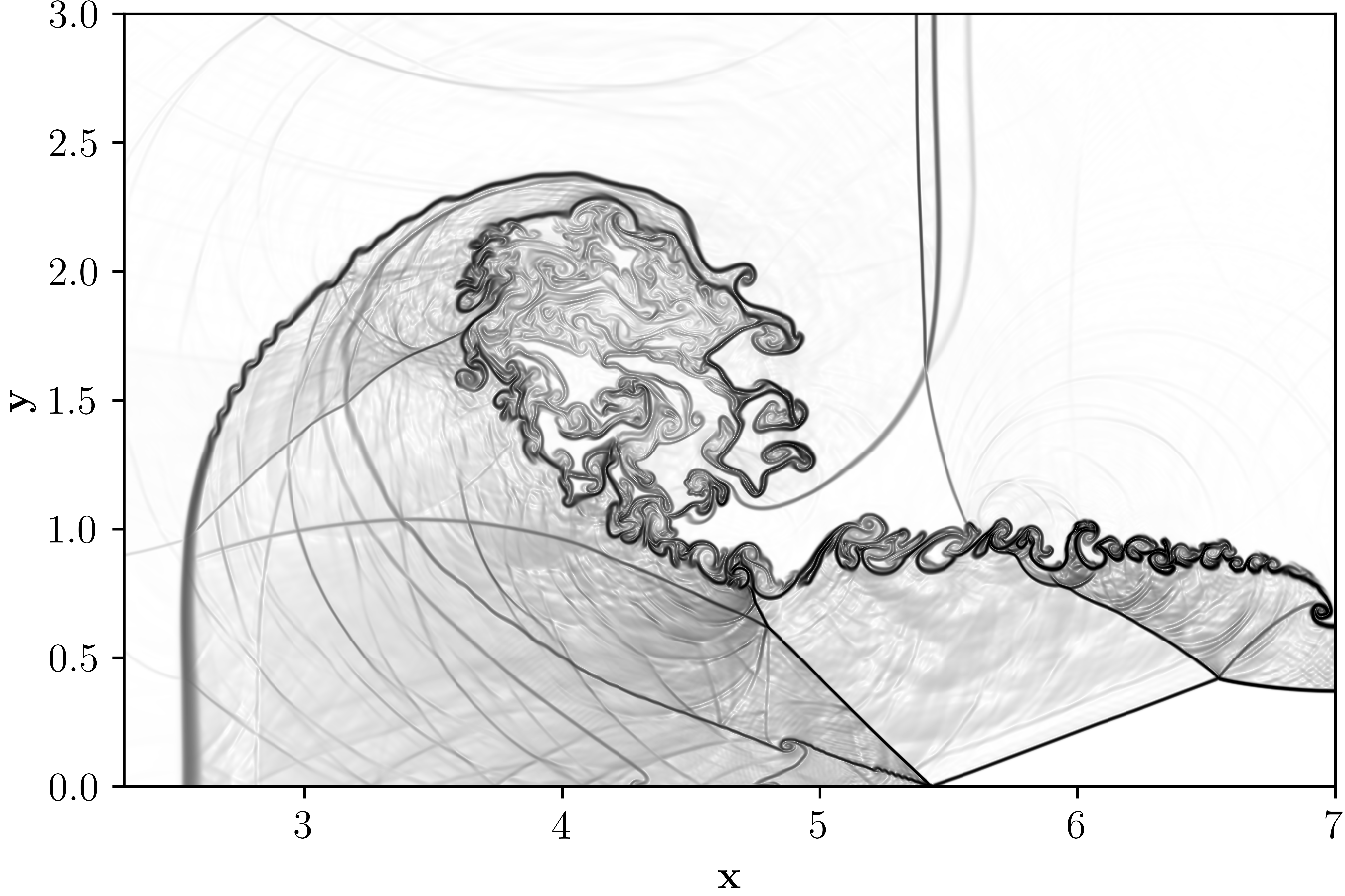}
\label{fig:weno_tp}}
\subfigure[\textcolor{black}{WENO-Z-THINC.}]{\includegraphics[width=0.48\textwidth]{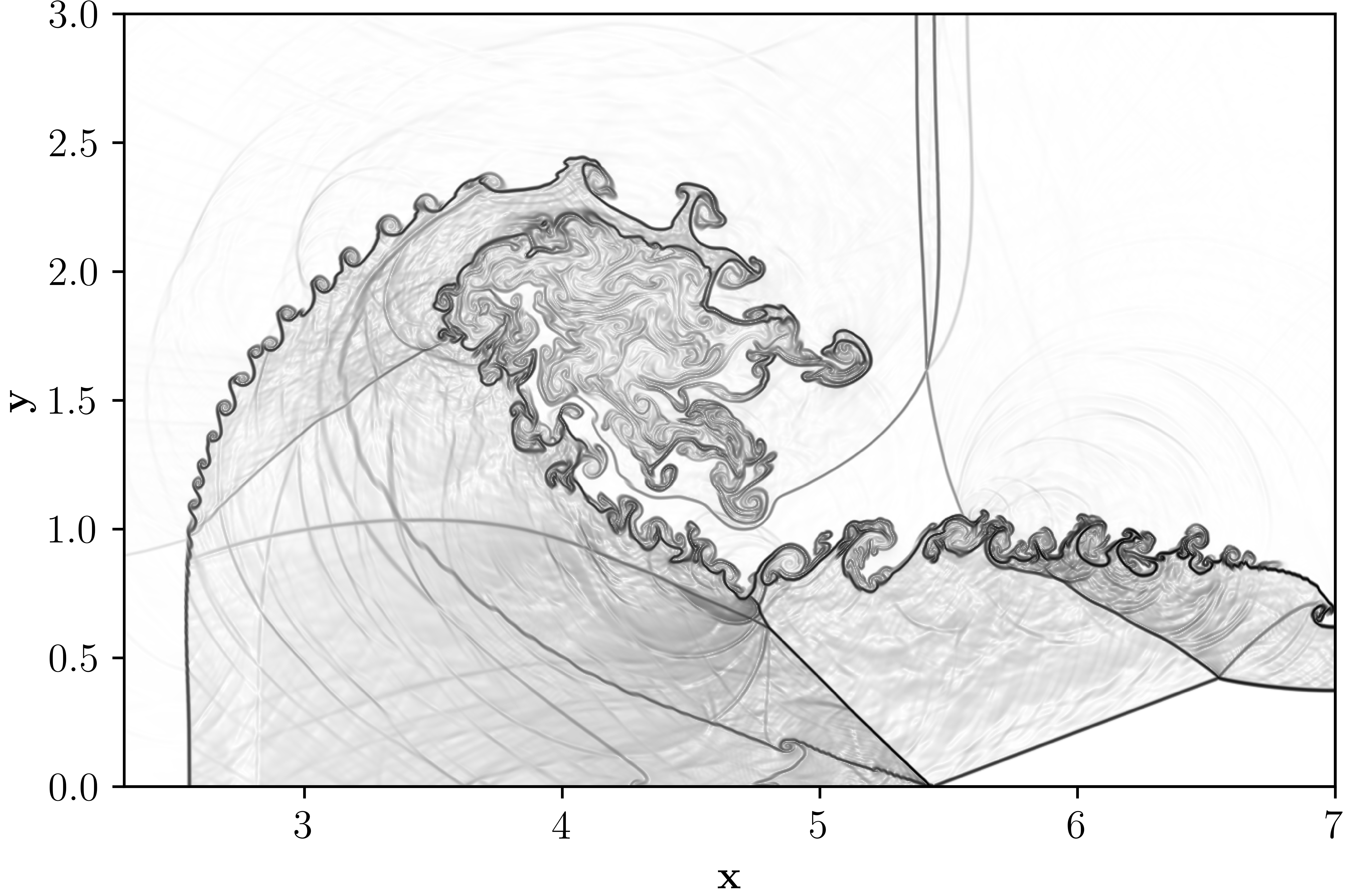}
\label{fig:wenot_tc}}
\caption{\textcolor{black}{Density gradient contours at time $t=5$ using various schemes, Example \ref{ex:triple}, on a grid resolution of 1792 $\times$ 768.}}
\label{fig_tapas_weno}
\end{figure}

 Fig.  \ref{fig_tapas_weno} shows the density gradient contours, inviscid scenario, obtained by the WENO-Z and WENO-Z THINC scheme at $t=5$ for Example \ref{ex:triple}. It can be observed that the WENO-Z-THINC scheme, Fig. \ref{fig:wenot_tc}, captured the material interface within a few cells compared to the base WENO-Z scheme, Fig. \ref{fig:weno_tp}, based on the contact discontinuity thickness. These results indicate the benefits of the proposed approach, where the THINC scheme can be used with two different approaches and produce oscillation-free and sharp numerical results.

%\subsection{\textcolor{black}{Analysis of the new discontinuity sensor}}\label{sec-appc}

\bibliographystyle{elsarticle-num}
\bibliography{contact_ref}

\end{document}